\documentclass[a4paper,onecolumn,10pt]{quantumarticle}
\pdfoutput=1
\usepackage[utf8]{inputenc}
\usepackage[english]{babel}
\usepackage[T1]{fontenc}
\usepackage{amsmath}
\usepackage{amssymb}
\usepackage{hyperref}
\usepackage{enumitem}
\usepackage[numbers,sort&compress]{natbib}
\usepackage{braket}
\usepackage{lineno}
\usepackage{cleveref}
\usepackage{tabularx}
\usepackage[dvipsnames]{xcolor}
\usepackage{tikz}
\usepackage{lipsum}
\usepackage{caption} 
\usepackage{subcaption}

\usetikzlibrary{positioning}
\usetikzlibrary{shapes.geometric, arrows}
\usetikzlibrary{matrix}

\usepackage{amsthm}
\usepackage{arydshln}

\let\mod\relax
\DeclareMathOperator{\mod}{mod}

\newtheoremstyle{break}%
    {}{}%
    {}{}%
    {\bfseries}{}
    {\newline}{}

\theoremstyle{break}

\newtheorem{example}{Example}[section]

\theoremstyle{remark}

\usepackage{algpseudocode}
\usepackage{algorithm}

\usepackage{multirow}
\usepackage{xspace}

\begin{document}
\author{Mark Webster}
\affiliation{Department of Physics \& Astronomy, University College London, London, WC1E 6BT, United Kingdom}
\thanks{\ Corresponding author: \url{mark.acacia@gmail.com}}
\author{Abraham Jacob}
\affiliation{Department of Physics \& Astronomy, University College London, London, WC1E 6BT, United Kingdom}
\author{Oscar Higgott}
\affiliation{Google Quantum AI}

\title{Distance-Finding Algorithms for Quantum Codes and Circuits}
\maketitle
\begin{abstract}
The distance of a classical or quantum code is a key figure of merit which reflects its capacity to detect errors. 
Quantum LDPC code families have considerable promise in reducing the  overhead required for fault-tolerant quantum computation, but calculating their distance is challenging with existing methods.
We generally assess the performance of a quantum code under  circuit level error models, and for such scenarios the circuit distance is an important consideration. 
Calculating circuit distance is in general more difficult than finding the distance of the corresponding code as the detector error matrix of the circuit is usually much larger than the code's check matrix.
In this work, we benchmark a wide range of distance-finding methods for various classical and quantum code families, as well as syndrome-extraction circuits. 
We consider both exact methods (such as Brouwer-Zimmermann, connected cluster, SAT and mixed integer programming) and heuristic methods which have lower run-time but can only give a bound on distance (examples include random information set, syndrome decoder algorithms, and Stim undetectable error methods).
We further develop the QDistEvol algorithm and show that it performs well for the quantum LDPC codes in our benchmark.
The algorithms and test data have been made available to the community in the codeDistance Python package.
\end{abstract}
\tableofcontents

\section{Introduction}\label{sec:introduction}
A fundamental figure of merit for any quantum error-correcting (QEC) protocol is its minimum distance, $d$. The distance of a QEC code or circuit is the minimum number of physical errors required to cause an undetectable logical failure. While the performance of a QEC protocol is ultimately determined by its behaviour under a hardware-realistic noise model and efficient decoding, the distance still provides a rigorous guarantee of its ability to suppress errors.

There is significant interest in the development of quantum LDPC codes \cite{Breuckmann_Eberhardt_2021}, which hold the promise of substantially reducing the resource overhead of fault-tolerant quantum computation. Moreover, asymptotically good QLDPC codes have been demonstrated \cite{breuckmann2021balancedproduct, LP_Panteleev}. Ultimately, the distance of any circuit implementation of these codes is upper-bounded by the distance of the codes themselves. Therefore, fast algorithms that establish an upper bound on the code distance simultaneously cap the potential performance of the circuit, making them very efficient for rapidly filtering out unpromising candidates. Conversely, algorithms capable of finding a tight lower bound on the code distance can provide a high degree of confidence that a particular code is worth pursuing, before substantial effort is invested into constructing fault-tolerant circuits or optimizing decoders.

Finding the distance of a QEC circuit can be even more challenging due to the substantially increased scale of the problem. As an example, verifying the $X$ or $Z$ distance of a distance 11 colour code involves considering only 91 unique error mechanisms, where each stabilizer detects at most 6 errors. In contrast, finding the distance of a 11-round memory experiment for the same code using the superdense circuit \cite{colour_code_circuits} requires considering 43,298 unique physical error mechanisms, with individual detectors that can be triggered by up to 270 distinct errors each. 
Given that the circuit-level distance problem can be over two orders of magnitude larger in scale, finding efficient algorithmic solutions (or approximations) becomes critical. Calculating the distance of a circuit is invaluable for designing better fault-tolerant primitives; it provides a decoder-agnostic figure of merit that can often be approximated much more efficiently than logical error rates, which otherwise require large-scale Monte Carlo simulations to resolve.

We emphasize that while distance-finding is a critical tool for informing the design of both codes and circuits, it is not a replacement for rigorous benchmarking. The ultimate evaluation of a QEC protocol requires full circuit-level simulations and, eventually, experimental demonstration to account for factors including hardware-specific constraints, practical decoding, and fault-tolerant logic. Nevertheless, robust distance-finding algorithms remain an indispensable component of the design pipeline, enabling the systematic discovery and evaluation of the next generation of quantum error correction schemes.

\section{Contribution of this Work}
In this work, we review the leading distance-finding algorithms for classical codes, quantum codes and quantum circuits. 
For analysis purposes, we divide these methods into exact algorithms and heuristic algorithms.
Exact algorithms tend to have a longer run-time, but on termination are guaranteed to give the  exact distance.
Heuristic algorithms have shorter run-time, but may only give a lower or upper bound on the distance.

We conducted benchmarking on a series of data sets which comprise various families of classical codes, quantum codes and syndrome extraction circuits.
The algorithms used for benchmarking are summarised in \Cref{tab:algorithm_summary}. 
The benchmarking methodology and detailed information on the accuracy and run-time of the algorithms are set out in \Cref{sec:benchmarks}.
The best performing exact and heuristic algorithms on each data set are listed in \Cref{tab:recommendations_exact,tab:recommendations_heuristic} respectively.

\begin{table}[h!]
\setlength\tabcolsep{2pt}
\fontfamily{lmss}\fontsize{8}{9}\selectfont{
\begin{center}										
\begin{tabular}{ |l|	p{3.7cm}|	l|	p{0.9cm}|	p{1.1cm}|	p{0.5cm}|	p{0.8cm}|	p{1.0cm}|	p{0.6cm}| p{0.5cm}|}	\hline
&	\textbf{Description}&	\textbf{Implementation}&	\textbf{Open Source}&	\textbf{Classical}&	\textbf{CSS}&	\textbf{Non-CSS}&	\textbf{Error Models}&	\textbf{Prob}&	\textbf{Par}\\	\hline
\multicolumn{10}{|l|}{\textbf{Exact Methods}}\\										\hline
Gurobi&	Mixed integer programming&	Python library &	N&	C&	Y&	2,3B&	Q&	N&	N\\	
MIP-SCIP&	Mixed integer programming&	Python library &	Y&	C&	Y&	3B&	Q&	N&	N\\	
CLISAT&	MaxSAT Solver &	Unix binary&	Y&	C&	Y&	3B&	Q&	N&	N\\	
m4riCC&	Connected Cluster&	C library&	Y&	Y&	Y&	3,4B&	D&	N&	N\\	
Magma&	Brouwer Zimmermann&	Mac/Unix binary&	N&	Y&	XZ&	2B&	D&	N&	N\\	\hline
\multicolumn{10}{|l|}{\textbf{Heuristic Methods}}\\										\hline
m4riRW&	Random Window (QDistRnd)&	C library&	Y&	Y&	Y&	3,4B&	D&	Y&	Y\\	
QDistRndMW&	QDistRnd with more options&	Python library &	Y&	Y&	Y&	2,3,4B&	E&	Y&	Y\\	
QDistEvol&	QDistRnd with evolutionary selection of permutations&	Python library &	Y&	Y&	Y&	2,3,4B&	E&	Y&	N\\	
BP-OSD&	Syndrome decoder using LDPC package \cite{Roffe_LDPC_Python_tools_2022} &	Python library &	Y&	C&	Y&	3B&	Any&	Y&	Y\\	
GEStim&	Stim Graphlike Error (max syndrome weight 2)&	Python library &	Y&	C&	Y&	3B&	D&	N&	N\\	
CCStim&	Stim Colour Code Error (max syndrome weight 3)&	Python library &	Y&	C&	Y&	3B&	D&	N&	N\\	
UEStim&	Stim Undetectable Error (increasing syndrome weight)&	Python library &	Y&	C&	Y&	3B&	D&	N&	N\\	\hline
\end{tabular}										
\end{center}										
}
\caption{Summary of Benchmarked Algorithms: in the Classical Codes column an entry of Y indicates the method natively handles classical codes, and C indicates that the algorithm is called with a basis of the complementary subspace  (see \Cref{sec:DEM_direct_calc}). In the CSS Codes column, Y indicates the method supports finding of X and Z distance separately and XZ for Magma indicates that the lowest of X and Z distances is returned. Distance-finding for non-CSS stabiliser codes is generally done by mapping the quantum code to a binary linear code using a two, three or four block representation. As accuracy and run time are sensitive to the representation chosen the Non-CSS column indicates which block representations are supported by the method. In the Error Models column, D indicates that the method can only be used to find the minimum distance, Q indicates that a biased error model can be handled using the `quantization' technique (see Stim \texttt{likeliest\_error\_sat\_problem} function), A indicates that the method supports arbitrary probabilities for each error mechanism and E indicates the method can easily be extended to handle arbitrary probabilities.
The Prob column indicates whether the algorithm is probabilistic or deterministic. 
The Par column indicates whether the algorithm is natively parallelisable.}
\label{tab:algorithm_summary}
\end{table}

\begin{table}[h!]
\setlength\tabcolsep{2pt}
\fontfamily{lmss}\fontsize{8}{9}\selectfont{
\begin{center}			
\begin{tabular}{ |l|	l|	l|}	\hline
&	\textbf{Recommendation}&	\textbf{Also Effective}\\	\hline
\multicolumn{3}{|l|}{\textbf{Classical Codes}}\\			\hline
\textbf{CodeTables GF(2)}&	Magma &	Gurobi, MIP-SCIP\\	
\textbf{Lifted Product GF(2)}&	Magma&	MIP-SCIP, Gurobi\\	\hline
\multicolumn{3}{|l|}{\textbf{Quantum Codes}}\\			\hline
\textbf{CodeTables Non-CSS}&	Magma&	Gurobi, MIP-SCIP\\	
\textbf{Hyperbolic Surface CSS}&	m4riCC&	Gurobi, MIP-SCIP, SAT\\	
\textbf{Hyperbolic Colour CSS}&	m4riCC&	Gurobi, MIP-SCIP, SAT\\	
\textbf{Lifted Product CSS}&	m4riCC&	Gurobi\\	
\textbf{Bivariate Bicycle CSS}&	m4riCC,Gurobi,MIP-SCIP&	-\\	\hline
\textbf{Quantum Tanner CSS}&	Gurobi&	m4riCC, Magma, MIP-SCIP, SAT\\	\hline
\multicolumn{3}{|l|}{\textbf{Quantum Syndrome Extraction Circuits}}\\			\hline
\textbf{Surface Code}&	m4riCC&	Gurobi, MIP-SCIP\\	
\textbf{Colour Code - Midout}&	m4riCC&	Gurobi, MIP-SCIP, SAT\\	
\textbf{Colour Code - Superdense}&	m4riCC&	Gurobi, MIP-SCIP, SAT\\	
\textbf{Bivariate Bicycle}&	Gurobi&	-\\	\hline
\end{tabular}			
\end{center}			
}
\caption{Recommended Distance-Finding Algorithms by Data Set - Exact Methods}
\label{tab:recommendations_exact}
\end{table}

\begin{table}[h!]
\setlength\tabcolsep{2pt}
\fontfamily{lmss}\fontsize{8}{9}\selectfont{
\begin{center}			
\begin{tabular}{ |l|	l|	l|}	\hline
&	\textbf{Recommendation}&	\textbf{Also Effective}\\	\hline
\multicolumn{3}{|l|}{\textbf{Classical Codes}}\\			\hline
\textbf{CodeTables GF(2)}&	QDistEvol&	QDistRnd\\	
\textbf{Lifted Product GF(2)}&	QDistEvol&	QDistRnd, BP-OSD\\	\hline
\multicolumn{3}{|l|}{\textbf{Quantum Codes}}\\			\hline
\textbf{CodeTables Non-CSS}&	QDistRnd&	QDistEvol\\	
\textbf{Hyperbolic Surface CSS}&	QDistEvol,BP-OSD&	Stim GE, Stim UE, Stim CC\\	
\textbf{Hyperbolic Colour CSS}&	QDistEvol,BP-OSD&	Stim UE, Stim CC, BP-OSD\\	
\textbf{Lifted Product CSS}&	QDistEvol&	QDistRnd\\	
\textbf{Bivariate Bicycle CSS}&	QDistEvol&	-\\	\hline
\textbf{Quantum Tanner CSS}&	QDistEvol&	-\\	\hline
\multicolumn{3}{|l|}{\textbf{Quantum Syndrome Extraction Circuits}}\\			\hline
\textbf{Surface Code}&	QDistEvol, QDistRnd, BP-OSD&	Stim GE\\	
\textbf{Colour Code - Midout}&	QDistEvol, QDistRnd, BP-OSD&	Stim GE\\	
\textbf{Colour Code - Superdense}&	QDistEvol, QDistRnd, BP-OSD&	Stim GE\\	
\textbf{Bivariate Bicycle}&	QDistEvol&	-\\	\hline
\end{tabular}			
\end{center}			
}
\caption{Recommended Distance-Finding Algorithms by Data Set - Heuristic Methods}
\label{tab:recommendations_heuristic}
\end{table}

Some of the novel contributions of this work are as follows. 
We refine the QDistEvol algorithm first introduced in \cite{QECC_evol} and our benchmarking results indicate that this algorithm has a significantly higher accuracy than other heuristic algorithms for our quantum LDPC code datasets.
We adapt and optimise existing methods for quantum CSS codes to classical codes, non-CSS quantum codes and quantum circuits. Examples include the connected cluster algorithm of \Cref{sec:connected_cluster}, the syndrome decoder-based methods of \Cref{sec:syndrome_decoder} and the mixed-integer programming and SAT solver methods of \Cref{sec:solvers}.
We also include an open-source python implementation of the Brouwer-Zimmermann algorithm that natively handles CSS and non-CSS quantum codes as well as classical codes.

We have made the algorithms and code test data available in the \href{https://github.com/m-webster/codeDistancePYPI}{codeDistance python package}.
The  package consists of wrappers for existing implementations of distance-finding algorithms, as well as new implementations of algorithms for use by the community.

\section{Background}\label{sec:background}

In this section we introduce background material on distance-finding in three different contexts. 
We first describe the distance problem for classical binary linear codes.
We then introduce quantum stabiliser codes and show how the classical concept of distance generalises to them.
Next, we discuss the distance problem for quantum circuits.
Finally we consider the computational complexity of distance-finding.

\subsection{Distance of Classical Linear Codes}\label{sec:dist_classical}

In this section we describe the construction of classical binary linear codes and define the distance problem in this context.
A binary linear code is defined by specifying a binary \textbf{generator matrix} $G$. 
For the purposes of this section, we assume that $G$ is a full rank $k \times n$ matrix.
The codewords of the code, denoted $\braket{G}$, are the set of binary linear combinations of the rows of $G$.
We encode a length $k$ vector $\mathbf{u}$ by  multiplying on the right by the generator matrix yielding an $n$-bit codeword ${\mathbf{u}_L} := \mathbf{u} G$ which is then transmitted over a noisy channel.

One strategy to detect and correct errors is syndrome decoding which we describe here.
Via Gaussian elimination, we can find a \textbf{check matrix} $H$ of rank $n-k$ which spans $\ker{G}$ and for which $GH^T=0$. 
The check matrix can be used to detect bit-flip errors as follows.
Assume that we transmit the codeword $\mathbf{u} G$ and a small number of bit-flips $E$ occur so that the received bit vector is $\mathbf{u} G + E$.
We calculate the syndrome by multiplying the received vector on the right by $H^T$ as follows:
$$
\mathbf{s} = (\mathbf{u} G + E)H^T = \mathbf{u} G H^T + E H^T = E H^T
$$
Providing $E$ is not a codeword, the syndrome vector is non-zero and the error can be detected. 
If $E$ is a codeword, it changes the encoded logical information and results in a \textbf{logical error}.
This is because in this case  $E = \mathbf{v}G$ for some vector $\mathbf{v}$ and so:
$$
\mathbf{u} G + E = \mathbf{u} G + \mathbf{v}G = (\mathbf{u}+\mathbf{v})G = (\mathbf{u}+\mathbf{v})_L
$$
The distance $d$ is defined as the minimum weight of a non-zero codeword - any errors of weight less than $d$ can be detected by calculating the syndrome vector.
We use the code parameters $[n,k,d]$ to denote that we encode $k$ logical bits in $n$ physical bits and that the code has distance $d$.

\subsection{Distance of Quantum Stabiliser Codes}\label{sec:dist_QECC}
In this section we show how the concept of distance for classical codes generalises to quantum stabiliser codes. 
For distance-finding, we will usually map quantum stabiliser codes to a binary linear code.
We also discuss different ways to represent stabiliser codes as binary linear codes (the two, three and four-block representations).
Choice of block representation can have significant consequences on the performance of distance-finding methods. 

A stabiliser code is defined by specifying a set of  Pauli operators acting on $n$ qubits referred to as stabiliser generators. 
The Pauli operators we consider are strings of single-qubit $I, X,Z$ and $Y:=iXZ$ operators with a sign of $\pm 1$ and have eigenvalues of $\pm 1$.
The codespace is the simultaneous +1 eigenspace of the stabiliser generators.
The group generated by the stabiliser generators is referred to as the \textbf{stabiliser group} and to define a non-trivial stabiliser code it must not contain $-I$.
The requirement that the stabiliser group not contain $-I$ is satisfied if and only if the stabiliser generators commute and each stabiliser generator has both $+1$ and $-1$ as eigenvalues.

The stabiliser generators play a similar role as checks do in classical codes.
Errors can be detected by measuring the stabiliser generators resulting in a binary syndrome vector $\mathbf{s}$  where the $i$th entry depends on whether the outcome of measuring $i$th stabiliser generator is $1$ or $-1$. 
As we explain in \Cref{sec:dist_QC}, the stabiliser generators are usually measured with a noisy quantum circuit that may be repeated, resulting in a larger syndrome vector and richer set of error mechanisms, but for now we consider this simpler ``code capacity'' setting, where we assume that stabiliser measurements are noiseless.

We model errors as unsigned Pauli strings $E$.
Providing $E$ anti commutes with at least one of the stabiliser generators $A$, it can be detected because in this case:
\begin{align*}
    A E\ket{\psi}_L = -EA\ket{\psi}_L = -E\ket{\psi}_L.
\end{align*}

If there are $r$ independent stabiliser generators, then we say the code has $k$ logical qubits and  we can identify $k:=n-r$ logical Pauli $X$ operators and $k$ logical Pauli $Z$ operators using the techniques in \Cref{sec:twoBlock}. 
The logical $X$ and $Z$ operators commute with the stabiliser generators and have the same commutation relations as physical $X$ and $Z$ operators on $k$-qubits (i.e. they anti-commute in pairs).
An error which is a product of stabilisers and logical Paulis commutes with the stabiliser generators and so is undetectable (it has a zero syndrome vector).
A non-trivial logical is such a product which is not in the stabiliser group - an error of this kind is undetectable and changes the logical state.

The distance $d$ of a quantum stabiliser code is the minimum weight of a Pauli string which commutes with all stabiliser generators  but is not in the stabiliser group.
The weight of a Pauli string is the number of non-trivial operators in the string.
A code on $n$ qubits with $k$ logical qubits and distance $d$ has code parameters $[[n,k,d]]$.

\subsubsection{Two-Block (Symplectic) Representation of Stabiliser Codes}\label{sec:twoBlock}
We now show how a quantum stabiliser code can be mapped to a binary linear code on $2n$ bits - this mapping will be very useful for distance-finding.
Pauli strings with a $\pm 1$ eigenvalues are of form $\pm HW(\mathbf{x|z})$ where $\mathbf{x}$ is a length $n$ binary vector called the X-component of the operator and $\mathbf{z}$ is the Z-component and:
\begin{align}
    HW(\mathbf{x|z}) = i^{\mathbf{x \cdot z} \mod 4}\prod_{0 \le i <n} X_i^{\mathbf{x}_i} Z_i^{\mathbf{z}_i}.\label{eq:pauli_HW_form}
\end{align}
We refer to the length $2n$ vector $\mathbf{(x|z)}$ as the \textbf{vector representation} of the Pauli operator.
Multiplication of Paulis corresponds to addition of the vector representations modulo 2, up to a phase correction. 
In the \textbf{symplectic representation} of a stabiliser code, $r$ independent stabiliser generators are represented (up to signs of $\pm 1$) as a two-block $r \times 2n$ binary check matrix $H = [H_X|H_Z]$ where the rows of $H_X$ are the X-components of the Paulis and $H_Z$ are the corresponding Z-components.
The Pauli operators in the stabiliser group are exactly those whose vector representations are in the binary linear code $\braket{H}$.

Commutation relations of Paulis are captured via the \textbf{symplectic inner product}:
\begin{align}
HW(\mathbf{x_1|z_1)}HW(\mathbf{x_2|z_2)} = (-1)^{(\mathbf{x_1|z_1})\Omega(\mathbf{x_2|z_2})^T} HW(\mathbf{x_2|z_2)} HW(\mathbf{x_1|z_1)},
\end{align}
where $\Omega:=\begin{bmatrix}0&I\\I&0\end{bmatrix}$ is the symplectic form which swaps the X and Z components when acting on the right so that $\mathbf{(x|z})\Omega = \mathbf{(z|x})$.
Using the vector representation of a Pauli operator $HW(\mathbf{x|z)}$ the weight of a Pauli string corresponds to the \textbf{symplectic weight}  $\text{wt}(\mathbf{x \lor z})$, where $\lor$ represents the bitwise or operator on vectors.

As in the classical case, we can construct a full rank $(2n-r) \times 2n$ generator matrix $G$ for the dual of the check matrix $H$. 
For quantum codes we consider the symplectic dual such that $G\Omega H^T = 0$.
Because the stabiliser generators commute, $\braket{H}$ is a subgroup of $\braket{G}$, so we can take the first $r$ rows of $G$ to be the rows of $H$. 
The remaining $2k := 2(n-r)$ rows represent Pauli operators which commute with the stabiliser generators, but are not in the stabiliser group. 
They correspond to the logical Pauli operators of \Cref{sec:dist_QECC}. 

\begin{example}{$[[5,1,3]]$ Quantum Error Correction Code.}\label{eg:5-1-3-generator-matrix} 
The  $[[5,1,3]]$ code is specified by the following independent stabiliser generators:
$$
\left[\begin{array}{c}
IXZZX\\
XIXZZ\\
ZXIXZ\\
ZZXIX
\end{array}\right]$$
These correspond to the $4\times 10$ binary check matrix:
$$H =  \left[\begin{array}{c|c}
01001&00110\\
10100&00011\\
01010&10001\\
00101&11000\\
\end{array}\right] $$
The symplectic dual code is generated by the $6\times 10$ generator matrix:
$$
G = \left[\begin{array}{c|c}
01001&00110\\
10100&00011\\
01010&10001\\
00101&11000\\
\hline
11111&00000\\
00000&11111\\
\end{array}\right]
$$
The last two rows correspond to the non-trivial logical operators $XXXXX$ and $ZZZZZ$ which together with $H$ generate the dual code $\braket{G}$.
\end{example}

\subsubsection{Three Block Representation Stabiliser Codes}\label{sec:detector_view}
Pauli operators can also be represented by length $3n$ binary strings, and this allows us to construct a three-block representation of stabiliser codes.
The map $\phi_{23}:\mathbf{(x|z)} \mapsto \mathbf{(x|z|x\oplus z)}$  maps two-block Pauli vector representations to three block representations and preserves addition of vectors modulo 2.
When finding distances of non-CSS quantum codes, certain methods are best used with the three-block representation. 
The three-block representation is also used to find automorphisms of  quantum codes \cite{Calderbank_Rains_Shor_Sloane_1998,autQEC}.

Given a set of stabiliser generators in symplectic form $H = [H_X|H_Z]$, the three block representation is $H_3:=[H_X|H_Z|H_X\oplus H_Z]$.
Once a low-weight codeword in three-block form is identified, it can be mapped back to the two-block representation using the  inverse map $\phi_{32}:\mathbf{(a|b|c)} \mapsto \mathbf{(b \oplus c|a \oplus c)}$.

Whether two Paulis anticommute can be determined by taking the ordinary dot product of the three-block vector representations.
Each row of the three-block representation has twice the weight of the corresponding Pauli operator and for this reason this representation is also referred to as the \textbf{isometric} representation \cite{White_Grassl_2006,Hernando_Quintana-Orti_Grassl_2024}.
In distance-finding methods, we often map a logical basis $L$ of the code to a 3-block representation $L_3$ whose rows also have even weight. 

Another important feature for distance finding is that the three blocks each detect a different type of Pauli error. 
In particular, a single-qubit $Z$ error anti-commutes with a row of $H$ if and only if the corresponding bit of $H_X$ is set, and contrariwise for $X$ errors and $H_Z$. A single-qubit $Y$ error anti-commutes with a row of $H$ if the corresponding bit in $H_X \oplus H_Z$ is set.

\begin{example}{Three-Block Representations of $[[5,1,3]]$ code.}
Given the generator matrix $G$ of \Cref{eg:5-1-3-generator-matrix}, the three block representation is as follows:
\begin{align}
H_3 &= \left[\begin{array}{c|c|c}
01001 & 00110 & 01111 \\
10100 & 00011 & 10111 \\
01010 & 10001 & 11011 \\
00101 & 11000 & 11101 \\
\end{array}\right]\\
L_3 &= \left[\begin{array}{c|c|c}
11111 & 00000 & 11111\\
00000 & 11111 & 11111\\
\end{array}\right]\label{eq:513_DEM}
\end{align}
By examining the second block, we see that an X error on the first qubit anti-commutes with the third and fourth stabiliser generator, as well as the Z-logical.
\end{example}

\subsubsection{Four-Block Representation of Stabiliser Codes}\label{sec:fourBlock}
Stabiliser codes also have a four-block representation as set out in \cite{fourBlockCSS}.
Given binary vectors $\mathbf{x,z}$ of length $n$ let $\mathbf{y:=x*z}$ be the bitwise multiplication of $\mathbf{x,z}$.
The map $\phi_{24}:\mathbf{(x|z)} \mapsto \mathbf{(x\oplus y|y|z\oplus y|x\lor z)}$ with inverse $\phi_{42}:\mathbf{(a|b|c|d)} \mapsto \mathbf{(a\oplus b|c\oplus b)}$ preserves addition of vectors modulo 2.
In the four-block representation of a Pauli string, entry $i$ in the first block is one only if the string has $X$ in the $i$th position.
Similarly, the entries in the second block correspond to $Y$ operators and the third block $Z$ operators. 
The final block indicates whether $i$ is in the support of the Pauli string.
Given a stabiliser code in symplectic form $H=[H_X|H_Z]$ and letting $H_Y:=H_X*H_Z$, the four-block representation is $H_4:= [H_X\oplus H_Y|H_Y|H_Z\oplus H_Y|H_X\lor H_Z]$.

As in the three-block representation, commutation relations can be determined by taking the dot product of the four-block representations, and the Hamming weight of the four-block representation is twice the Pauli weight.
The four-block representation is also self-orthogonal.
Appending the matrix $[I_n|I_n|I_n|I_n]$ to $H_4$ (which forces all logicals to be of even weight) and using the result for both the X and Z-checks yields a $[[4n,2k,2d]]$ CSS code. 
This makes the four-block representation useful for distance-finding algorithms which only take CSS codes as input.

\subsubsection{CSS Codes}\label{sec:dist_CSS}
CSS (Calderbank–Shor–Steane) codes \cite{CSS_codes} are an important subclass of quantum stabiliser codes.
For CSS codes, the stabiliser generators can be written as a set of operators composed entirely of Paul $X$ operators and those composed of entirely  $Z$ operators.
These can be expressed as  $n$-column binary matrices  $H_X$ and $H_Z$, where one in the $i$th position in a row of $H_X$  indicate that the stabiliser generator has an $X$ operator acting on qubit $i$ (and similarly for $H_Z$ and $Z$ operators).
We can always find a basis for the logical X operators composed entirely of Pauli X operators, and we use a full-rank binary $k \times n$ matrix $L_X$ to represent these (and we have a similar logical $Z$-basis written $L_Z$).
For certain code families, it may be known that the $Z$-distance, that is the lowest weight logical operator composed entirely of $Z$ operators, is lower than the $X$-distance. 
Where this is the case, we need only the $Z$-distance to find the minimum distance.

Developing new families of high-rate quantum LDPC (Low Density Parity Check) codes has been an active area of research recently \cite{Breuckmann_Eberhardt_2021,BB_IBM,LP_Xu}.
These are CSS code families where $H_X$ and $H_Z$ can be written in such a form that the row and column weights are less than or equal to a fixed constant. 
Due to the sparse nature of the check matrices, they are often represented as bipartite \textbf{Tanner graphs} where there are two types of nodes, corresponding to rows and columns of the check matrix, and there is an edge wherever the row-column entry is 1.
For LDPC code families, the degree of each row and column node of the Tanner graph are bounded by a fixed constant. 
Much of our algorithm pseudocode is written for the case of CSS codes as such code can generally be extended to classical codes, non-CSS quantum codes or quantum circuits in a fairly straightforward way.

\subsection{Distance of Quantum Circuits}\label{sec:dist_QC}
In this section, we show how the concept of distance can be generalised to quantum circuits. 
We consider a class of quantum error correction circuits defined by a stabilizer circuit, a noise model (specified as Pauli error channels at various circuit locations), as well as detector and observable annotations.
Such a circuit can be defined by a Stim circuit~\cite{Gidney2021stimfaststabilizer}.
A detector is a parity of measurement outcomes in the circuit that is deterministic in the absence of noise, and can be considered a generalization of a stabilizer measurement.
A logical observable is usually defined as a parity of measurement outcomes corresponding to the measurement of a logical operator of the code (or codes).

Such a quantum error correction circuit can be represented as a Detector Error Model (DEM), which can be constructed automatically from a circuit by Stim's error analyzer~\cite{Gidney2021stimfaststabilizer}. 
A DEM can be conveniently represented as a \textit{detector check matrix} $H$ and \textit{logical observable matrix} $L$.
Both $H$ and $L$ are binary matrices.
Row $i$ of $H$ corresponds to detector $i$ in the circuit and row $l$ of $L$ corresponds to logical observable $l$.
Column $j$ of $H$, as well as column $j$ of $L$, both correspond to error mechanism $j$ in the DEM.
In general many error mechanisms in the circuit may be merged into a single error mechanism in the DEM, since DEM errors are specified by their detector and observable symptoms (i.e.~multiple physical errors may have the same symptoms).
By definition, $H[i,j]$ is 1 if and only if error mechanism $j$ flips detector $i$, and is 0 otherwise.
Similarly, $L[l,j]$ is 1 if and only if error mechanism $j$ flips logical observable $l$, and is 0 otherwise.

A logical error is any set of DEM error mechanisms that flips one or more observables, but does not flip any detectors.
In other words, an error is any binary vector in the kernel of $H$, but not in the kernel of $L$.
We denote the set of logical errors by $B := \{ \mathbf{e} \in \mathbb{F}_2^m \mid \mathbf{e} \in \ker H, \mathbf{e} \notin \ker L \}$, where here $m$ is the number of unique DEM error mechanisms.
The \textit{distance} of a circuit is the minimum weight of any logical error, $d = \min_{\mathbf{e} \in B} |\mathbf{e}|.$.
In general, error mechanisms may have varying probabilities, specified by a vector of probabilities $\mathbf{p} \in \mathbb{R}^m$, and so a generalization of the distance is instead the \textit{most probable error}, which reduces to a minimum distance error when all error mechanisms have the same probability. 
Not all distance-finding algorithms allow for this more general notion of distance.

Note that the circuit-level distance finding problem is a generalization of all the other distance finding problems considered in this work.
For example, the distance of a stabilizer code can be defined as the distance of a \textit{circuit} that measures all the stabilizers noiselessly twice, with single-qubit depolarization placed between the two rounds of consecutive stabilizer measurements, and with all logical observable Pauli operators annotated\footnote{Note that as of Stim v1.15, observable annotations can be given Pauli targets, which makes it more straightforward to define observables that do not commute. i.e.~All logical operators can be straightforwardly defined in a single Stim circuit, without the need to introduce auxiliary qubits to enforce commutativity of measurements.}.
The distance of a classical binary linear code is the distance of a stim circuit where each qubit is initialized in the $Z$ basis, followed by a single-qubit $X$ error channel on each qubit, then a $Z$ Pauli operator measurement for each parity check of the linear code (where each Pauli $Z$ operator measured has the same support on qubits as the parity check has on bits), finally followed by a separate logical observable for the measurement of each qubit in the $Z$ basis.
Such a circuit's error model is represented by a detector check matrix $H$ which is exactly equal to the parity check matrix of the corresponding binary linear code, and the logical observable matrix $L$ can simply be set to the identity matrix $I$ (and note that just a subset of the rows of $I$ defining the complementary subspace to $H$ is sufficient, see 
\Cref{sec:DEM_direct_calc}).
Therefore, to solve any of the distance finding problems described in this work, it is sufficient to be able to find the minimum weight binary vector in the kernel of a binary matrix $H$, but not in the kernel of another matrix $L$.
We refer the reader to \cite{Higgott2025sparseblossom,Derks2025designingfault,higgott2024practical} for more background on detector error models.

\subsection{Direct Calculation of Detector Matrices for Classical and Quantum Codes}\label{sec:DEM_direct_calc}
Several distance-finding algorithms use detector and observable matrices - for instance methods which rely on error enumeration and solvers (see \Cref{sec:alg_enumeration_errors,sec:solvers}).
When applying such methods to classical and quantum codes, it may be more convenient to calculate detector and observable matrices  directly from the code rather than via a quantum circuit.

When calculating the distance of quantum stabiliser codes, our error mechanisms are Pauli $X, Y$ and $Z$ operators on each qubit. The 3-block form $H_3$ of \Cref{sec:detector_view} corresponds to the detector matrix - the first block detects $Z$ errors, the second  $X$ errors and the third $Y$ errors. 
We use the three-block logical basis $L_3$ for the observable matrix.

When calculating Z-distance of a CSS code, the error mechanisms are Pauli Z on each qubit and we can use $H_X$ for the detector matrix and  a logical basis $L_X$ for the observable matrix.

Classical codes on the other hand do not have a set of logical operators for use in the algorithm.
For an $[n,k]$ classical code we use the check matrix $H$ as the detector matrix.
For the observable matrix, we use a basis for the complementary subspace to $H$ (i.e. a set of vectors $L$ which together with $H$ generate the entire vector space).
We find $L$ by calculating the RREF of $H$ which results in a rank $n-k$ matrix of form $[I|C]$ up to column permutations. 
The rows of the rank $k$ matrix $L:=[0|I]$ span the complementary space.
By construction, these are not in the span of $H$ and so any codeword will have a non-zero dot product with any non-trivial linear combination of the basis of the complementary space.

\subsection{Computational Complexity of Distance Finding}
Finding the exact distance of an arbitrary classical binary linear code or a quantum stabiliser code is an NP-complete problem (\cite{vardy2002intractability,kapshikar2023hardness}).
Estimating code distance within a constant multiple via a heuristic algorithm is also NP-complete (\cite{dist_const_factor}). 
Naive algorithms which enumerate all codewords (see \Cref{sec:alg_enum_codewords}) or all errors by increasing weight (\Cref{sec:alg_enumeration_errors}) have exponential complexity. 
For particular code families, however, we may be able to reduce the exponent in the complexity.

The Brouwer-Zimmermann algorithm is generally regarded to be the most efficient algorithm for finding the exact distance of a randomly generated binary linear or quantum stabiliser code (see \Cref{sec:brouwer_zimmermann}), but for particular code families, other methods may have lower complexity exponents. 
A good survey of these techniques can be found in Refs.~\cite{Dumer_Kovalev_Pryadko_2014,Dumer_Kovalev_Pryadko_2016}.
In these works, the authors present a range of algorithms which reduce the complexity exponent for low distance codes (via the matching bipartition algorithm see \Cref{sec:meet_in_the_middle}) and low density parity check codes (via the irreducible/linked/connected cluster algorithm \Cref{sec:connected_cluster}).
They also present a random information set algorithm which finds correct distances with high probability (see \Cref{sec:QDistRnd}).

Aside from the asymptotic complexity of an algorithm, the run-time of an algorithm may depend on implementation-specific factors. 
Often choices must be made in how to implement the algorithm in detail and these can have a significant impact on performance. 
Processing time and accuracy of distance-finding for non-CSS stabiliser codes are highly dependent on the block representation used when mapping the quantum code to a binary linear code (see \Cref{sec:twoBlock,sec:detector_view,sec:fourBlock}).
For instance, the Brouwer-Zimmermann algorithm was developed for distance-finding of binary linear codes and there has been considerable literature on the best mapping to use for quantum stabiliser codes (see \Cref{sec:brouwer_zimmermann}).
Another example is the use of the `method of four Russians' for Gaussian elimination in the dist-m4ri C implementation of the random information set algorithm. 
This changes the asymptotic complexity of the algorithm and is evident for large code sizes.
Finally, whether the algorithm is implemented in a high-level language (such as Python or GAP) or a low-level language (such as C) can also affect processing time and memory usage, though we do not expect this to affect asymptotic scaling.

Distance-finding is related to the decoding problem. 
For instance, syndrome decoders can be applied to find distances of codes (see \Cref{sec:syndrome_decoder}). 
The proof in \cite{vardy2002intractability} that distance finding for binary linear codes is NP-complete works via a reduction to maximum likelihood decoding.
For quantum stabiliser codes, due to code degeneracy, the correspondence breaks down – quantum maximum likelihood decoding is \#P-complete, whereas distance-finding remains NP-complete (see \cite{iyer2015hardness,kapshikar2023hardness}). 

\section{Distance-Finding Algorithms}\label{sec:distance_finding_algorithms}
In this section we set out the various distance-finding algorithms we consider in our benchmarking work. 
We start by setting out  desirable characteristics of distance-finding algorithms.
We then review methods which involve enumerating codewords - that is linear combinations of the rows of the generator matrix of a classical code, non-trivial logical Pauli operators of a quantum stabiliser code or linear combinations of the observable and detector matrices.
Next, we review methods which involve enumerating errors and checking whether these represent codewords or non-trivial logicals.
Finally, we review SAT solver and mixed-integer programming methods.

\subsection{Desiderata and Characteristics of Distance-Finding Algorithms}\label{sec:alg_desiderata}

Here we set out in more detail the desirable features and characteristics of distance-finding algorithms that we consider in our benchmarking exercise.
We focus on two main use-cases for distance-finding with a different trade-off between accuracy and computational complexity.
The first scenario is where we are exploring code families and wish to get a rough first-pass estimate of the distance of a large number of code candidates. 
We may wish to analyse large codes and circuits so minimal use of computational resources (memory and processing time) is important. 
In this case, we may use \textbf{heuristic algorithms} which give an upper bound on distance relatively quickly, but do not guarantee that there are no lower-weight codewords.
The second use case is where we are considering a single code and wish to know the distance exactly. For this use case, accuracy is paramount and we may be willing to wait several days or more for the calculation - examples include Brouwer-Zimmermann and Mixed-Integer Programming  methods (see \Cref{sec:brouwer_zimmermann,sec:MIP}).
As the run-time for these \textbf{exact algorithms} can be very long, we may set a maximum run-time and seek to terminate the algorithm early. 
If the algorithm is terminated early, it is desirable that it return a distance upper bound and if possible a lower bound as well.

Methods should also be applicable to a wide range of codes and circuits.
Ideally the method should work for classical codes, CSS and non-CSS quantum codes  as well as quantum circuits.
Where possible, methods  should also cater for different error models - for example biased error models where the probability of a Pauli Z error may be higher than the probability of a Pauli X or Y error. 
In this case, we assign probabilities to each error mechanism and the algorithm should return the highest probability error. 
Whilst we focus on distance-finding in this paper, we indicate whether each method can be adapted to handle different error models.

Finally, certain methods are probabilistic and require a  number of iterations to improve confidence in the result. 
Other methods are deterministic and will have the same outcome no matter how many iterations are used.

\subsection{Enumeration and Sampling of Codewords}\label{sec:alg_enum_codewords}
In this section, we review distance-finding methods which involve enumerating or sampling the codewords of the generator matrix or detector error model.
We first describe exhaustive methods for classical binary linear codes and quantum codes.
We then set out the Brouwer-Zimmermann algorithm which in many cases can run faster than the exhaustive method, but still provides an upper and lower bound on distance.
Next, we describe the QDistRnd algorithm which is a heuristic, probabilistic algorithm which gives an upper bound on the distance.
Finally we set out the QDistEvol algorithm which uses an evolutionary search algorithm and for many code families is more accurate than QDistRnd with a modest increase in run-time.

\subsubsection{Exhaustive Codeword Methods}\label{sec:exhaustive_codewords}
The distance of classical binary linear codes can be calculated by exhaustively listing all codewords of the code (ie all vectors in the linear space $\braket{G}$ where $G$ is the generator matrix of the code).
The distance is the minimum weight of a non-zero codeword.
For a binary linear code encoding $k$ bits, the number of codewords is $2^k$.
The dominating factor in runtime for the exhaustive algorithm is how many additions modulo 2 are completed in generating the entire code.
The number of additions can be minimised by the use of Gray codes (\cite{Gray_codes,Walsh_2003}) which allow for sequential generation of codewords by adding a single element of the generator matrix at each step, reducing run-time with minimal memory overhead. 

Where $k > n-k$, the complexity can also be reduced by enumerating the codewords of the dual code and using the MacWilliams Identity \cite{Macwilliams_Identity}.
The weight enumerator of the code is defined as the polynomial:
\begin{align}\label{eq:weight_enumerator}
W_G(a,x) &:= \sum_{{\mathbf{u}}\in \braket{G}} a^{n-\text{wt}({\mathbf{u}})}x^{\text{wt}({\mathbf{u}})}
\end{align}
The coefficient of the term $a^{n-w}x^w$ is the number of codewords of weight $w$.
The weight enumerator of the code can be calculated from the weight enumerator of the dual code using the MacWilliams identity:
\begin{align}
W_G(a,x) = W_H(a+x,a-x)/|\braket{H}|.
\end{align}
For quantum codes, the naive approach involves enumerating all logical operators,  excluding stabilisers, then finding the minimum weight non-trivial logical operator. 
This can be done by using the binary symplectic form as described in \Cref{sec:dist_QECC} and calculating  $\braket{G} - \braket{H}$.
If the quantum code has $n$ physical qubits and $k$ logical qubits, then there are $2^{n+k} - 2^{n-k}$ non-trivial logical operators.
As for classical codes, we can reduce the complexity of this calculation to $\mathcal{O}(2^{n-k})$ by calculating $\braket{H}$ then using the quantum MacWilliams identity \cite{Shor_Laflamme_1997}. 
To use the quantum MacWilliams identity for a non-CSS quantum code, we construct a generator matrix by stacking the symplectic representations of the stabiliser and logical generators to form a classical code $G$ as in \Cref{eg:5-1-3-generator-matrix}.
Recall from \Cref{sec:dist_QECC} that the symplectic dual of $G$ is the check matrix $H$. 
The weight enumerator for $G$ is the same as set out in \Cref{eq:weight_enumerator} except that the weights of the codewords are calculated using the symplectic weight. 
The weight enumerator of $\braket{G}$ can be calculated from the weight enumerator of $\braket{H}$ as follows:
\begin{align}\label{eq:quantum_macwilliams}
W_G(a,x) = W_H(a+3x,a-x)/|\braket{H}|.
\end{align}
We can then find the weights of the non-trivial logical Pauli operators by calculating $W_G(a,x) - W_H(a,x)$.
The generalisations of the quantum MacWilliams identities set out in  \cite{Hu_Yang_Yau_2018} allow us to find the complete weight distribution (counting codewords by the number of $I,X,Y$ and $Z$ operators in each) and are useful for biased noise error models.
\begin{example}[Quantum MacWilliams Identity]
In this example, we apply the quantum MacWilliams identity to the $[[5,1,3]]$ code of \Cref{eg:5-1-3-generator-matrix}.
We first calculate $W_H(a,x) = a^5 + 15ax^4$ - this represents a single codeword of weight zero and 15 of weight 4 with total code size $|\braket{H}| = 16$. 
Applying \Cref{eq:quantum_macwilliams}, we find that $W_G(a,x) = a^5 + 30a^2x^3 + 15ax^4+18x^5$ with $|\braket{G}| = 64$. 
To find the weights of the non-trivial logical operators, we calculate $W_G(a,x) - W_H(a,x) = 30a^2x^3 +18x^5$.
Hence, we have 30 non-trivial logical operators of weight 3 and 18 of weight 5 and so the minimum distance of the code is 3. 
By using the quantum MacWilliams identity, we can find the code distance by enumerating the 16 codewords of $\braket{H}$ rather than the 64 codewords of $\braket{G}$.

\end{example}
\subsubsection{Brouwer-Zimmermann Algorithm}\label{sec:brouwer_zimmermann}
The Brouwer-Zimmermann Algorithm can in many cases be faster than exhaustive codeword methods whilst returning the exact code distance. 
We first set out the Brouwer-Zimmermann algorithm for classical codes, following the description in Section 1.8 of \cite{Error_Correcting_Linear_Codes_2006} and Section of \cite{Grassl_BZ}.

Pseudocode for the Brouwer-Zimmermann algorithm is set out in \Cref{alg:BZ_algorithm}.
The input to the algorithm is the generator matrix of the code $G$.
The first step is to identify a series of information sets and partial information sets for $G$ by calculating the reduced row echelon form (RREF) of $G$ a number of times.
We call this operation $\text{ISBasis}$ and the algorithm is set out in \Cref{alg:BZ_IS}.
We start by identifying the non-zero columns of $G$ and storing the column indices in the list \texttt{cList}.
We compute the RREF of $G$ by restricting eliminations to columns in the ordered list \texttt{cList} as set out in the \texttt{RREFZ2} function of \Cref{alg:RREFZ2}. 
This gives a set of pivots which form the first information set and we denote this operation \texttt{G, pivots := RREFZ2(G, cList)}.
We remove the pivots from \texttt{cList} and again compute \texttt{RREFZ2(G, cList)}.
The number of pivots returned is either the same as the first set (in which we have a second information set) or it may be lower (representing a partial information set). 
We save the pivots and reduced matrices into the arrays \texttt{ISList} and \texttt{GList} later use. 
In the material below, we refer to the results of the $i$th RREF calculation as $G_i$ and $\text{IS}_i$ respectively, and the number of columns in information set $\text{IS}_i$ as $k_i$.
\begin{figure}[H]
\begin{algorithm}[H]
\caption{ISBasis: Information Sets of Classical Code}\label{alg:BZ_IS}
\begin{algorithmic}
\State\textbf{Input:} A classical code defined by $r\times n$ binary generator matrix $G$
\State\textbf{Output:}  
\State A set of generator matrices $\text{GList}$ and (partial) information sets $\text{ISList}$ 
\State\textbf{Method:}
\State Let $\text{cList}$ be a list of columns with non-zero weight in $G$
\State $\text{GList} := \text{list}()$
\State $\text{ISList} := \text{list}()$
    \While{$\text{len(cList)} > 0$}
        \State $G, \text{pivots} := \text{RREFZ2}(G,\text{cList})$
        \State $\text{GList.append}(G)$
        \State $\text{ISList.append(pivots)}$
        \State $\text{cList}:=\text{cList} - \text{pivots}$
    \EndWhile
\State $\text{return GList, ISList}$
\end{algorithmic}
\end{algorithm}
\end{figure}
\begin{figure}[H]
\begin{algorithm}[H]
\caption{RREFZ2: Reduced Row Echelon of Binary Matrix using Ordered Column List}\label{alg:RREFZ2}
\begin{algorithmic}
\State\textbf{Input:} 
\begin{itemize}
\item An $r \times n$ binary matrix G.
\item cList: an ordered list of columns of G.
\end{itemize}
\State\textbf{Output:}  
\State Reduced Row Echelon form of G and list of pivots
\State\textbf{Method:}
\State pivots = list()
\State currRow:=0
\For {currCol in cList}
    \State i:=currRow
    \While{G[i,currCol] = 0 and i < r}
        \State i:=i+1
    \EndWhile
    \If{i < r}
        \If{i > currRow}
            \State SWAP(G[currRow],G[i])
        \EndIf
        \For{i in [0..r-1]}
            \If{i != currRow and G[i,currCol]=1}
                \State G[i] := G[i] $\oplus$ G[currRow]
            \EndIf
        \EndFor
        \State pivots.append(currCol)
        \State currRow := currRow + 1
    \EndIf
\EndFor
\State return G, pivots
\end{algorithmic}
\end{algorithm}
\end{figure}
The next step in the algorithm is to generate linear combinations of $t$ rows of each of the basis matrices $G_i$.
We start by setting $t:=1$ and calculating the weight of single rows of the first generator matrix, giving us an upper bound on the code distance.
As explained below, this also gives us a lower bound on the minimum weight of codewords which have not yet been enumerated, restricted to columns in the first information set. 
We iterate over all generator matrices and increase $t$ by one until the lower and upper bounds converge - pseudocode for the algorithm is set out in \Cref{alg:BZ_algorithm}.

The lower bound is calculated by maintaining a list \texttt{mList} where the $i$th entry is the minimum weight of the columns in information set $\text{IS}_i$ of codewords which have not yet been enumerated.
The list \texttt{mList} is initially all zero, and here we show how the $i$th entry is updated by enumerating $t$ rows of the generator matrix $G_i$.
If $G_i$ has a full information set $\text{IS}_i$ of size $r:= \text{Rank}(G) = k_i$, then $G_i$ restricted to columns $\text{IS}_i$ takes the form of an identity matrix $I_{r}$. 
Forming a linear combination of $t$ rows implies that the resulting vector restricted to $\text{IS}_i$ is of weight $t$.
If $G_i$ has a partial information set $\text{IS}_i$ of size $k_i < r$ then $G_i$ restricted to $\text{IS}_i$ is of form $\begin{bmatrix}I_{k_i}\\0_{(r-k_i) \times k_i}\end{bmatrix}$.
Hence a linear combination of $t$ rows of $G_i$ has weight at least $\text{Max}(0, t - r + k_i)$ when restricted to $\text{IS}_i$.
Once all combinations of up to $t$ rows of $G_i$ have been enumerated, the minimum weight of codewords which have not yet been enumerated is $\text{Max}(0, t + 1 - r + k_i)$ and we update entry $i$ of \texttt{mList} accordingly.
The overall lower bound of codewords not yet enumerated is the sum of the entries of \texttt{mList}.
If the lower bound is greater than or equal to the upper bound, then the algorithm returns the upper bound.

\begin{figure}[H]
\begin{algorithm}[H]
\caption{Brouwer-Zimmermann Distance-finding - Classical Code}\label{alg:BZ_algorithm}
\begin{algorithmic}
\State\textbf{Input:} A classical code defined by $r\times n$ full rank binary generator matrix $G$
\State\textbf{Output:}  
\State The minimum distance of the code, UB 
\State\textbf{Method:}
\State $\text{GList}, \text{ISList} := \text{ISBasis}(G)$
\State $\text{kList} := [\text{len(IS) for IS in ISBasis}]$
\Comment{Min weight of cols in IS of codewords not yet enumerated}
\State $\text{mList} := [\text{0 for IS in ISBasis}]$ 
\State $\text{UB} := n$
\State $\text{LB} := 1$
\State $t:=1$
    \While{$t \le r$}
        \For{$G_i \text{ in GList}$}
            \For{binary vectors $\mathbf{u}$ of length $r$ and weight $t$ }
                \State $\text{UB} := \text{Min}(\text{UB}, \text{wt}(\mathbf{u}G_i))$
            \EndFor
            \Comment{Min weight of cols in IS[i] - combinations of t+1 or more rows of $G_i$}
            \State $\text{mList}[i] := \text{Max}(0,t+1-r+\text{kList}[i])$
            \State $\text{LB}:=\sum_i \text{mList}[i]$
            \If{$\text{UB} \le \text{LB}$}
                \State return UB
            \EndIf
        \EndFor
        \State $t := t+1$
    \EndWhile
    \State return UB
\end{algorithmic}
\end{algorithm}
\end{figure}
There are a number of refinements to the basic Brouwer-Zimmermann algorithm which reduce run-time. 
Firstly, where the generator matrix has rows of even weight, then all codewords are of even weight. This means that for even codes we can increase $LB$ by 1 if it is odd,  leading to faster convergence with $UB$.
Where the code is doubly-even or triply even, the codeword weights are multiples of 4 or 8 respectively and the lower bounds can be adjusted accordingly.
Whether a code is even, doubly-even or triply even can be determined with complexity polynomial in the rank of the generator matrix (eg see \cite{triply_even}).
Secondly, the information set finding process described above depends on the order of the columns for the generator matrix, and may result in different sizes of information sets. 
The Brouwer-Zimmermann algorithm works best when there are as few information sets as possible and these are of maximal size (see \cite{Grassl_BZ}). 
This can be ensured by applying \texttt{ISBasis} using a number of different random column permutations and taking the best outcome, or by using the matroid partitioning technique of \cite{Lisonek_Trummer_2016}.
Thirdly, as we improve the upper bound on the code distance, we may be able to reduce the number of rows $t$ we need to consider for linear combinations. 
After each iteration, if \texttt{UB} has decreased, we calculate $t_\text{Max}$ as large as possible such that $\text[UB] \ge \sum_i \text{Max}(0,t_\text{Max} - r + \text{kList}[i])$. 
The algorithm is guaranteed to terminate after enumerating combinations of $t_\text{Max}$ rows because after this the upper and lower bound will have converged.
Any partial information sets where $t_\text{Max} - r + k_i \le 0$ can be ignored because it will never have any contribution to minimum distance and this further reduces complexity.

It is relatively straightforward to extend the Brouwer-Zimmermann algorithm to CSS quantum codes. 
In this case, the input to the algorithm would be the $X$-check matrix $H_X$ and logical X-basis $L_X$ (or equivalently $Z$-checks and logicals). 
We stack these to form the generator matrix $G$ to generate linear combinations. 
Only codewords corresponding to non-trivial logical operators count towards the minimum distance upper bound - this can be determined calculating the dot product of each codeword with a basis $L_Z$ of the logical $Z$ operators, or by appending an identity matrix to the logical $X$ operators so that the generator matrix  of the  form described below.
Any linear combinations of this matrix representing non-trivial logicals will be non-zero in the columns with index greater than the number of qubits $n$:
\begin{align}
    G := \begin{bmatrix}
        H_X & 0\\
        L_X & I
    \end{bmatrix}.
\end{align}
To apply the Brouwer-Zimmermann algorithm to non-CSS codes is more complicated, and there have been a number of works on how best to do this (see \cite{White_Grassl_2006,Hernando_Quintana-Orti_Grassl_2024}).
There are two main ways discussed in the literature for encoding a non-CSS quantum code as a binary code for use by the Brouwer-Zimmermann algorithm.
The first option is to use the two-block symplectic representation of \Cref{sec:dist_QECC}.
In  \cite{White_Grassl_2006}, the authors suggest thinking of the matrix as having $n$ columns with entries from $GF(4)$ represented by length 2 binary vectors (i.e. one from each block of the symplectic form), corresponding a single-qubit Pauli operator.
In this view, there may be up to two pivots for each column, with either one or two `pseudo generators'. 
A single pivot means that all entries in that column are either identity or one of the Paulis ($X,Y$ or $Z$). 
Two pivots indicates that column entries can be any element from $I,X,Y$ and $Z$ and any non-trivial linear combination of the two pseudo generators corresponding to the pivot rows is guaranteed to contribute weight one to the information set $\text{IS}_i$.

An alternative approach is to encode the checks and logical Paulis of non-CSS quantum codes using the three block representation of \Cref{sec:detector_view}.
This results in  binary codes of even weight and we apply the method for CSS codes outlined above.

The Brouwer-Zimmermann algorithm is implemented in GAP, the QubitSerf C package and the proprietary Magma computational algebra package \cite{GAP4,qubitserf,Magma}.
We used Magma's implementation of the Brouwer-Zimmermann algorithm for benchmarking.
Two different Magma distance-finding algorithms can be called via our \href{https://github.com/m-webster/codeDistancePYPI}{codeDistance package} - magmaMinWeight and magmaMinWord. 
The magmaMinWeight function outputs the minimum distance of the code, whereas magmaMinWord also provides an example of a minimum weight codeword (though it runs more slowly).
Our package allows the user to specify a maximum processing time for Magma.
Where the algorithm is terminated early, an upper and lower bound on distance is usually available and this can be useful.
Magma can be used for classical and non-CSS quantum codes, but does not facilitate calculating the X and Z distance of CSS codes separately. 
Instead, both X and Z stabilisers must be provided to Magma in the form of a stabiliser code and the lowest of the X and Z distance is returned.
For non-CSS quantum stabiliser codes, Magma appears to use the pseudo generator method of \cite{White_Grassl_2006}.

\begin{figure}[h!]
\centering
  \includegraphics[width=0.50 \linewidth]{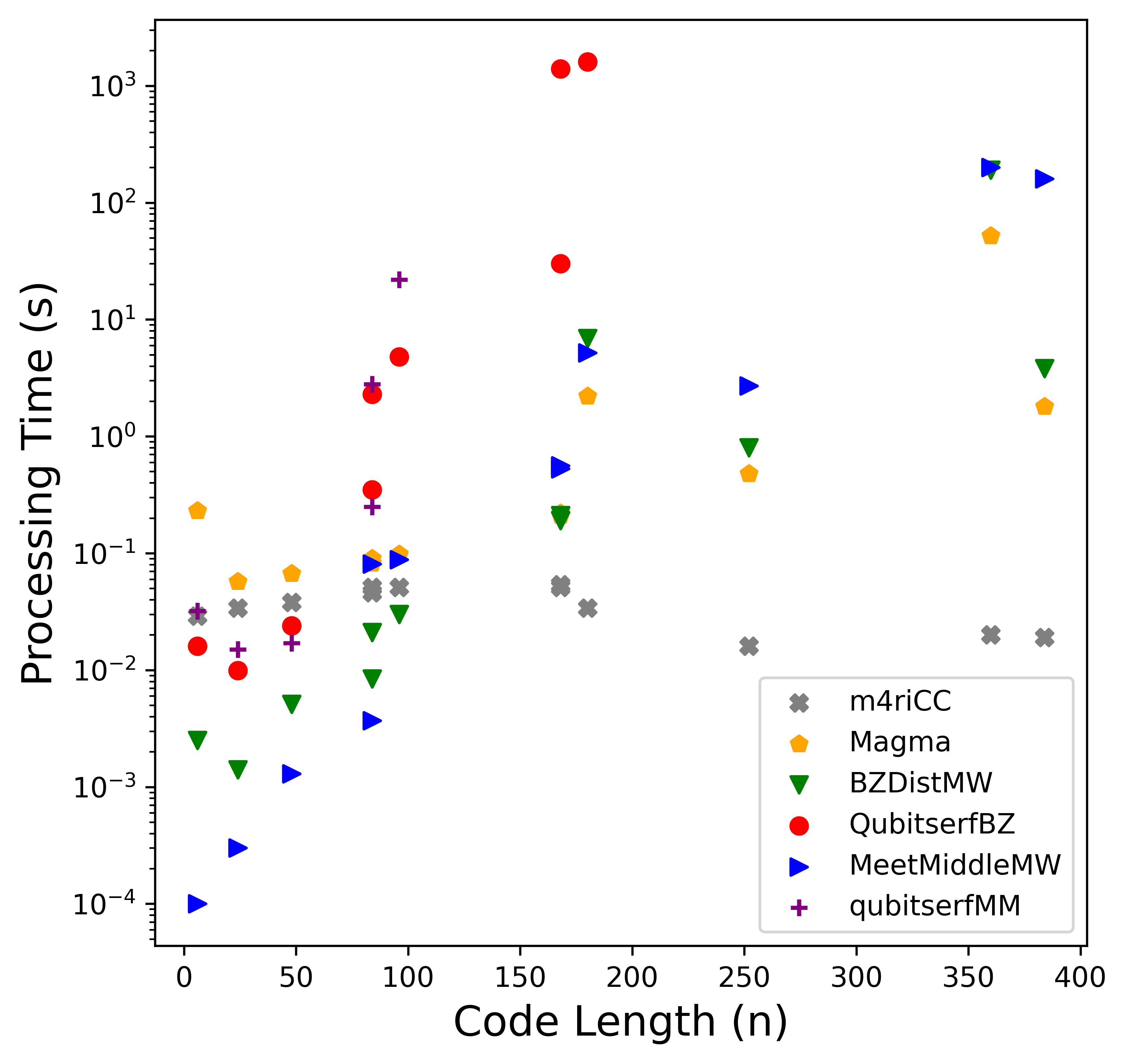}
     \caption{Benchmarking of Meet-in-the-Middle and Brouwer-Zimmermann algorithms for hyperbolic surface codes. We benchmark against m4riCC which is the fastest exact algorithm for the dataset. Note that the QubitserfMM algorithm gave results for codes up to $n=96$ and  QubitserfBZ up to $n=168$}
     \label{fig:BenchmarkBZMM}
\end{figure}

We have developed a python implementation of the Brouwer-Zimmermann algorithm, \texttt{BZDistMW}.
This implementation uses Gray codes to reduce run-time, and can use two, three or four-block representations of non-CSS codes (see \Cref{sec:twoBlock,sec:detector_view,sec:fourBlock}).
It can be applied natively to CSS codes and for these has competitive run-time versus Magma and Qubitserf as illustrated in \Cref{fig:BenchmarkBZMM} for hyperbolic surface codes.

\captionsetup[subfigure]{margin=5pt}

\subsubsection{Random Information Set/QDistRnd Algorithm}\label{sec:QDistRnd}
The random information set algorithms of \cite{Leon_1988,kruk89,cof90} involve randomly permuting the columns of the generator matrix $G$ of a classical code, then calculating the reduced row echelon form (RREF) to generate low-weight logicals/codewords.
The basis for this algorithm is that there always exists at least one column permutation such that the bottom row of the reduced row echelon form of $G$ with permuted columns is a codeword of minimum weight. 
In practice, it is simpler to change the order in which columns are considered for elimination by using  \texttt{RREFZ2} as set out in  \Cref{alg:RREFZ2}.
By choosing a large number of random column permutations, we find an upper bound on the code distance and we consider the weights of all rows of the reduced matrix to increase the number of candidates.

The QDistRnd algorithm, available in GAP and C versions \cite{QDistRNDGAP,dist-m4ri}, is a variation of the random information set algorithm. 
It involves calculating the RREF of the check matrix $H$ of a classical code using a permutation of the column indices.
This results in a matrix of form $K:=(I|C)$ up to column permutations. 
Hence $G$ has the same span as $(C^T|I)$ up to column permutations because $(I|C) (C^T|I)^T = 0 \mod 2$. 
Again, we can calculate the kernel via the \texttt{KerZ2} function which takes as input a binary matrix and an ordered list of columns which contain at least a full information set (see \Cref{alg:KerZ2}).

To extend QDistRnd to quantum CSS codes, the $X$ check matrix $H_X$ is used (or equivalently $H_Z$) to calculate a kernel basis $K_X$. 
The span of $H_Z$ and $L_Z$ is the same as the span of $K_X$.
Only non-trivial logical operators contribute to the upper bound on distance, and whether a row of $K_X$ is trivial is determined by multiplying $K_X$ by $L_X^T$. The QDistRnd algorithm is set out in \Cref{alg:QDistRnd}.

QDistRnd can also be extended to non-CSS codes by using the two, three or four-block representation of the check matrix $H$ and a logical basis $L$ (see \Cref{sec:twoBlock,sec:detector_view,sec:fourBlock}).
We then calculate a kernel basis $K$ from $H$ using a random permutation.
To check whether each row of $K$ corresponds to a non-trivial logical operator, we use the appropriate means for determining commutation relations the rows of $L$ (symplectic inner product for 2-block and vector dot product for 3 and 4-block).
The Pauli weight is calculated using the method appropriate for the block encoding (i.e. symplectic weight for 2-block and half the Hamming weight for 3 and 4-block representations).
To map the lowest weight logical back to the symplectic form, we use the appropriate inverse mapping function. 

The QDistRnd algorithm is available as a GAP package \cite{QDistRNDGAP,QDistRndGAPGithub} with two distance-finding methods - DistRandCSS for CSS codes and DistRandStab for stabiliser codes. 
Both qubit codes and prime-dimensional qudit codes can be analysed using the GAP package.
The algorithm is also available as the `random window' algorithm within the dist-m4ri C library \cite{dist-m4ri} which we refer to here as {m4riRW}.
The C implementation generally runs more quickly than the GAP package though it does not natively handle non-CSS quantum codes. 
When calling dist-m4ri, we use either the three-block or four-block representation of the stabiliser generators (see \Cref{sec:detector_view,sec:fourBlock}).

The GAP version of QDistRnd sets out a \href{https://qec-pages.github.io/QDistRnd/doc/chap3.html#X7CCA4B9B834960EE}{useful methodology} for calculating the probability of the algorithm returning the correct minimum distance.
This involves tracking the lowest weight codewords found by the algorithm, and recording how often each of these is encountered.
The average number of times each codeword has been encountered gives an estimate for the probability that the correct distance has not been found.
Assume we have found $m$ different codewords of the same low weight, and the $i$th codeword has been found $n_i$ times. Then the probability that the correct distance has not been found can be estimated as $P_{\rm fail} < e^{-\braket{n}}$ where $\braket{n}:=\frac{1}{m}\sum_{0\le i < m} n_i$.

QDistRnd is available via our \href{https://github.com/m-webster/codeDistancePYPI}{codeDistance package} in the form of a wrapper to the GAP and C versions, and as a python-coded version QDistRndMW which has additional functionality.
Whilst the C version of QDistRnd does not collate statistics on the number of codewords of lowest weight, QDistRndMW does so and estimates $P_{\rm fail}$.

Neither the GAP or C versions of QDistRnd allow for different error probabilities and only the distance can be calculated. 
This is not an inherent limitation and the algorithm can easily be modified to handle arbitrary probabilities for each error mechanism.


\begin{figure}[H]
\begin{algorithm}[H]
\caption{QDistRnd - CSS Quantum Codes}\label{alg:QDistRnd}
\begin{algorithmic}
\State\textbf{Input:} 
\begin{itemize}
\item CSS code defined by $r\times n$ binary X-check matrix $H_X$ and $k \times n$ X-logical basis $L_X$.
\item iterCount: number of iterations to run
\end{itemize}

\State\textbf{Output:}  
\State An upper bound on the Z-distance of the code
\State\textbf{Method:}
\State UB := n
\For{i in [1..iterCount]}
    \State cList := randomPermutation(n)
    \State $K := \text{KerZ2}(H_X,\text{cList})$
    \State $C := K  L_X^T$
    \For{$[i: \text{wt}(C[i])>0]$}
        \State UB := min(UB, $\text{wt}(K[i])$
    \EndFor
\EndFor
\State return UB
\end{algorithmic}
\end{algorithm}
\end{figure}

\begin{figure}[H]
\begin{algorithm}[H]
\caption{KerZ2: Kernel of Binary Matrix using Ordered Column List}\label{alg:KerZ2}
\begin{algorithmic}
\State\textbf{Input:} 
\begin{itemize}
\item An $r \times n$ binary matrix G.
\item cList: an ordered list of $s \ge \text{Rank}(G)$ columns representing a (possibly over-complete) information set of G.
\end{itemize}
\State\textbf{Output:}  
\State A basis of the kernel of G
\State\textbf{Method:}
\State C, pivots := RREFZ2(G, cList)
\State KT := ZeroMatrix(n,n-len(pivots))
\For{i in [0..len(pivots)-1]}
    \State KT[pivots[i]] := C[i]
\EndFor
\State nonPivots := [0..n-1] - pivots
\For{i in [0..len(nonPivots)-1]}
    \State KT[nonPivots[i],i] :=1
\EndFor
\State return transpose(KT)
\end{algorithmic}
\end{algorithm}
\end{figure}

\subsubsection{Evolutionary Algorithms}\label{sec:QDistEvol}
In this section, we introduce evolutionary distance-finding algorithms which may have improved accuracy compared to random information set algorithms.
The QDistRnd algorithm of \Cref{sec:QDistRnd} can be thought of as a search algorithm where our objective is to find column permutations which result in minimum weight codewords. 
Where the number of permutations giving minimum weight codewords is very small compared to the total number of possible permutations, it may take a very large sample to find the minimum distance. 
In \cite{distGenetic}, the distance of a classical code is calculated via a modified version of the random information set algorithm using column permutations generated by a genetic algorithm.  
The algorithm starts with an initial population of random permutations.
For each member of the population the columns of the generator matrix are permuted and the RREF calculated. 
The lowest weight row of the resulting matrix is used as the fitness function evaluate the permutations in the population.
The fittest parents are selected and used to populate the next generation based on random mutation or crossing parents or both.
The authors show that this leads to improved results for a selection of GF(2) and GF(8) classical codes.
For an introduction to genetic algorithms and other meta heuristic optimisation techniques see \cite{metaheuristics}.

QDistEvol is a modified version of the genetic algorithm in \cite{distGenetic} and is based on our previous work on optimising stabiliser codes via evolutionary algorithms \cite{QECC_evol}.
In this work, we optimise QDistEvol and show that it works well for a range of classical and quantum codes.
We now explain the QDistEvol algorithm in detail and how it differs from the algorithm in \cite{distGenetic} - pseudocode  is set out in \Cref{alg:QDistEvol}.
As for the algorithm in \cite{distGenetic}, QDistEvol starts with a population of random column permutations.  
The size of the population ($\lambda$) and the number of parents selected to propagate the next generation ($\mu$) are key parameters to the algorithm. 
 
Low weight codewords for each permutation are found by calculating the matrix $K$ in a similar way to QDistRnd, rather than calculating the RREF of the generator matrix. 
The fitness function in \cite{distGenetic} has discrete values whereas the one used by QDistEvol has continuous values and so is better able to distinguish similar solutions. QDistEvol's fitness function is a tuple - the minimum weight together with the average weight of (non-trivial) codewords in $K$. 
  
\begin{figure}[h!]
\centering
\begin{subfigure}[t]{.33\textwidth}
  \centering
  \includegraphics[width=\linewidth]{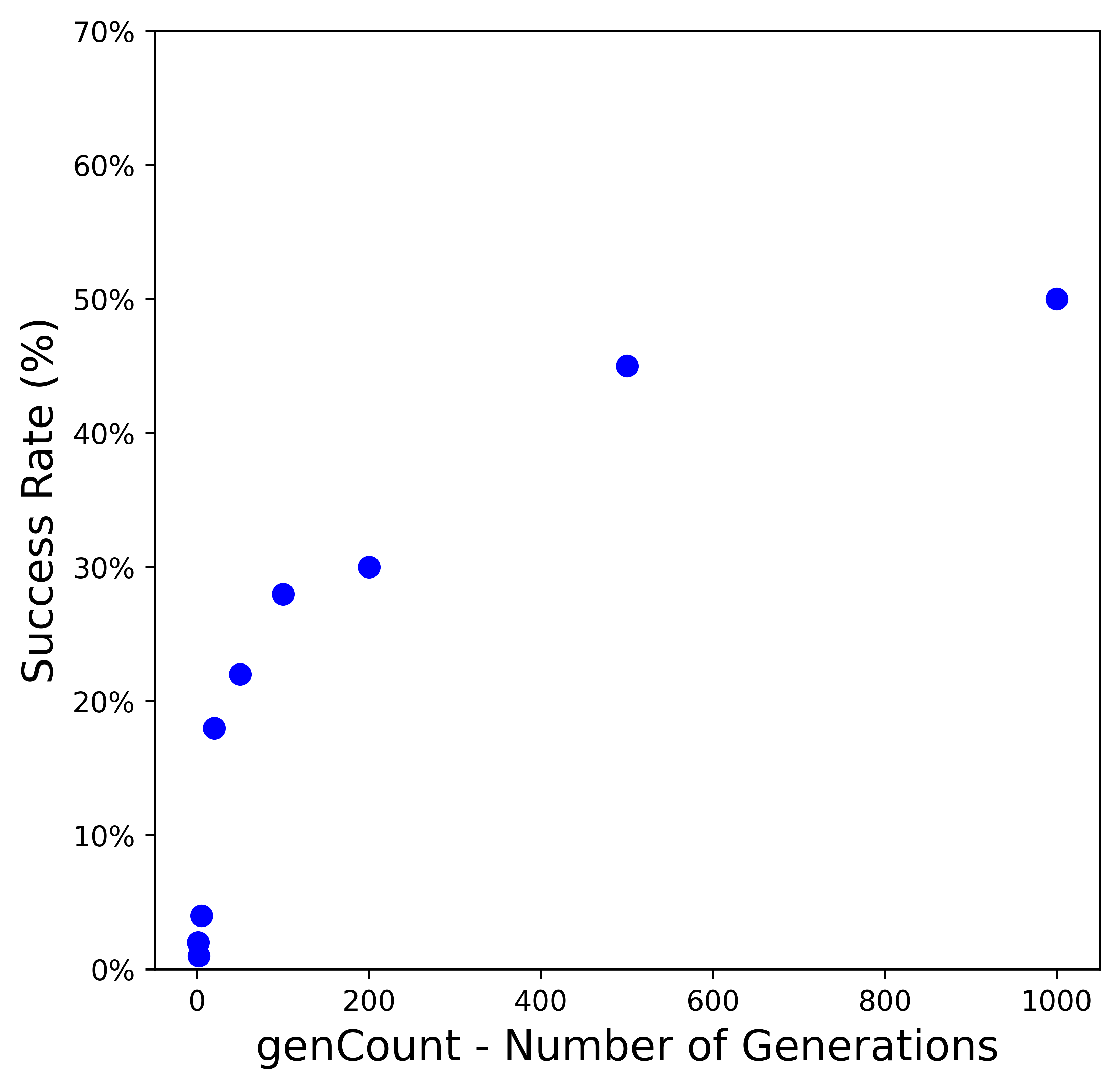}
  \subcaption{Sensitivity to Number of Generations}\label{fig:BB756_nGen}
    \end{subfigure}%
\begin{subfigure}[t]{.33\textwidth}
  \centering
  \includegraphics[width=\linewidth]{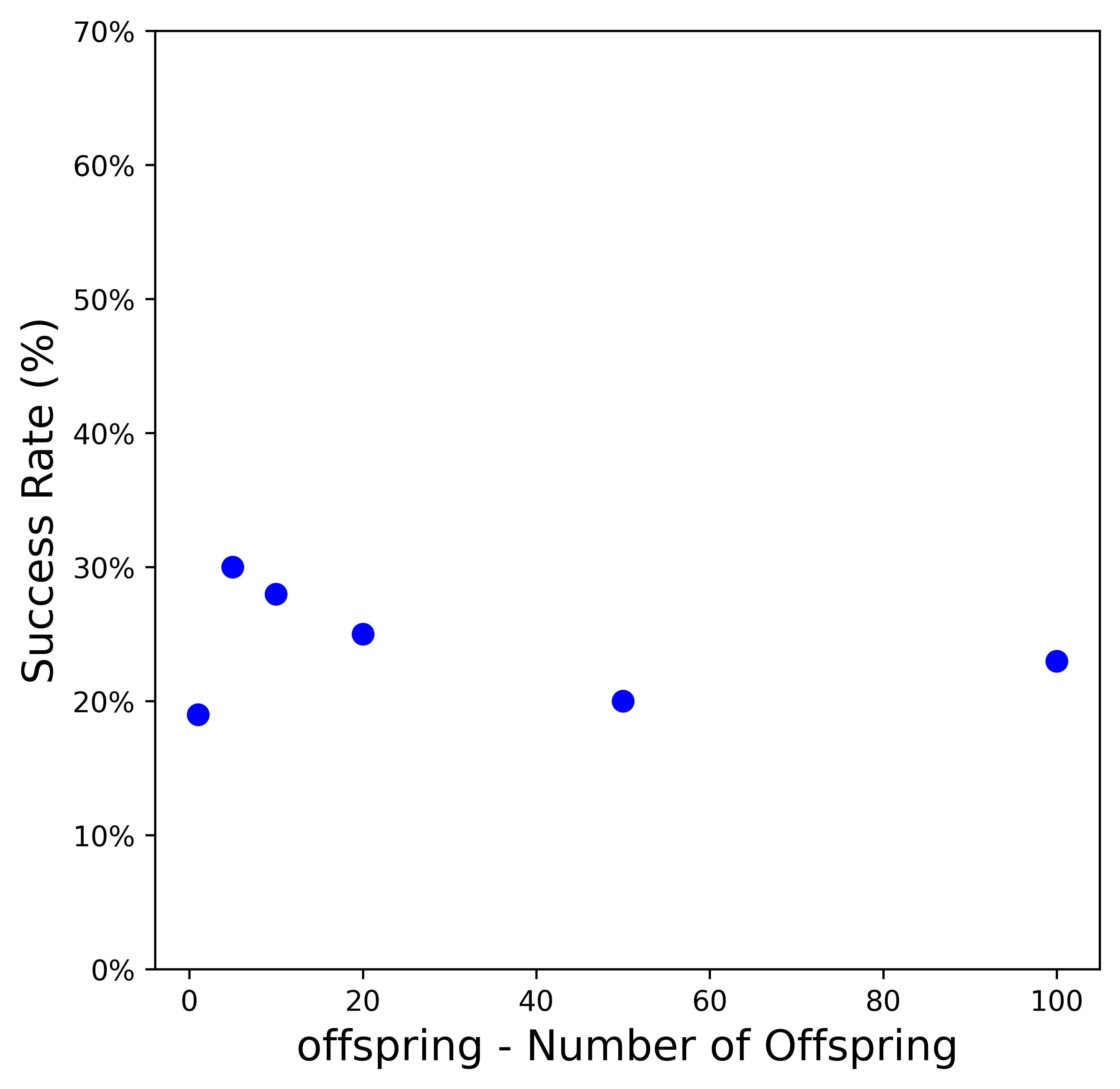}
  \subcaption{Sensitivity to Number of Offspring}\label{fig:BB756_offspring}
    \end{subfigure}%
\begin{subfigure}[t]{.33\textwidth}
  \centering
  \includegraphics[width=\linewidth]{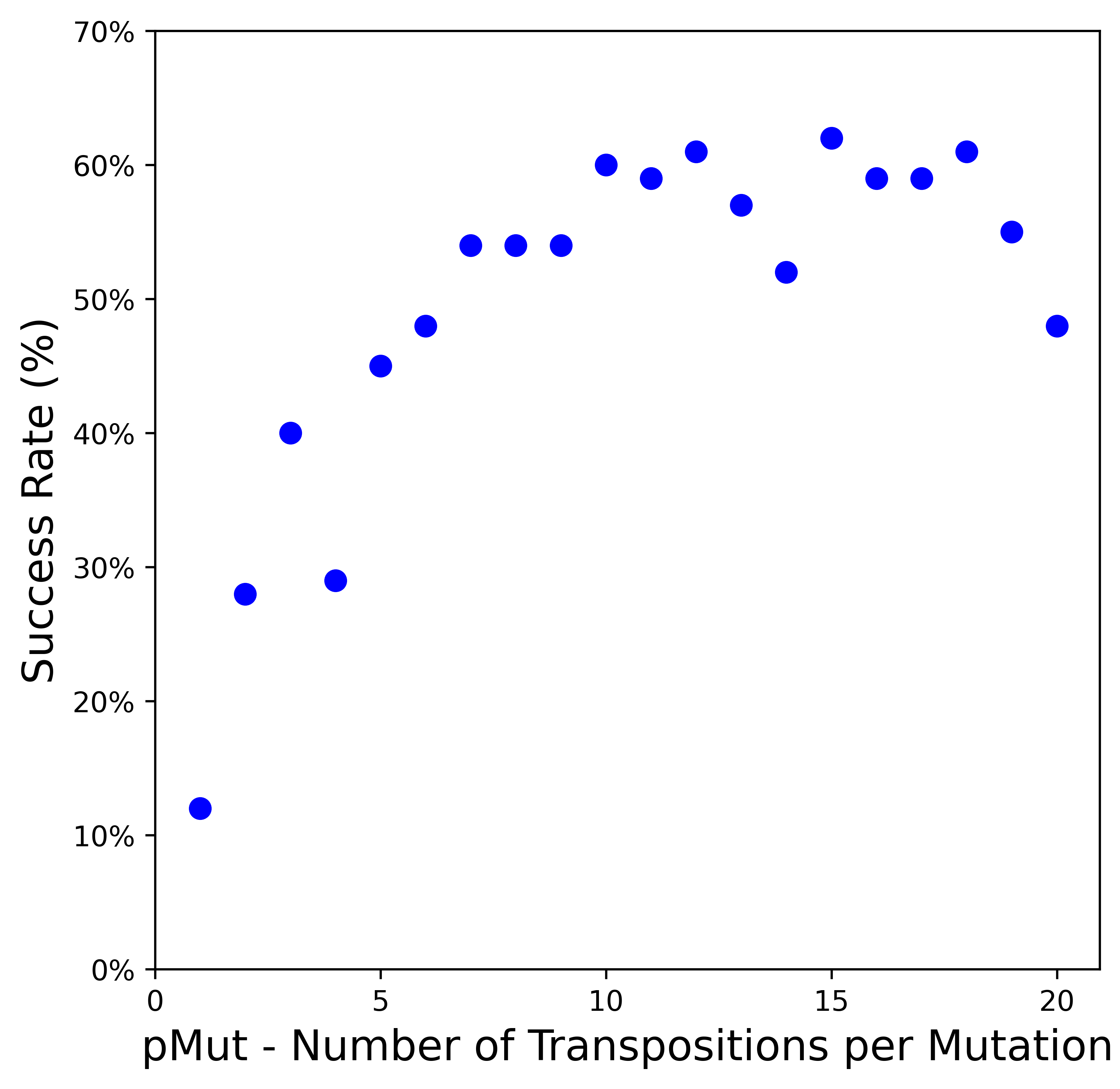}
  \subcaption{Sensitivity to Number of Transpositions per Mutation}\label{fig:BB756_pMut}
    \end{subfigure}%
\caption{QDistEvol Sensitivity Analysis for 756-qubit bivariate bicycle code. Here we plot the success rate for QDistEvol returning a distance of 34 when varying the number of generations (\texttt{genCount}), number of offspring per parent ($\mu$) and number of transpositions per mutation (\texttt{pMut}) parameters with \texttt{iterCount}=10000 RREF calculations per trial. We used 250 trials for this analysis with  default settings \texttt{genCount}=100, $\mu$ = 10 and \texttt{pMut}=2 and \texttt{swapPivot}=\texttt{TRUE}.}
\label{fig:BB756_parameters}
\end{figure}

Using the fitness function, the best $\mu$ permutations are selected as parents of the next generation,  each of which is used to generate $\lambda/\mu$ child permutations.
The child permutation is generated by applying random transpositions to the parent permutation. 
The number of transpositions given by $p_\text{Mut}(1 + s_\text{Mut}v)$ where $p_\text{Mut}, s_\text{Mut}$ are parameters which can be set by the user and $v$ is a random number between 0 and 1 for each mutation. 
In QDistEvol we track the pivots which result from the RREF operation to produce the matrix $K$.
Using the boolean \texttt{swapPivot} parameter, users can restrict to mutations which which swap between pivots and non-pivots.
In our benchmarking, we saw that this increases the accuracy of the method (see \Cref{fig:BB756_parameters}).
The algorithm in \cite{distGenetic} also uses a cross operation to produce the next generation, but this is not the case for QDistEvol and only mutation is used.

The QDistEvol algorithm terminates after a specified  number of generations. 
For benchmarking purposes, we use the same total number of RREF calculations for both QDistRnd and QDistEvol (10000 unless stated otherwise).

There is considerable scope for tuning the parameters $\lambda, \mu, p_\text{Mut}, s_\text{Mut}$ and the number of generations to improve the performance of QDistEvol for particular code families.
By default we use 100 generations each with a population of $\lambda:=100$ and select the best $\mu:=10$ parents. 
The algorithm is very sensitive to the choice of mutation process.
We found that increasing the average number of transpositions by increasing $p_\text{Mut}$ with the block size $n$ of the code gave good results for a wide range of code families. 
By default our algorithm uses $p_\text{Mut} := n/50$ and   $s_\text{Mut}:=1/5$. 
By default we also set \texttt{swapPivot:=TRUE}.

In \Cref{fig:BB756_parameters}, we analyse different parameter settings for QDistEvol when applied to the 756-qubit bivariate bicycle code of \cite{BB_IBM}.
The best estimate for the distance for the 756-qubit code is 34.
In a test of 100 runs of QDistRndMW each with 10000 trials a distance of 34 was returned only twice, giving a base success rate of 2\% for this code.
We then tested QDistEvol using 250 runs for each parameter setting with 10000 trials each and tracked the success rate for each parameter setting.
In \Cref{fig:BB756_nGen} we see that increasing the number of generations to 1000 (whilst maintaining the same number of RREF calculations at 10000) increased the success rate to around 50\%. 
When varying the number of offspring, we saw that a relatively low number of 5 offspring gave the best results, though the maximal success rate peaked at 37\%.
Increasing the value of \texttt{pMut}, which controls the number of transposition per mutation, had the most significant effect. We found that a value of \texttt{pMut} = 15  gave a success rate of over 60\% per trial. When testing for the entire bivariate data set and other data sets, we found that scaling \texttt{pMut} with the code block size gave good results and the figure of $1/50 \approx 15/756$ was a good default scaling factor.
We also tested the binary \texttt{swapPivot} parameter parameter and found that setting this to \texttt{TRUE} had a significant impact with a success rate of 12\% for \texttt{FALSE} versus a success rate of 28\% for \texttt{TRUE} using default parameter settings.

\captionsetup[subfigure]{margin=5pt}

There are also a number of choices for non-CSS stabiliser codes which can affect performance of QDistEvol. 
We found that reordering the columns of the check  matrix so that columns corresponding to the same qubits are next to each other led to more accurate results - this is controlled via the boolean \texttt{regroupPerm} parameter.
The block representation of the checks can be changed by varying the \texttt{GF4blockRep} parameter. 
We analyse the effect of varying the parameter settings  \texttt{swapPivot}, \texttt{regroupPerm} and \texttt{GF4blockRep} in \Cref{fig:Q80BlockSensEvol}.

\begin{figure}[p]
\begin{algorithm}[H]
\caption{QDistEvol - CSS Quantum Codes}\label{alg:QDistEvol}
\begin{algorithmic}
\State\textbf{Input:} 
\begin{itemize}
    \item CSS code defined by $r\times n$ binary X-check matrix $H_X$ and $k \times n$  X-logical basis $L_X$.
    \item nGens: number of generations to run
    \item $\lambda$: population size
    \item $\mu$: number of parents to select from each generation
    \item $p_\text{Mut}$: average number of transpositions for each mutation
    \item $s_\text{Mut}$: variance in number of transpositions per mutation
\end{itemize}
\State\textbf{Output:}  
\State An upper bound on the Z-distance of the code
\State\textbf{Method:}
\State population := [randomPerm(n) for i in [0..$\lambda-1$]]
\State UB := n
\For{i in [1..nGens]}
    \State mList = list()
    \For{cList in population}
        \State $K := \text{KerZ2}(H_X,\text{cList})$
        \State $C := K L_X^T$
        \State wList := list()
        \For{$[i: \text{wt}(C[i])>0]$}
            \State wList.append($\text{wt}(K[i])$)
        \EndFor
        \State mList.append(min(wList),average(wList)
    \EndFor
    \State popOrder := argsort(mList)
    \State wMin, avgMin := mList[popOrder[0]]
    \State UB := min(UB,wMin)
    \State populationNew := list()
    \For {i in [0..$\mu-1$]}
        \State cListParent := population[popOrder[i]]
        \For {j in [0..$\lambda/\mu-1$]}
            \State cListNew := [0..n-1]
            \State $v := $ randomNumber(0..1)
            \State transCount := round($p_\text{Mut}(1 + s_\text{Mut}v)$)
            \For {l in [1..transCount]}
                \State cListNew :=  randomTransposition(n) $\circ$ cListNew 
            \EndFor
            \State populationNew.append(cListParent $\circ$ cListNew)
        \EndFor
    \EndFor
    \State population := populationNew
\EndFor
\State return UB
\end{algorithmic}
\end{algorithm}
\end{figure}

 \subsubsection{Syndrome Decoder Method}\label{sec:syndrome_decoder}
The syndrome decoder method has been used in \cite{BB_IBM} to estimate the distance of large quantum LDPC CSS codes  and the algorithm is set out in \Cref{alg:syndrome_decoder}.
To find low-weight logical $Z$ operators, we  select a random logical operator of $X$-type. 
We then use a  decoder to find a correction for a syndrome which is zero for all of the $X$ checks, and 1 for the randomly selected logical operator. 
The correction corresponds to a non-trivial logical $Z$ operator because it commutes with all of the $X$-checks and anti-commutes with the randomly selected logical $X$ operator.
As is the case for random information set algorithms, the syndrome decoder method  is typically run for a large number of iterations.
For each randomly selected logical $X$ operator, the chance of it anti-commuting with the lowest weight logical $Z$ operator is $\frac{1}{2}$.
Assuming the decoder identifies a minimum weight correction each time, the likelihood of not finding the lowest weight logical operator after $m$ iterations is $2^{-m}$.
The accuracy and processing time of this method depends on whether the decoder is well-suited to the code or circuit family of interest. 

The method can also be used for non-CSS quantum codes using a random, non-trivial logical Paulis and a similar syndrome vector as for CSS codes. 
For non-CSS codes, we found that using the three-block representation of the stabiliser generators gave better results than the two or four-block representations (see \Cref{sec:detector_view,sec:fourBlock}).
To apply the method to classical binary linear codes, we use the check matrix $H$ and non-trivial linear combinations of the basis of the complementary subspace for the logical vector (see \Cref{sec:DEM_direct_calc}).

The syndrome decoder  algorithm also works well for circuits and in this case we choose a random linear combination of the observables from the detector error matrix.
Many decoders can take detector error models as input and often allow different probabilities to be set for each error mechanism, allowing for a range of error models to be handled. 

In \cite{BB_IBM}, the BP-OSD decoder \cite{ldpc_decoders} was used to estimate an upper bound for the bivariate bicycle family of CSS codes and for syndrome extraction circuits. 
In fact any syndrome decoder can be used in this algorithm, and in our \href{https://github.com/m-webster/codeDistancePYPI}{codeDistance package} we allow the user to select various decoders and the settings of the decoder can also be varied (for instance the number of BP iterations and type of post-processing).

Given a decoder which is guaranteed to give a minimum weight correction, applying the decoder to find a correction for each member of a logical basis and taking the minimum correction weight yields an exact distance-finding algorithm (this method is proposed in  \cite{beverland2025failfasttechniquesprobe}). 
We tested this approach and found that it leads either to an inaccurate distance estimate (if using a heuristic decoder such as Tesseract) or long processing times (if using an exact decoder) and in practice does give an advantage over the solver methods discussed in \Cref{sec:solvers}.

For benchmarking, we use the BP-OSD decoder from the LDPC python package \cite{Roffe_LDPC_Python_tools_2022}. 
We used the following settings for benchmarking the BP-OSD algorithm. 
We used 100 iterations of product-sum belief propagation with a parallel schedule.
We used combination sweep ordered statistics decoding with order 1.

The syndrome decoder algorithm in our \href{https://github.com/m-webster/codeDistancePYPI}{codeDistance package} has several options for sampling of random logical operators which can enhance performance in some circumstances.
For CSS codes with one logical qubit, $L_X$ has only one row. 
If we restrict ourselves to linear combinations of the generators, this results in only a single logical operator being sampled with far from optimal results.
Sampling can be improved by randomly permuting the columns of the check matrix or DEM before decoding.
Additionally, one can add random combinations of stabilisers to generate random logicals. 
We analysed the results of distance-finding using BP-OSD for the $[[48,5,10]]$ best-known-distance code from codetables.de.
Finding the correct distance of this code is quite challenging.
The BP-OSD syndrome decoder method without DEM permutations or adding random combinations of stabilisers over 100 trials did not return the correct distance of 10 even once.
Adding stabilisers did not change the success rate, but applying random permutations to the DEM increased success rate to 50\%. 
Applying both random stabilisers and DEM permutations resulted in a 100\% success rate.

\begin{figure}[H]
\begin{algorithm}[H]
\caption{Distance via Syndrome Decoder - CSS Quantum Code}\label{alg:syndrome_decoder}
\begin{algorithmic}
\State \textbf{Inputs:}
\State $H_X: r\times n$ X-check matrix
\State $L_X: k \times n$  X-logical basis
\State iterCount: desired number of iterations
\State \textbf{Output:}
\State An upper bound on the Z-distance
\State \textbf{Method:}
\State $\mathbf{s} := [0] * r + [1]$
\State $\text{UB}:=n$
    \For{ $i$ in $[1..\text{iterCount}]$}
        \State Let $\mathbf{a}$ be a random binary vector of length $r$
        \State Let $\mathbf{b}$ be a random, non-zero binary vector of length $k$
        \State The vector $\mathbf{x}:=\mathbf{a}H_X + \mathbf{b}L_X$ is a random non-trivial logical
        \State $P:=\text{randomPermutation}(n)$
        \State Find a correction $F$ for $\mathbf{s}$ using a syndrome decoder on $\begin{bmatrix}
            H_X\\
            \mathbf{x}
        \end{bmatrix}P$
        \State $\text{UB} := \text{min(UB, wt(F))}$
    \EndFor
    \State return $\text{UB}$
\end{algorithmic}
\end{algorithm}
\end{figure}

 \subsection{Enumeration of Errors}\label{sec:alg_enumeration_errors}
 An alternate strategy to enumerating codewords is to enumerate errors. 
  Algorithms in this family generally give an increasing lower bound on distance, in contrast to the codeword enumeration algorithms of \Cref{sec:alg_enum_codewords} which give a reducing upper bound estimate.
 The exhaustive method for finding the $Z$ distance of a length $n$ CSS code involves generating binary strings of length $n$ of increasing weight and terminating when an error satisfies all X-checks but violates at least one X-logical, and is set out in \Cref{alg:error_enumeration_exhaustive}.
The complexity of the exhaustive error enumeration algorithm is $\mathcal{O}((ne/d)^d)$ for CSS codes where $d$ is the minimum distance.
For non-CSS quantum codes, we enumerate all Pauli strings on $n$ qubits of increasing weight and check if the strings commute with all stabiliser generators and anti-commute with at least one logical Pauli
 and has complexity $\mathcal{O}((3ne/d)^d)$.
As in the case of exhaustive codeword enumeration, exhaustive error enumeration can only be used for codes with low distance and/or small block size.
\begin{figure}[H]
\begin{algorithm}[H]
\caption{Distance via Exhaustive Error Enumeration - Quantum CSS Code}\label{alg:error_enumeration_exhaustive}
\begin{algorithmic}
\State\textbf{Input:} A CSS code defined by:
    \State $H_X: r\times n$ X-check matrix
    \State $L_X: k \times n$ logical X basis
\State\textbf{Output:}  
\State The Z-distance of the code
\State\textbf{Method:}
\For{LB in [1..n]}
    \For{binary vectors $E$ of length $n$ and weight LB}
        \If{$EH_X^T  = 0$ and $EL_X^T \ne 0$}
        \State return LB
        \EndIf
    \EndFor
\EndFor
\end{algorithmic}
\end{algorithm}
\end{figure}

 \subsubsection{Matching Bipartition}\label{sec:meet_in_the_middle}
 The matching bipartition algorithm (also referred to as a `meet in the middle' algorithm) is a way of reducing the complexity of error enumeration.
 We explain how the algorithm find the $Z$-distance of a quantum CSS code defined by X-checks $H_X$ and X-logical basis $L_X$.
 We generate binary strings of length $n$  of increasing weight \texttt{LB} which correspond to strings of Pauli $Z$ operators. 
 We first consider the case where \texttt{LB} is even. 
For each row ${P}$ in $L_X$ we iterate through each string $E$ of weight $w := \text{LB}/2$.
We calculate the syndrome $\mathbf{s} := EH_X^T$.
If $PE = 0$, then $E$ commutes with ${P}$, and we add $\mathbf{s}$ to the set $C_w$ and otherwise to the set $A_w$.
If there is a common element of $C_w$ and $A_w$, then we know that there are errors $E$ and $F$ both of weight $w$ which have the same syndrome but where $E$ commutes with ${P}$ and $F$ anti-commutes. As a result, the combined error $E\oplus F$ satisfies the $X$-checks but anti-commutes with ${P}$ and so is a non-trivial logical of weight at most \texttt{LB}.
For \texttt{LB} odd, we set $w_1 :=  (\text{LB}+1)/2$ and $w_2:=\text{LB}-w_1$ and calculate $C_{w_1}, C_{w_2}, A_{w_1}$ and $A_{w_2}$ and check $C_{w_1} \cap A_{w_2}$ and $A_{w_1} \cap C_{w_2}$.
The complexity of the algorithm is $\mathcal{O}((2ne/d)^{d/2})$ for CSS codes where $n$ is the number of physical qubits and $d$ is the minimum distance. 
Pseudocode for the algorithm for CSS codes is set out in \Cref{alg:meet_in_the_middle}.

It is straightforward to extend the above algorithm to non-CSS quantum code - in this case, we select binary strings of length $2n$ which have increasing symplectic weight, choose a basis of the logical Paulis and use the symplectic inner product to determine commutation relations.
For non-CSS codes, the complexity of the algorithm is $\mathcal{O}((6ne/d)^{d/2})$.

 The algorithm has been implemented for non-CSS stabiliser codes as part of the Qiskit QEC package \cite{qiskit_qec} and both C and python versions are available, though these methods only work on codes with under 63 qubits. 
 For CSS codes, the algorithm has been implemented in the QubitSerf C package \cite{qubitserf}. 
We have made the QubitSerf  version available as part of our \href{https://github.com/m-webster/codeDistancePYPI}{codeDistance package} and generalised it to non-CSS codes using the method of \cite{fourBlockCSS} to map non-CSS codes to CSS codes.
We also include a python implementation of the meet in the middle algorithm, \texttt{MeetMiddleMW}, which works for classical, CSS and non-CSS quantum codes.
Performance of \texttt{MeetMiddleMW} is competitive with the other implementations and it does not have a limit on maximum code size as illustrated in \Cref{fig:BenchmarkBZMM} for hyperbolic surface codes.

\begin{figure}[H]
\begin{algorithm}[H]
\caption{Distance via Meet in the Middle Error Enumeration - Quantum CSS Code}\label{alg:meet_in_the_middle}
\begin{algorithmic}
\State\textbf{Input:} A CSS code defined by:
    \State $H_X: r\times n$ binary X-check matrix
    \State $L_X: k \times n$ binary logical X-basis
\State\textbf{Output:}  
\State The minimum Z-distance
\State\textbf{Method:}
\For{LB in [1..n]}
    \State wList = []
    \Comment{List of error weights. For even LB, we in fact only need to calculate errors of weight LB/2}
    \State wList.append(LB//2)
    \State wList.append(LB - wList[0])
    \For{P in $L_X$}
        \State CAList := []
        \For{w in wList}
        \Comment Set of syndromes which commute/anti-commute with P
            \State CA := [set(),set()]
            \For{binary vectors E of length n and weight w}
                \State $s:= EH_X^T$
                \State $p:= EP^T$
                \State CA[p].add(s)
            \EndFor
            \State CAList.append(CA)
        \EndFor
        \If{(CAList[0][0] $\cap$ CAList[1][1]) != $\emptyset$) or (CAList[1][0] $\cap$ CAList[0][1]) != $\emptyset$)}
        \State return LB
        \EndIf
    \EndFor
\EndFor
\end{algorithmic}
\end{algorithm}
\end{figure}

 \subsubsection{Connected Cluster Algorithm}\label{sec:connected_cluster}
The connected cluster algorithm (also referred to as the irreducible cluster or linked cluster algorithm) of \cite{Dumer_Kovalev_Pryadko_2014,Dumer_Kovalev_Pryadko_2016} is designed for use with LDPC (Low Density Parity Check) codes where the row and column weights of the check matrix are bounded.
It uses the fact that low weight errors form connected components on the Tanner graphs of LDPC codes.
The algorithm is an exhaustive breadth-first search along the Tanner graph and pseudocode for the algorithm is set out in \Cref{alg:connected_cluster}.
For an LDPC code whose check matrix $H$ has row weight $m$, the complexity of this algorithm is $\mathcal{O}([2(m-1)]^d)$ (see \cite{Dumer_Kovalev_Pryadko_2016}).
For benchmarking, we used the connected cluster algorithm in the C m4ri-dist GitHub repository accessed via the \texttt{dist\_m4ri\_CC} function in our  \href{https://github.com/m-webster/codeDistancePYPI}{codeDistance package}.
We have also implemented a python version of the algorithm via the \texttt{ConnectedClusterMW} function.

The algorithm can be applied to non-CSS stabiliser codes using either the three or four-block representation (see \Cref{sec:detector_view,sec:fourBlock}).
When applied to the codeTables best-known-distance codes dataset of \Cref{sec:codeTables_QECC}, we found that the three block representation led to much faster run time than the four block representation.

The main drawback of the connected cluster method is that the number of codewords to be assessed in the breadth-first search may grow very quickly before the distance is found.
For codes which have Tanner graphs of low degree (eg surface and colour codes), the algorithm performs quite well.
The C implementation of the algorithm in m4ri-dist also allows the user to restrict the search to logical operators to those which have support on a specified qubit.
For symmetric codes where each qubit supports the same number of low-weight logical operators, this simplification can allow for larger codes to be analysed. 
One disadvantage of the C implementation is that when execution is terminated before completion, partial lower bound results are not available.
\begin{figure}[H]
\begin{algorithm}[H]
\caption{Distance via Connected Cluster}\label{alg:connected_cluster}
\begin{algorithmic}
\State\textbf{Input:} A quantum CSS code defined by:
    \State $H_X: r\times n$ binary X-check matrix
    \State $L_X: k \times n$ binary X-logical basis
\State\textbf{Output:}  
\State Minimum Z-distance
\State\textbf{Method:}
\Comment Matrix of errors related to each other by a path via one check - matrix multiplication over the integers
\State $H_2 :=  H_X^TH_X$
\State Q0 := [1..n]
\For{LB in [1..n]}
    \State Q1 = []
    \For{E in Q0}
        \If{$EH_X^T$ = 0 AND $EL_X^T$ != 0}
            \State return LB
        \EndIf
        \For{j in supp($E H_2$) - E}
            \State Q1.add(E $\cup $ \{j\})
        \EndFor
    \EndFor
    \State Q0 = Q1
\EndFor
\end{algorithmic}
\end{algorithm}
\end{figure}

 \subsubsection{Stim Undetectable Error Algorithm}\label{sec:stim_UE}
The Stim package has a number of functions for distance-finding which are designed to work well on topological codes and circuits where the degree of the Tanner graph is small. 
They are similar to the connected cluster algorithm of \Cref{sec:connected_cluster} but the search is truncated to reduce the memory requirement and increase processing speed.

We first describe the {search\_for\_undetectable\_logical\_errors} function (referred to in this work as \texttt{UEStim}) which takes as input the check detector matrix $H$ and the logical detector matrix $L$. 
The function uses several parameters, based on the syndrome weight (i.e. number of detectors or rows of $H$ flipped by errors) to truncate the search as follows:
\begin{itemize}
    \item dont\_explore\_edges\_with\_degree\_above:  excludes error mechanisms with syndrome weight greater than the set value (weight of single columns of $H$) - referred to below as \texttt{sMaxE} below for brevity. 
    \item  dont\_explore\_detection\_event\_sets\_with\_size\_above: combinations of errors with syndrome weight greater than the set value are not explored (determined by the weight of the sum of the corresponding columns of $H$) - \texttt{sMax}.
    \item dont\_explore\_edges\_increasing\_symptom\_degree: if set to True, any errors which increase the initial syndrome weight are not explored - \texttt{sNonInc}.
\end{itemize}

We now explain how these parameters are used by the {UEStim} function in our \href{https://github.com/m-webster/codeDistancePYPI}{codeDistance package}.
The function runs the Stim undetectable error function using increasing values of \texttt{sMax}, starting with 2.
Once the same minimum distance has been found twice, the algorithm terminates and returns a codeword/logical at that distance.
The \texttt{sMaxE} parameter is set to the number of detectors, and so is not used to restrict the search. 

The Stim {shortest\_graphlike\_error} function (called using GEStim in our package) can be thought of as a special case of the Stim undetectable errors function but where \texttt{sMaxE} and \texttt{sMax} are both set to 2. 
The method is intended to be used on codes and circuits where errors can only flip at most two detectors (eg in the surface code and its syndrome extraction circuit).
Our package also includes the function {CCStim} which is intended for use on colour codes where each error can flip up to 3 detectors. 

The advantage of this family of algorithms is that it can be used on topological codes where the maximum number of detectors flipped by any error is known in advance.
Furthermore, the GEStim method is an exact method for distance-finding problems where each error flips either one or two detectors (such that shortest errors are guaranteed to be stringlike).
Distance finding problems with this property include Euclidean and hyperbolic surface codes (or any CSS code where the column weights of the X and Z check matrices are at most two).
Otherwise, the method gives an upper bound on distance only and may not be effective on all code families, which is why we classify these methods as heuristic algorithms in general.

\subsection{Solver-Based Methods}\label{sec:solvers}
One might want to find the exact distance of a code; in such cases heuristic methods will not suffice. We refer to algorithms that are able to certify that a given solution is the global optimum as \textit{exact} methods or algorithms. In this section we discuss two methods which are able to provide such a guarantee: linear programming and maximum satisfiability. 

\subsubsection{Mixed Integer Linear Programming}\label{sec:MIP}
Mixed Integer Linear Programming (MILP) can be used to find the minimum distance of a code. Here we describe how to recast the distance problem as a constraint optimisation problem which can be passed to mature existing solvers, following a similar approach to Ref.~\cite{landahl2011fault}.\\
\\
The general linear programming (LP) problem minimises an objective function $Z$ subject to linear constraints:
\begin{equation}
    \min Z = c^T x \quad \text{s.t.} \quad Ax \leq b, \quad x \geq 0.
\end{equation}
Here, $c$ represents the cost coefficients, while $A$ and $b$ are the coefficients and bounds of the linear constraints. Continuous LP is solvable in polynomial time~\cite{khachiyan1980polynomial}, imposing integrality renders the problem NP-hard~\cite{karp2009reducibility}, due to the feasible region becoming non-convex. MILP solvers accommodate integer constraints using a \textit{Branch-and-Bound} approach~\cite{land2009automatic}. This method \textit{relaxes} the integrality constraints and solves the problem over a continuous solution space. It then iteratively "branches" on fractional results, creating sub-problems for $\lfloor x_i \rfloor$ and $\lceil x_i \rceil$ and "prunes" suboptimal solutions. Two sub-problems are created for each integer variable; hence, the MILP problems have exponential complexity in the number of integer variables.\\
\\
To utilise modern solvers, the binary distance finding problem must be cast into the MILP framework.
The approach we describe here is an adaptation of the formulations for MILP decoding of Refs.~\cite{landahl2011fault,breuckmann2017localdecoders,surface_code_CX,takada2024Ising}, with explicit constraints added to ensure anti-commutation with at least one logical operator~\cite{breuckmann2023privatecommunication,BB_IBM}.
The objective function is to minimise the Hamming weight of an error operator $E \in \mathbb{F}_2^{n}$, subject to the parity check condition $HE=0 \mod 2$ and logical operator anti-commutation $LE \neq 0 \mod 2$. Standard solvers do not natively support modular arithmetic; hence, we map the problem to real variables by introducing integer \textit{slack} variables. The slack variables cancel out multiples of the modulus. The distance finding problem can be recast as:
\begin{align}
    \begin{bmatrix}
        H & 0 & 2I & 0 \\
        L & I & 0 & 2I \\
        0 & I & 0 & 0 
    \end{bmatrix} 
    \begin{bmatrix}
        E \\
        P  \\
        {S}_H \\
        {S}_L
    \end{bmatrix}
    = 
    \begin{bmatrix}
       \mathbf{w}(HE + 2{S}_H) \\
        \mathbf{w}(LE + {P} + 2{S}_L) \\
        \mathbf{w}({P})
    \end{bmatrix} 
    = 
    \begin{bmatrix}
        0 \\
        0\\
        \geq 1
    \end{bmatrix} 
\end{align}
Here, $E \in \mathbb{F}^{n}_{2}$ is the error vector and $\mathbf{w}$ returns the Hamming weight. The vector $P \in \mathbb{F}^{k}_{2}$ acts as a binary slack variable, ensuring $E$ anti-commutes with at least one logical operator, which is enforced by the third row constraint $\mathbf{w}(P) \geq 1$. The vectors $S_H \in \mathbb{Z}^{n-k}$ and $S_L \in \mathbb{Z}^{k}$ are composed of slack variables which offset even-weight contributions in the weight function. \\
\\
For CSS codes the above implementation is sufficient for finding the $X-$ and $Z$- distances. One merely needs to replace $H/ L$ with $H_Z/L_Z$ to compute the $X-$ distance and $H_X/L_X$ for the $Z-$distance. For non-CSS codes, one can construct the detector matrix (DEM) as described in Sec.~\ref{sec:detector_view} and perform an analogous  operation on the logical observables $L_O$. Then one can replace $H/L$ above with DEM/$L_O$. Note that one is solving for an expanded error vector $E \in \mathbb{F}^{3n}_2$. To get the true error configuration corresponding to the minimum distance, compute $(L_z \oplus L_y | L_x \oplus L_y).$ Additionally, for non-CSS codes the MIP-SCIP implementation only supports the three-block representation of the problem. Whereas Gurobi supports the two-block representation and the symplectic weight as the cost function. We find that using the two-block representation in Gurobi leads to faster run-times than the three-block representation see Fig. \ref{fig:QBlockSensGurobi}.


MW:

\begin{figure}[H]
\begin{algorithm}[H]
\caption{Distance via Mixed Integer Programming - CSS Codes}\label{alg:MIP}
\begin{algorithmic}
\State \textbf{Inputs:}
\State $H_X: r\times n$ X-check matrix
\State $L_X: k \times n$ X-logical basis
\State \textbf{Outputs:}
\State The Z-distance of the CSS code
\State \textbf{Method:}
\State Construct MIP model using following variables and constraints: 
\begin{itemize}
    \item Error $E$: length $n$ binary vector - the objective of the solver is to minimise the weight of $E$
    \item Detector slack variable $S_H$: length $r$ integer vector with constraint $H_X E = 2 S_H$
    \item Logical parity $P$: length $k$ binary vector with constraint $\text{wt}(P) >0$
    \item Logical slack variable $S_L$ length $k$ integer vector with constraint $L_X E = 2 S_L + P$
\end{itemize}
\State Apply a MIP solver and return $\text{wt}(E)$
\end{algorithmic}
\end{algorithm}
\end{figure}

\subsubsection{MaxSAT Solvers}\label{sec:MaxSAT}
An alternative exact method for finding the distance of a quantum code is to use MaxSAT solvers. Maximum Satisfiability (MaxSAT) extends standard Boolean Satisfiability by distinguishing between \textit{hard} clauses, which must be satisfied, and \textit{soft} clauses, which incur a penalty if unsatisfied. For our experiments, we use CASHWMaxSAT-CorePlus (CWM)~\cite{pan2025efficient}, a complete solver that utilises a core-guided strategy with conflict-driven clause learning~\cite{biere2009handbook}. While effective, CWM binaries are only available on UNIX devices; for flexibility, we integrate the PySAT solver~\cite{imms-sat18, itk-sat24} into the \href{https://github.com/m-webster/codeDistancePYPI}{codeDistance package}.\\
\\
To translate the distance finding problem into a solver-compatible format, we express the problem in Conjunctive Normal Form (CNF), using the MaxSAT formulation of distance finding introduced in~\cite{shutty2024stimpr703}. The elements of the error vector $E$ become Boolean variables, with soft clauses that penalise the solver for introducing errors. Additionally, the solver must satisfy the stabiliser constraints ($HE=0 \mod 2$) and logical operator constraints ($LE \neq 0 \mod 2$). This is achieved through the following clause structure. We apply uniformly weighted soft clauses that prefer $x_i = \text{False}$ for all variables. Any variable assigned True (indicating an error) incurs a penalty, ensuring the solver finds the minimum weight vector. The hard clauses enforce the parity constraints of the linear code. \\
\\CNF is based on logical ORs; we implement the $\mathbb{F}_2$ linear algebra using Tseitin encoding~\cite{tseitin1983complexity}. This introduces auxiliary variables to capture the XOR outputs of a sum. For an operation $C = A \oplus B$, we strictly forbid assignments that violate the XOR truth table using four hard clauses:

\begin{align}
    (A \lor B \lor \neg C) \quad & \text{forbids } 0 \oplus 0 = 1 \\
    (A \lor \neg B \lor C) \quad & \text{forbids } 0 \oplus 1 = 0 \\
    (\neg A \lor B \lor C) \quad & \text{forbids } 1 \oplus 0 = 0 \\
    (\neg A \lor \neg B \lor \neg C) \quad & \text{forbids } 1 \oplus 1 = 1
\end{align}
By enforcing these hard XOR constraints across the parity check matrix and logical operators, any valid assignment corresponds to a valid logical error, and the MaxSAT optimisation ensures it is the lowest weight instance. Stim has inbuilt functionality for generating WCNF encodings for Stim circuits; this also supports more complex noise models.\\
\\
The general method described above can be specialised for both CSS codes and non-CSS codes. For CSS codes, identifying $H = H_X$ and $L = L_X$ will find the $Z$ distance of the code, one can find the $X$ distance analogously. For non-CSS the code must be converted to a 3-block DEM as described in Sec.~\ref{sec:detector_view} and a similar transformation applied to the logical operators. Once these transformations are complete, the above general approach can be applied, taking care to convert the logical observables back to operators, $(L_z \oplus L_y | L_x \oplus L_y)$, if the logical operator structure is required. The pseudo-code for the algorithm can be found in App.~\ref{alg:SAT}.


\section{Algorithm Benchmarks}\label{sec:benchmarks}

In this section, we present our benchmarking results.
We first set out the methodology we  use in the benchmarking work, then discuss results for families of classical codes, quantum codes and quantum circuits.

\subsection{Benchmarking Methodology}\label{sec:benchmark_methodology}
Benchmarking was conducted on the UCL Physics Hypatia cluster using a single node with 40 cores and 8GB of memory per core. 
The benchmarking methodology is slightly different for exact distance-finding methods than for heuristic distance-finding methods.

Because the exact algorithms in the benchmark typically have long processing times, we set a maximum execution time of 8 hours for these. 
Where the algorithm was terminated early, we retrieved partial results where possible - for Magma, MIP and SAT solvers this usually gives an upper and/or lower bound on distance.
For exact algorithms, the main figures of merit are processing time, and the overall success rate.
The overall success rate is calculated from the number  of codes or circuits which could be completed within the 8 hour maximum plus those where partial results gave the lowest distance.
As the codetables datasets include a large number of codes where exact methods took over 8 hours to complete, we benchmarked exact methods against a reduced selection of codes for these datasets.

For heuristic methods, the key figure of merit was the overall success rate - the number of codes in the family where the algorithm returned the correct minimum distance.  
Where the minimum distance was not previously known, we used the lowest distance which was returned by any method in our benchmark.
Many heuristic algorithms are probabilistic and require a number of repeated trials to improve accuracy. 
For probabilistic methods, a maximum of 10,000 trials was used for all data sets.
We recorded the number of trials which returned the lowest known distance for each member of the data set. 
Deterministic algorithms on the other hand require only a single trial. 
To compare both types of methods, we use the processing time per successful trial to estimate the expected processing time to correctly calculate the distance.
A maximum run time of 48 hours for each code or circuit was set as this is limit for the UCL cluster. 

For the C random information set algorithm implementation {m4riRW}, it is not possible to calculate the time per successful trial.
This is because the method returns a single upper bound on distance, and does not give statistics on the distance returned in each trial. 
Accordingly, we used the QDistRndMW python function which implements the same algorithm to return the number of trials at the lowest known distance. 
Note that the python implementation has a different run-time complexity to the C implementation, particularly for large/sparse matrices as it does not use the `method of the four Russians' method for Gaussian elimination.

For deterministic methods, we use a single run for each code/circuit family member. 
For the solver-based MIP and SAT methods, we have a single trial. 
For the deterministic Stim Undetectable Error and Graphlike Error methods, we consider each choice of \texttt{sMax} to be a trial and report on the number of trials which return the lowest distance for the method (see \Cref{sec:stim_UE}).

Our presentation of benchmarking results includes one table each for exact and heuristic methods which shows results in aggregate for each of the data sets.
We also include three charts which show processing time per successful trial and success rate with increasing code length which indicate how each method scales.

\subsection{Classical Codes}\label{sec:benchmark_classical}
We used two classical code data sets in our benchmarking exercise - a set of relatively small binary linear codes from codetables.de and a set of larger classical lifted product LDPC codes.

\subsubsection{GF(2) Linear Codes from codetables.de}\label{sec:codeTables_GF2}
Our first data set for benchmarking is a set of binary linear codes from codetables.de \cite{codetables}.
This is a set of 121 best-known-distance codes with between 5 and 128 bits.
The most challenging classical linear codes have encoding rates of around $\frac{1}{2}$ (see \cite{Dumer_Kovalev_Pryadko_2014} page 2).
Accordingly, we applied the QDistRndMW algorithm to CodeTables codes with encoding rates between $\frac{1}{3}$ and $\frac{2}{3}$ with  $n \le 128$ bits.
We selected the code for each value of $n$ which had the lowest proportion of successful trials or, where the correct distance is not found, the largest difference between the correct distance and the estimated distance.

In \Cref{tab:codetables_GF2_exact} we compare  exact methods for distance finding for this data set.
The row `No Result' indicates codes where the the algorithm did not return any distance estimate, even after 8 hours of run time. 
Due to the long processing time of exact algorithms, we considered only codes with up to 100 qubits (93 codes for this data set).

The MagmaMinWeight algorithm completed execution within this timeframe for all codes (up to $n=128$) and also had the lowest average processing time per trial.
The Gurobi mixed integer programming algorithm completed 75 codes within the 8-hour window,  the SCIP mixed integer programming algorithm completed 67, the SAT command line interface algorithm 66 and the m4riCC connected cluster algorithm 51.
Outside the 8 hour maximum execution time, we also consider the algorithm to have succeeded if the partial result gives the correct distance. 
For the m4riCC algorithm, partial results are not available if execution terminates early and this impacts the overall success rate.

\begin{table}[h!]
\setlength\tabcolsep{2pt}
\fontfamily{lmss}\fontsize{8}{9}\selectfont{
\begin{center}							
\begin{tabular}{ |l|	r|	r|	r|	r|	r|	}	\hline
 &	\textbf{Gurobi} &	\textbf{MIP-SCIP} &	\textbf{CLISAT} &	\textbf{m4riCC} &	\textbf{Magma}	\\	\hline
\textbf{Result Returned} &	93 &	93 &	93 &	51 &	93	\\	
\textbf{Completed < MaxTime} &	75 &	67 &	66 &	51 &	93	\\	
\textbf{Completed > MaxTime} &	18 &	26 &	27 &	0 &	0	\\	
\textbf{No Result} &	0 &	0 &	0 &	42 &	0	\\	
\textbf{At lowest distance} &	93 &	93 &	78 &	51 &	93	\\	
\textbf{Overall success rate} &	100.0\% &	100.0\% &	83.9\% &	54.8\% &	100.0\%	\\	\hline
\textbf{Total Time} &	6.2E+05 &	8.8E+05 &	8.3E+05 &	1.3E+06 &	2.8E+03	\\	
\textbf{Time/Trial} &	6.7E+03 &	9.4E+03 &	8.9E+03 &	1.4E+04 &	3.1E+01	\\	
\textbf{Time/Successful Trial} &	6.7E+03 &	9.4E+03 &	1.1E+04 &	2.5E+04 &	3.1E+01	\\	\hline
\end{tabular}							
\end{center}									
}
\caption{Distance finding benchmark for codetables.de GF(2) codes - Exact Methods.}
\label{tab:codetables_GF2_exact}
\end{table}

In \Cref{tab:codetables_GF2_heuristic}, we display results for heuristic methods. We see that QDistRnd, QDistEvol and BP-OSD methods all performed well on this dataset, finding the correct distance in all instances.
The C implementation of the QDistRnd algorithm had the lowest overall run time.
The Stim GraphlikeError family of algorithms had low success rate on this dataset as the codes do not in general have string-like errors.
\begin{table}[h!]
\setlength\tabcolsep{2pt}
\fontfamily{lmss}\fontsize{8}{9}\selectfont{
\begin{center}									
\begin{tabular}{ |l|	r|	r|	r|	r|	r|	r|	r|	}	\hline
 &	\textbf{m4riRW} &	\textbf{QDistRndMW} &	\textbf{QDistEvol} &	\textbf{BP-OSD} &	\textbf{GEStim} &	\textbf{CCStim} &	\textbf{UEStim}	\\	\hline
\textbf{Result Returned} &	121 &	121 &	121 &	121 &	2 &	12 &	32	\\	
\textbf{No Result} &	0 &	0 &	0 &	0 &	119 &	109 &	89	\\	
\textbf{At lowest distance} &	121 &	121 &	121 &	121 &	2 &	12 &	32	\\	
\textbf{Overall Success Rate} &	100.0\% &	100.0\% &	100.0\% &	100.0\% &	1.7\% &	9.9\% &	26.4\%	\\	\hline
\textbf{Total Time (s)} &	2.3E+01 &	3.1E+02 &	3.8E+02 &	5.9E+03 &	1.3E+00 &	1.3E+00 &	4.5E+00	\\	
\textbf{Trials} &	NA &	1,210,000 &	1,210,000 &	1,210,000 &	121 &	121 &	970	\\	
\textbf{Trials at lowest distance} &	 &	598,412 &	941,933 &	520,735 &	2 &	12 &	63	\\	
\textbf{Trial Success Rate} &	 &	49.5\% &	77.8\% &	43.0\% &	1.7\% &	9.9\% &	6.5\%	\\	
\textbf{Time/Trial} &	 &	2.5E-04 &	3.1E-04 &	4.8E-03 &	1.0E-02 &	1.1E-02 &	4.6E-03	\\	
\textbf{Time/Successful Trial} &	 &	5.1E-04 &	4.0E-04 &	1.1E-02 &	6.3E-01 &	1.1E-01 &	7.2E-02	\\	\hline
\end{tabular}									
\end{center}																
}
\caption{Distance finding benchmark for codetables.de GF(2) codes - Heuristic Methods.}
\label{tab:codetables_GF2_heuristic}
\end{table}

\begin{figure}[h!]
\centering
\begin{subfigure}[t]{.33\textwidth}
  \centering
  \includegraphics[width=\linewidth]{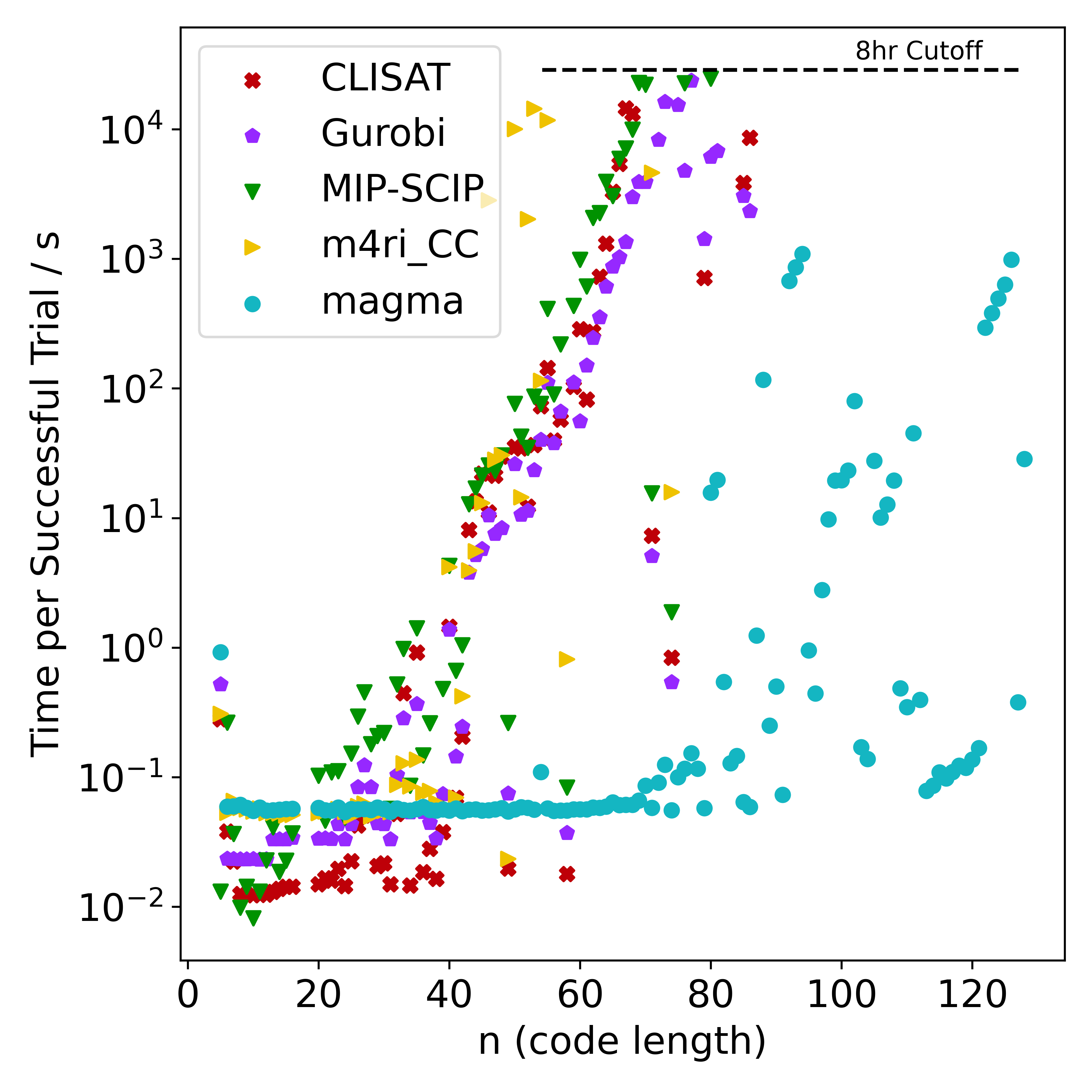}
  \subcaption{Time per Successful Trial \\ (Exact Algorithms)}\label{fig:01_codeTables_GF2_exact_ttd_time_exact}
    \end{subfigure}%
\begin{subfigure}[t]{.33\textwidth}
  \centering
  \includegraphics[width=\linewidth]{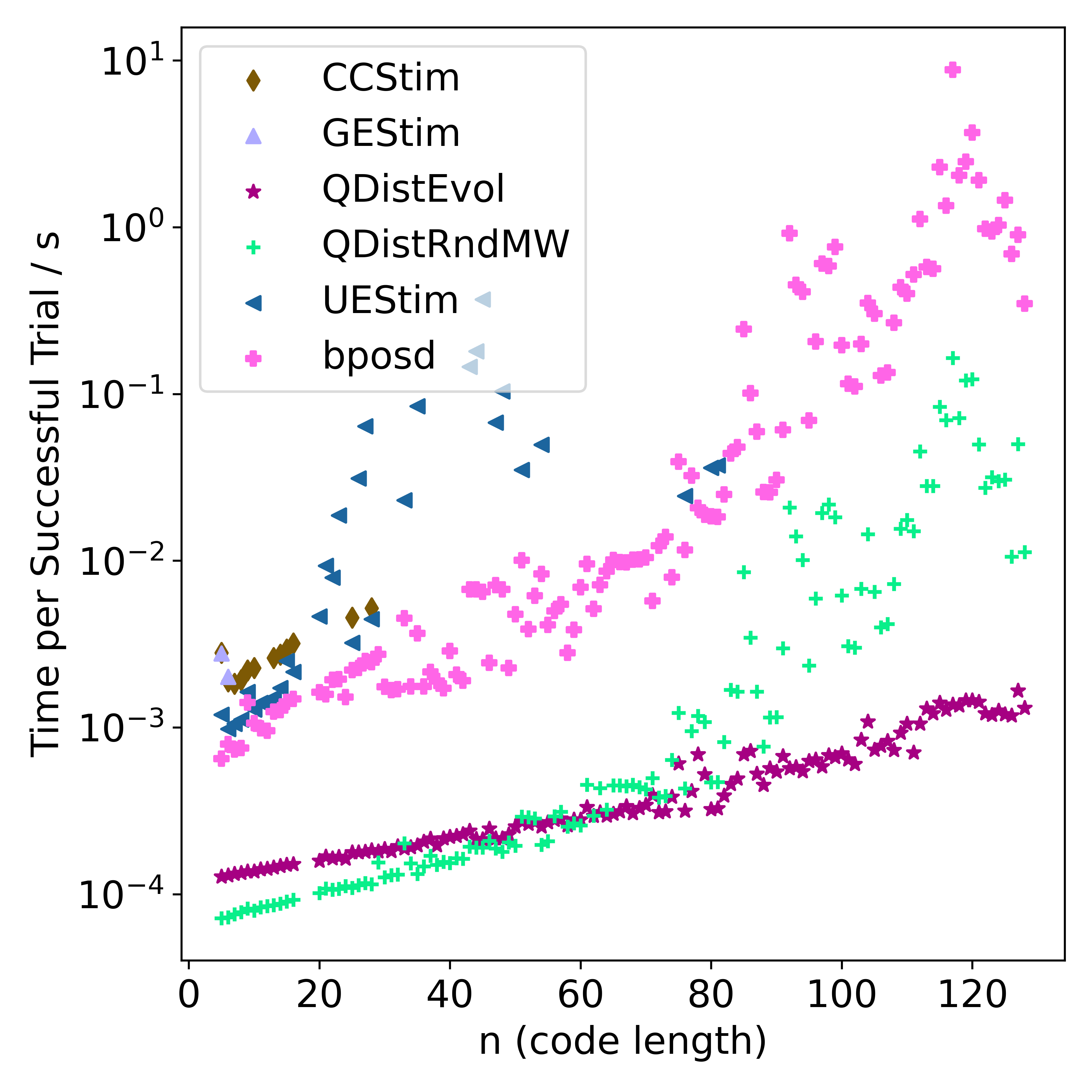}
  \subcaption{Time per Successful Trial \\ (Heuristic Algorithms)}\label{fig:01_codeTables_GF2_heuristic_ttd}
    \end{subfigure}%
\begin{subfigure}[t]{.33\textwidth}
  \centering
  \includegraphics[width=\linewidth]{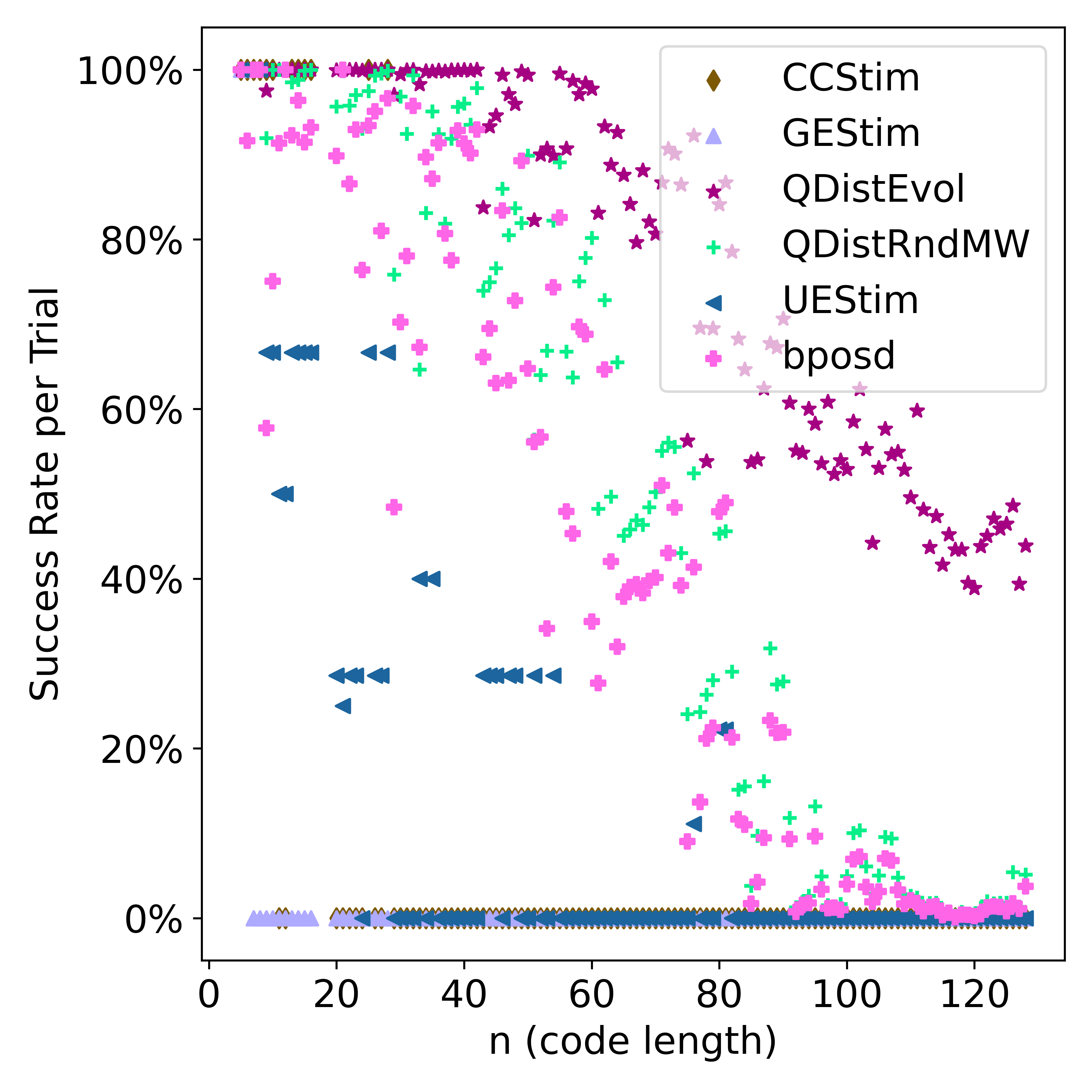}
  \subcaption{Success Rate per Trial \\ (Heuristic Algorithms)}\label{fig:01_codeTables_GF2_heuristic_success_rate}
    \end{subfigure}%
\caption{Benchmark data by code length - codetables.de GF(2) codes}
\label{fig:01_codeTables_GF2}
\end{figure}

In \Cref{fig:01_codeTables_GF2} we plot the time per successful trial for exact algorithms for all algorithms in our benchmark. 
We see that the Magma Brouwer-Zimmermann algorithm has the best performance for larger codes, though this is not necessarily  the case for codes of length under 60. 
In \Cref{fig:01_codeTables_GF2_exact_ttd_time_exact} we see that the QDistRnd algorithm  has the lowest processing time per successful trial for smaller codes but that this increases significantly for larger members of the data set.
After code sizes of around 60, QDistEvol has better processing time per successful trial and we do not see as significant a deterioration in performance for larger codes. 
This behaviour is driven by QDistEvol having favourable success rate per trial and this is set out in  \Cref{fig:01_codeTables_GF2_heuristic_success_rate}. 
Whilst success rate for other methods are  close to zero for codes with length over 100, success rate for QDistEvol is still close to 40\% even for the largest codes.
\captionsetup[subfigure]{margin=5pt}

\subsubsection{Classical Lifted Product Codes}\label{sec:LP_classical}

The next data set is a family of classical LDPC codes which are used to construct a family of quantum lifted product codes \cite{LP_Panteleev,LP_Xu}.
This is a family of 19 codes with between 5 and 475 bits. 

In \Cref{tab:lifted_product_GF2_exact} we compare exact methods for this code family.
Magma completed 9 of the 19 codes within the 8 hour maximum execution time whereas the mixed integer (Gurobi and SCIP) and SAT algorithms completed 6 and  m4riCC completed 5.
The accuracy of partial results after early termination varies widely - we find that MIP-SCIP and Magma found the lowest distance in all cases, but the accuracy of the Gurobi and SAT methods was lower.
\begin{table}[h!]
\setlength\tabcolsep{2pt}
\fontfamily{lmss}\fontsize{8}{9}\selectfont{
\begin{center}							
\begin{tabular}{ |l|	r|	r|	r|	r|	r|	}	\hline
 &	\textbf{Gurobi} &	\textbf{MIP-SCIP} &	\textbf{CLISAT} &	\textbf{m4riCC} &	\textbf{Magma}	\\	\hline
\textbf{Result Returned} &	19 &	19 &	17 &	5 &	19	\\	
\textbf{Completed < MaxTime} &	6 &	6 &	6 &	5 &	9	\\	
\textbf{Completed > MaxTime} &	13 &	13 &	11 &	0 &	10	\\	
\textbf{No Result} &	0 &	0 &	2 &	14 &	0	\\	
\textbf{At lowest distance} &	16 &	19 &	7 &	5 &	19	\\	
\textbf{Overall success rate} &	84.2\% &	100.0\% &	36.8\% &	26.3\% &	100.0\%	\\	\hline
\textbf{Total Time} &	3.8E+05 &	3.8E+05 &	3.8E+05 &	4.0E+05 &	2.9E+05	\\	
\textbf{Time/Trial} &	2.0E+04 &	2.0E+04 &	2.0E+04 &	2.1E+04 &	1.5E+04	\\	
\textbf{Time/Successful Trial} &	2.3E+04 &	2.0E+04 &	5.4E+04 &	8.1E+04 &	1.5E+04	\\	\hline
\end{tabular}							
\end{center}												
}
\caption{Distance finding benchmark classical lifted product codes - Exact Methods.}
\label{tab:lifted_product_GF2_exact}
\end{table}
In \Cref{tab:lifted_product_GF2_heuristic} we compare heuristic methods for this data set.
We again find that m4riRW, QDistRndMW, QDistEvol and BP-OSD algorithms performed very well finding the lowest distance in all cases.
The Stim distance-finding methods had significantly lower success rates for this data set.
QDistEvol had the lowest overall time per successful trial, driven by a higher success rate per trial. 

\begin{table}[h!]
\setlength\tabcolsep{2pt}
\fontfamily{lmss}\fontsize{8}{9}\selectfont{			\begin{center}									
\begin{tabular}{ |l|	r|	r|	r|	r|	r|	r|	r|	}	\hline
 &	\textbf{m4riRW} &	\textbf{QDistRndMW} &	\textbf{QDistEvol} &	\textbf{BP-OSD} &	\textbf{GEStim} &	\textbf{CCStim} &	\textbf{UEStim}	\\	\hline
\textbf{Result Returned} &	19 &	19 &	19 &	19 &	1 &	2 &	4	\\	
\textbf{No Result} &	0 &	0 &	0 &	0 &	18 &	17 &	15	\\	
\textbf{At lowest distance} &	19 &	19 &	19 &	19 &	1 &	2 &	4	\\	
\textbf{Overall Success Rate} &	100.0\% &	100.0\% &	100.0\% &	100.0\% &	5.3\% &	10.5\% &	21.1\%	\\	\hline
\textbf{Total Time (s)} &	3.1E+01 &	3.8E+02 &	3.8E+02 &	4.6E+04 &	1.2E+00 &	1.4E+00 &	4.2E+00	\\	
\textbf{Trials} &	NA &	190,000 &	190,000 &	190,000 &	19 &	19 &	154	\\	
\textbf{Trials at lowest distance} &	 &	53,904 &	136,633 &	40,404 &	1 &	2 &	8	\\	
\textbf{Trial Success Rate} &	 &	28.4\% &	71.9\% &	21.3\% &	5.3\% &	10.5\% &	5.2\%	\\	
\textbf{Time/Trial} &	 &	2.0E-03 &	2.0E-03 &	2.4E-01 &	6.4E-02 &	7.5E-02 &	2.7E-02	\\	
\textbf{Time/Successful Trial} &	 &	7.1E-03 &	2.8E-03 &	1.1E+00 &	1.2E+00 &	7.1E-01 &	5.2E-01	\\	\hline
\end{tabular}									
\end{center}									
}
\caption{Distance finding benchmark classical lifted product codes - Heuristic Methods.}
\label{tab:lifted_product_GF2_heuristic}
\end{table}

\begin{figure}[h!]
\centering
\begin{subfigure}[t]{.33\textwidth}
  \centering
  \includegraphics[width=\linewidth]{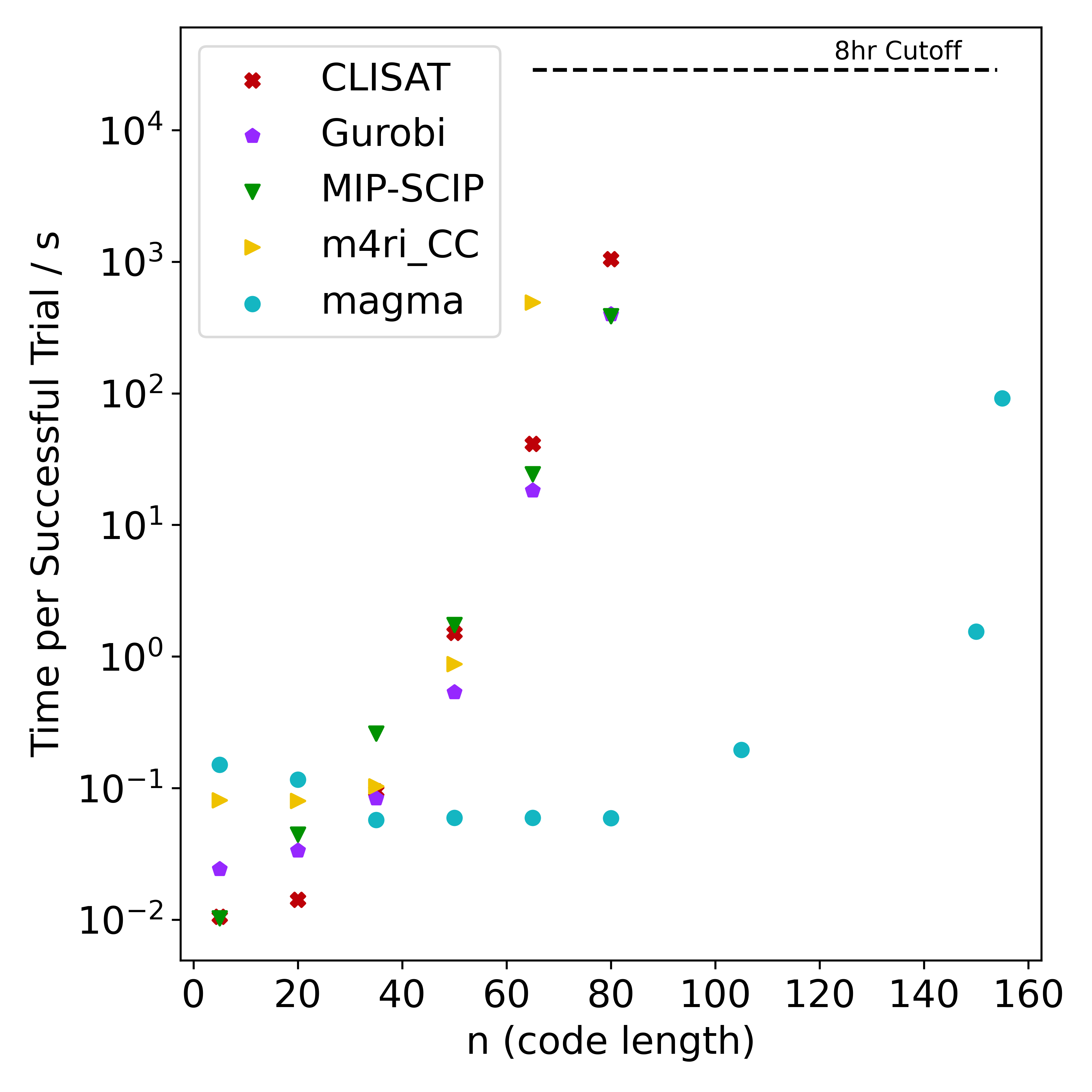}
  \subcaption{Time per Successful Trial \\ (Exact Algorithms)}\label{fig:02_lifted_product_GF2_exact_ttd_time_exact}
    \end{subfigure}%
\begin{subfigure}[t]{.33\textwidth}
  \centering
  \includegraphics[width=\linewidth]{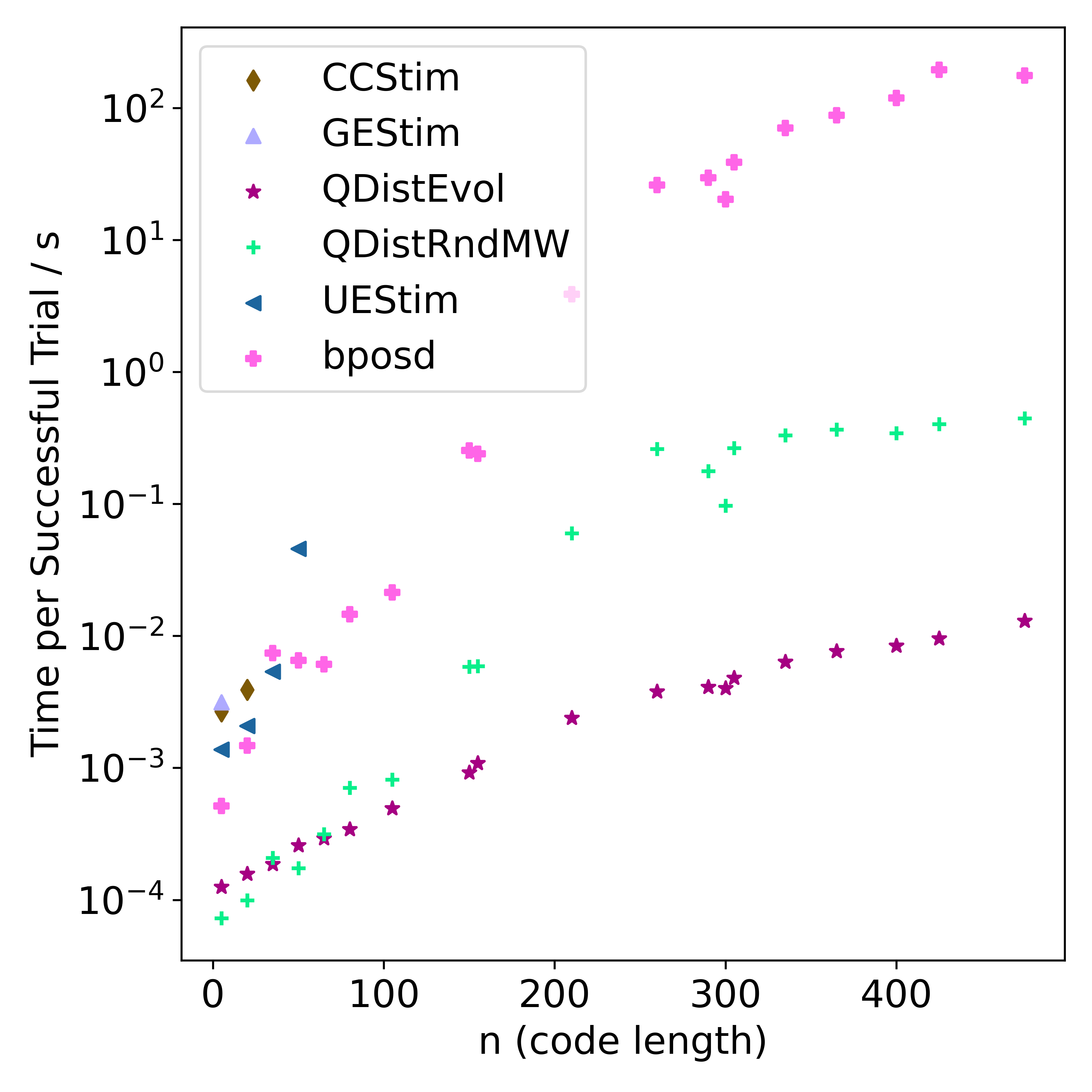}
  \subcaption{Time per Successful Trial \\ (Heuristic Algorithms)}\label{fig:02_lifted_product_GF2_heuristic_ttd}
    \end{subfigure}%
\begin{subfigure}[t]{.33\textwidth}
  \centering
  \includegraphics[width=\linewidth]{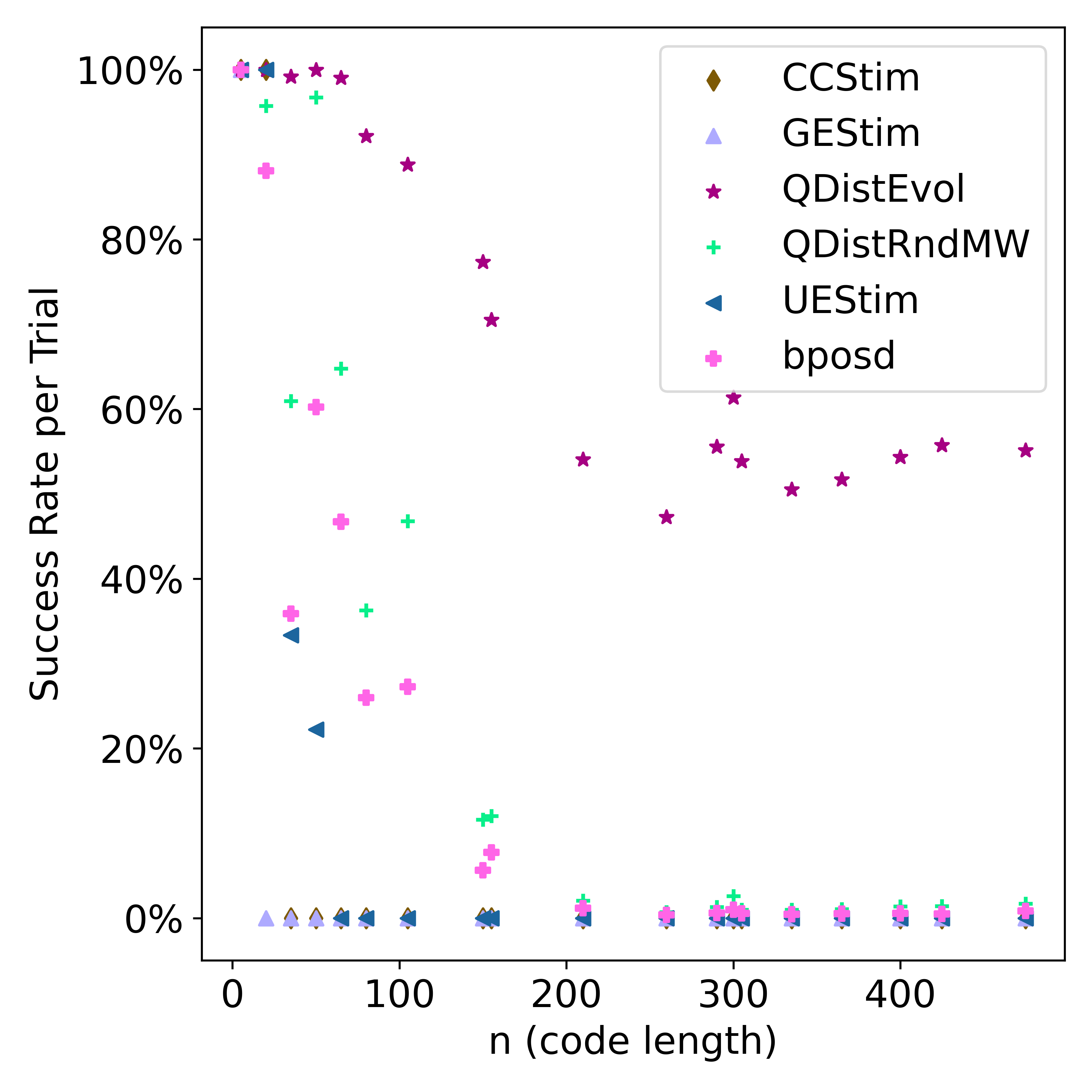}
  \subcaption{Success Rate per Trial \\ (Heuristic Algorithms)}\label{fig:02_lifted_product_GF2_heuristic_success_rate}
    \end{subfigure}%
\caption{Benchmark data by code length - classical lifted product codes}
\label{fig:02_lifted_product_GF2}
\end{figure}
In \Cref{fig:02_lifted_product_GF2_exact_ttd_time_exact} we plot the time per successful trial for exact algorithms for all algorithms in our benchmark - we  see that Magma has  lower execution time than the other exact methods from around code size 30.
In \Cref{fig:01_codeTables_GF2_heuristic_ttd} we see that QDistEvol had the lowest time per successful trial, apart for codes of size less than 60 where QDistRnd had better performance.  This was again driven by QDistEvol's higher success rate per trial. 

\subsection{Quantum Stabiliser Codes}\label{sec:benchmark_QECC}
We have chosen a variety of different stabiliser code families for benchmarking.
These represent a series of non-CSS codes from the best-known-distance codes from codetables.de, as well as a range of LDPC CSS code families.

\subsubsection{Non-CSS Quantum Codes: codetables.de}\label{sec:codeTables_QECC}
The most challenging non-CSS quantum codes for which to calculate the minimum distance are those with encoding rate close to zero (see \cite{Dumer_Kovalev_Pryadko_2014} page 2). 
We took a selection of codes with encoding rate $k/n$ between $0$ and $\frac{1}{3}$ and $n \le 128$ from codetables.de \cite{codetables} using the same methodology as \Cref{sec:codeTables_GF2}.
This resulted in a family of 121 codes with between 3 and 123 physical qubits.

\begin{figure}[h!]
\centering
\begin{subfigure}[t]{.33\textwidth}
  \centering
  \includegraphics[width=\linewidth]{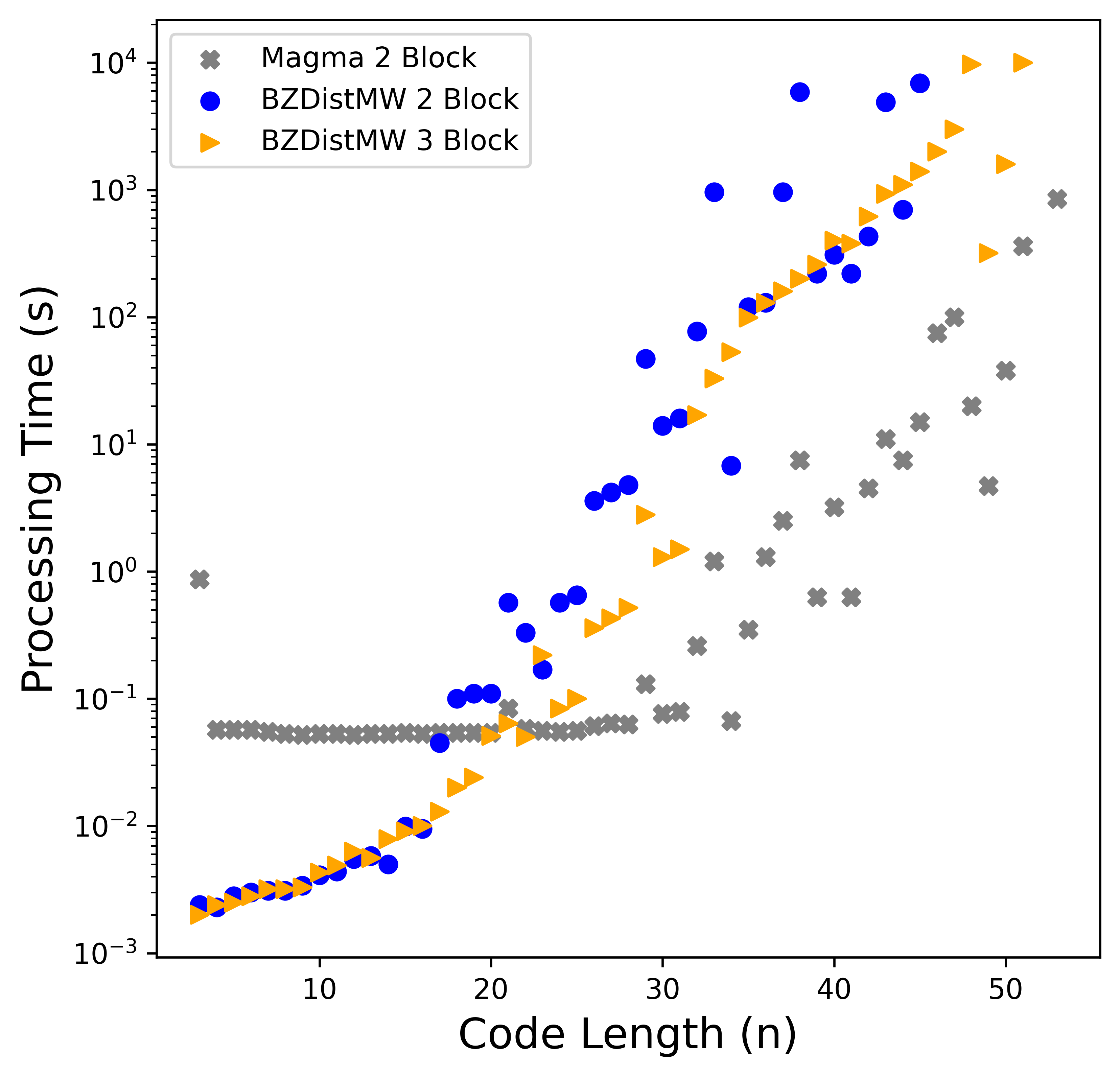}
  \subcaption{Processing Time - Brouwer Zimmermann MW - 2 versus 3 block encoding}\label{fig:QBlockSensBZ}
    \end{subfigure}%
\begin{subfigure}[t]{.33\textwidth}
  \centering
  \includegraphics[width=\linewidth]{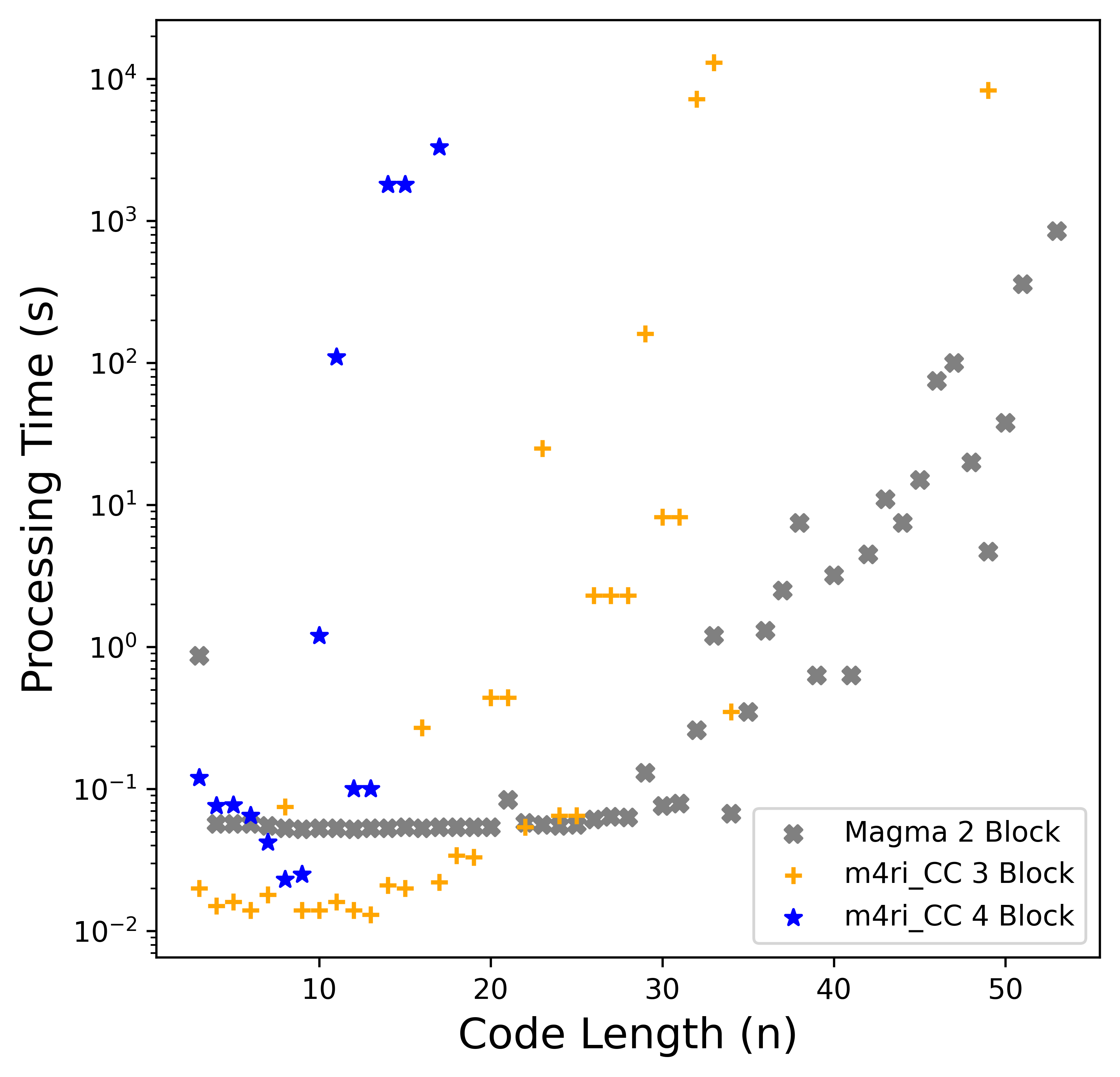}
  \subcaption{Processing Time - m4riCC 3 versus 4 block encoding}\label{fig:QBlockSensM4RICC}
    \end{subfigure}%
\begin{subfigure}[t]{.33\textwidth}
  \centering
  \includegraphics[width=\linewidth]{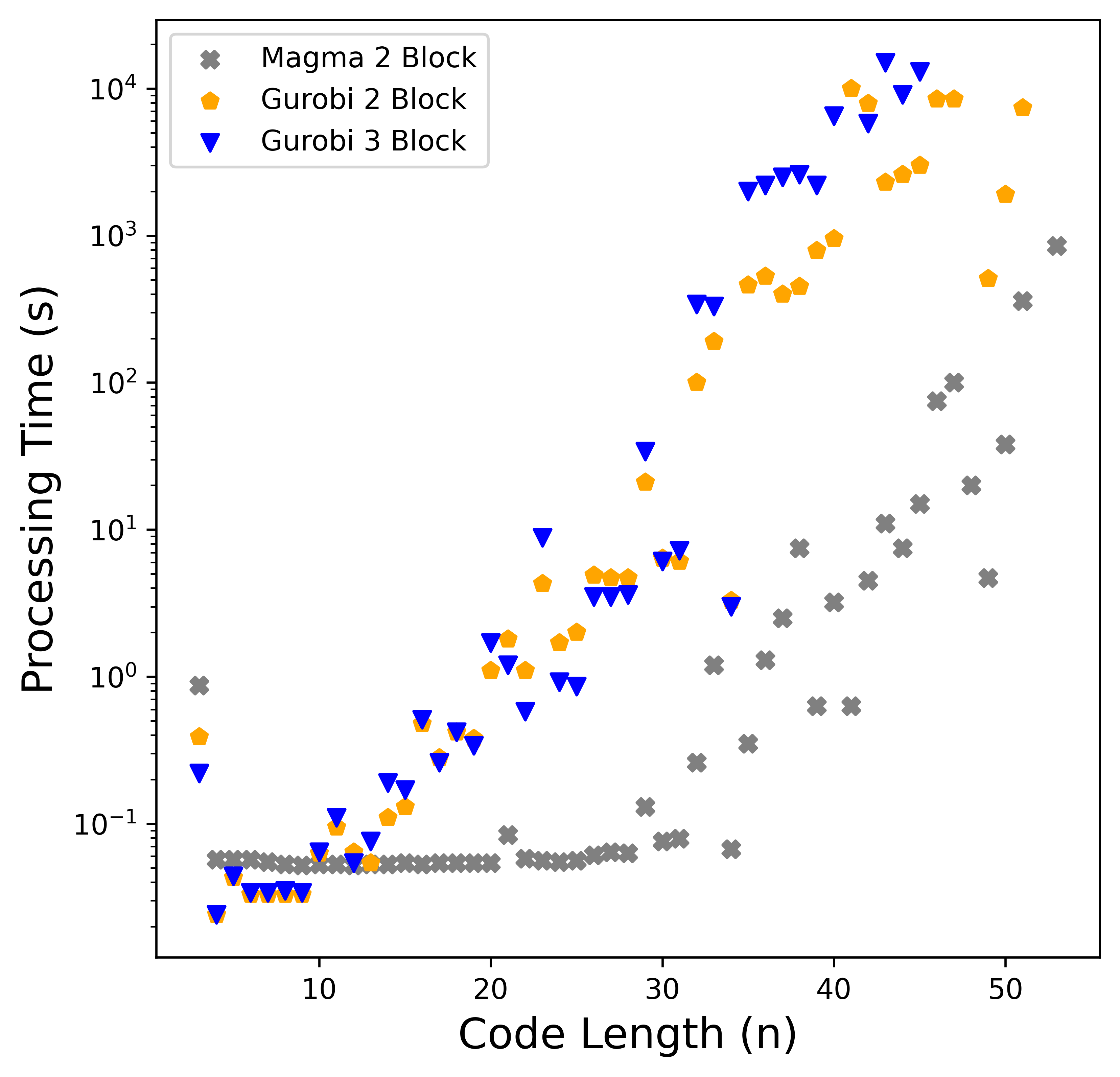}
  \subcaption{Processing Time - Gurobi 2 versus 3 block encoding}\label{fig:QBlockSensGurobi}
    \end{subfigure}%
\caption{Sensitivity of Exact Distance-Finding Methods to Choice of Block Encoding. We compare processing times for the non-CSS codetables data set for different block encoding choices for the Brouwer Zimmermann MW, m4riCC and Gurobi distance-finding algorithms. We include the processing time for Magma's two-block Brouwer-Zimmermann implementation as a base line.}
\label{fig:QBlockSens}
\end{figure}

We found that both exact and heuristic methods were very sensitive to the choice of block encoding of the stabiliser code (see \Cref{sec:twoBlock,sec:detector_view,sec:fourBlock}).
In \Cref{fig:QBlockSens}, we compare processing times for the non-CSS codetables data set for different block encoding choices for the Brouwer-Zimmermann MW, m4riCC and Gurobi distance-finding algorithms.
We find that the three-block encoding gives faster processing time for the Brouwer-Zimmermann algorithm - this is consistent with the results in \cite{Hernando_Quintana-Orti_Grassl_2024}. 
The m4riCC algorithm had better run-time with the three block encoding versus the four-block.
Finally, the two-block encoding has better performance than three-block for the Gurobi method.

\begin{figure}[h!]
\centering
\begin{subfigure}[t]{.33\textwidth}
  \centering
  \includegraphics[width=\linewidth]{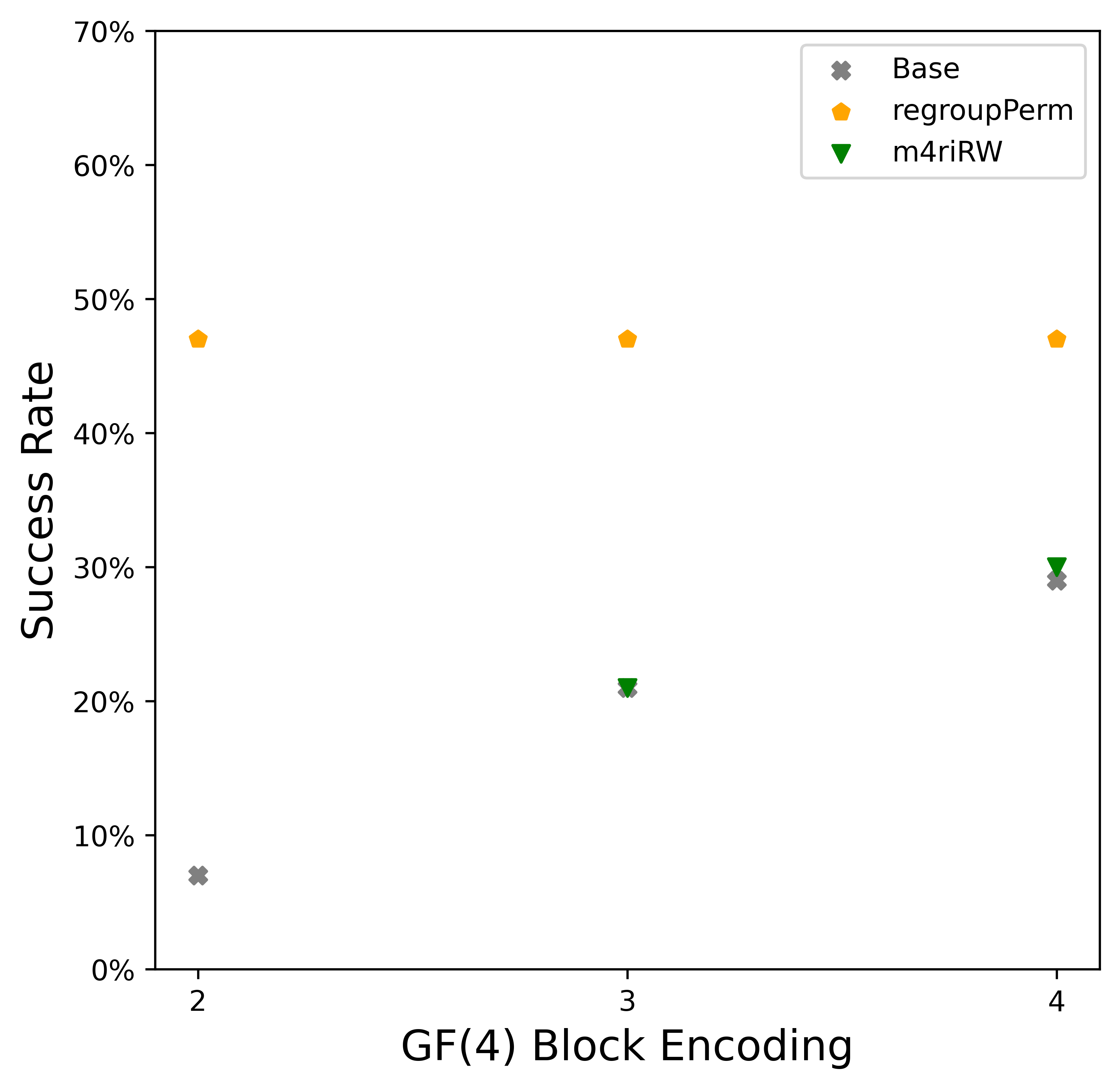}
  \subcaption{Success rate - QDistRndMW}\label{fig:Q80BlockSensRnd}
    \end{subfigure}%
\begin{subfigure}[t]{.33\textwidth}
  \centering
  \includegraphics[width=\linewidth]{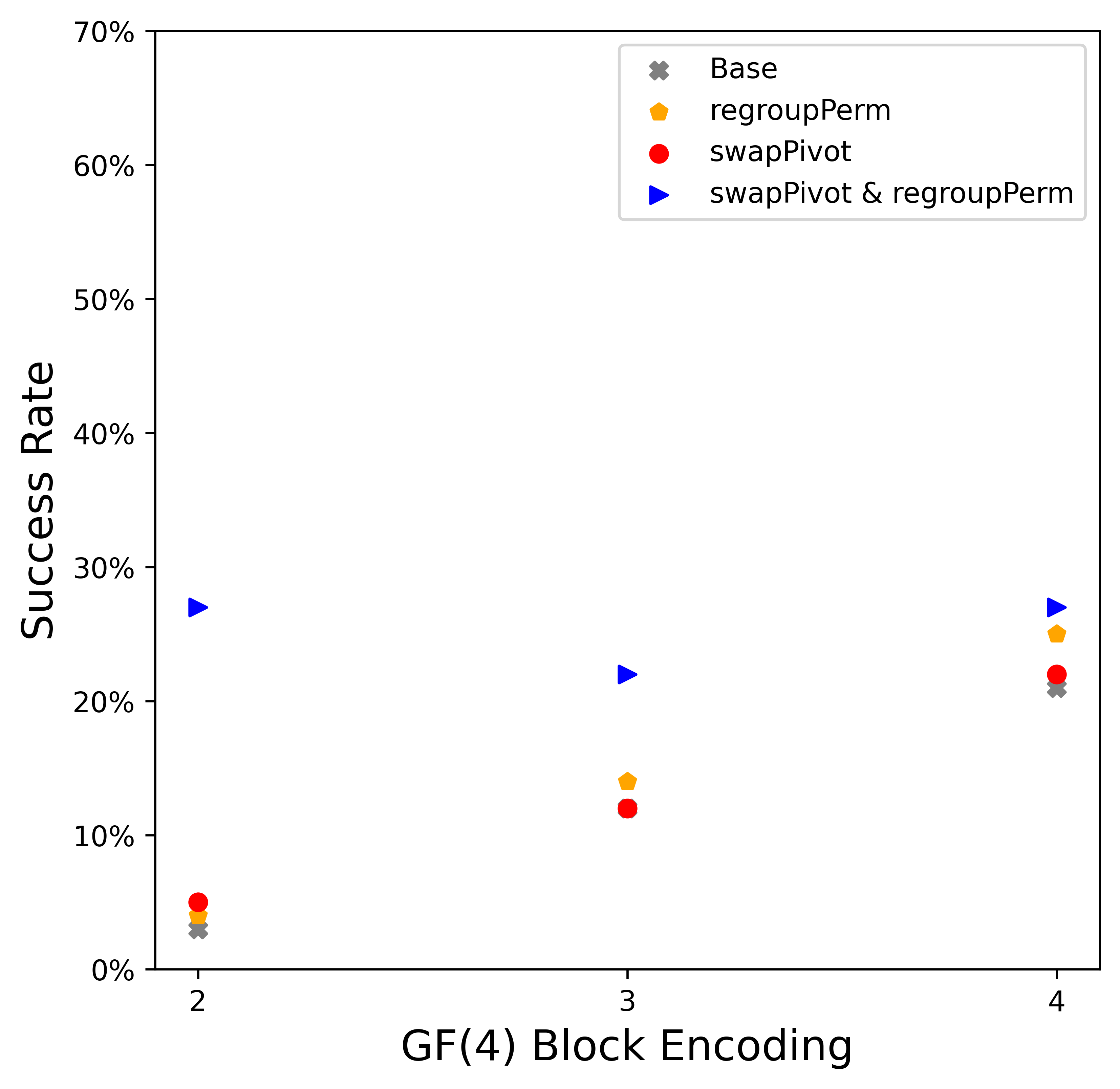}
  \subcaption{Success rate - QDistEvol}\label{fig:Q80BlockSensEvol}
    \end{subfigure}%
\caption{Sensitivity of QDistRndMW and QDistEvol Heuristic Distance-Finding Methods to Choice of Block Encoding. For the $[[80,7,15]]$ best-known-distance stabiliser code, using 1000 QDisRndMW runs of 10000 trials each, we compare success rates for the  algorithm using two, three and four-block encodings, and also toggling the \texttt{regroupPerm} parameter.
We also include results for m4riRW using three and four-block encodings - note that there is no equivalent for the \texttt{regroupPerm} in the C implementation.
For QDistEvol, we also toggle the \texttt{swapPivot} parameter.}
\label{fig:Q80BlockSens}
\end{figure}

Turning now to the heuristic random information set algorithms, we noted during testing that the two-block encoding led to faster processing times than the three or four-block encoding.
However, the accuracy of the methods did change with different block encodings and they were also sensitive to the \texttt{regroupPerm} and \texttt{swapPivot} parameter settings (see \Cref{sec:QDistEvol}).
In \Cref{fig:Q80BlockSens}, we compare success rates for the $[[80,7,15]]$ best-known-distance code for m4riRW, QDistRndMW and QDistEvol using two, three and four-block encodings.  We also test sensitivity to the \texttt{regroupPerm} parameter for QDistRndMW and both \texttt{regroupPerm} and \texttt{swapPivot} for QDistEvol.
We applied each algorithm for each parameter setting 1000 times and used 10000 iterations each time.
For QDistRndMW and m4riRW and with \texttt{regroupPerm:=FALSE}, we find that the four-block encoding gives the most accurate results and that the accuracy of QDistRndMW and m4riRW tracks very closely as expected. If we now set  \texttt{regroupPerm:=TRUE}, accuracy jumps significantly, and now the success rate for all four block encodings is around 47\%.
As the \texttt{regroupPerm} option is not available for the m4riRW C implementation, we have used a four block encoding and \texttt{regroupPerm:=FALSE} for our QDistRndMW algorithm to align it with m4riRW's success rate.

For QDistEvol, we note that when both \texttt{regroupPerm:=FALSE} and  \texttt{swapPivot:=FALSE} accuracy is also best for the four-block encoding.
This time, setting either \texttt{regroupPerm:=TRUE} or \texttt{swapPivot:=TRUE}  results in only a modest improvement.
When setting both parameters to \texttt{TRUE}, we see that the accuracy of the two-block encoding increases to 27\%.
Note that this success rate is still lower than the success rate for QDistRnd with optimal settings and this suggests that there is not enough structure in the codetables.de codes for QDistEvol to perform well versus randomly chosen column permutations.
For benchmarking QDistEvol, we use the two-block encoding and set both \texttt{regroupPerm:=TRUE} and \texttt{swapPivot:=TRUE}.

In \Cref{tab:codetables_QECC_exact} we compare exact methods for this code family.
We only considered codes on up to 80 qubits for exact methods due to the long processing times involved.
Magma completed 55 of the 78 codes within the maximum time frame of 8 hours, and the other exact methods completed significantly fewer codes. 
Where the algorithm was terminated after the 8 hour time frame, Magma's distance estimate was correct in almost all cases. 
The accuracy of partial results of the Gurobi and SCIP mixed integer programming methods were lower than Magma but noticeably higher than the SAT method.

\begin{table}[h!]
\setlength\tabcolsep{2pt}
\fontfamily{lmss}\fontsize{8}{9}\selectfont{
\begin{center}							
\begin{tabular}{ |l|	r|	r|	r|	r|	r|	}	\hline
 &	\textbf{Gurobi} &	\textbf{MIP-SCIP} &	\textbf{CLISAT} &	\textbf{m4riCC} &	\textbf{Magma}	\\	\hline
\textbf{Result Returned} &	78 &	78 &	78 &	33 &	78	\\	
\textbf{Completed < MaxTime} &	46 &	40 &	39 &	33 &	56	\\	
\textbf{Completed > MaxTime} &	32 &	38 &	39 &	0 &	22	\\	
\textbf{No Result} &	0 &	0 &	0 &	45 &	0	\\	
\textbf{At lowest distance} &	67 &	67 &	49 &	33 &	77	\\	
\textbf{Overall success rate} &	85.9\% &	85.9\% &	62.8\% &	42.3\% &	98.7\%	\\	\hline
\textbf{Total Time} &	9.8E+05 &	1.2E+06 &	1.2E+06 &	1.3E+06 &	6.9E+05	\\	
\textbf{Time/Trial} &	1.3E+04 &	1.5E+04 &	1.6E+04 &	1.7E+04 &	8.9E+03	\\	
\textbf{Time/Successful Trial} &	1.5E+04 &	1.8E+04 &	2.5E+04 &	4.0E+04 &	9.0E+03	\\	\hline
\end{tabular}							
\end{center}															
}
\caption{Distance finding benchmark codetables.de non-CSS quantum codes - Exact Methods.}
\label{tab:codetables_QECC_exact}
\end{table}

In \Cref{tab:codetables_QECC_heuristic} we compare heuristic methods for this data set.
We find that the random information set methods performed well, though they did not find the correct minimum distance in all cases. 
QDistEvol had a higher success rate per trial for this data set than QDistRndMW and m4riRW.
Note  that the use of permutation regrouping, pivot swapping and the two-block symplectic representation in the QDistEvol algorithm may also play an important role in increasing accuracy and processing time. 
These enhancements are not available for m4riRW so we did not use them for the  QDistRndMW results in this analysis.
The BP-OSD method was significantly slower than the random information set methods and also less accurate.
The Stim undetectable error family of algorithms returned results for some but not all codes - the relatively poor performance due to the fact that codetables.de codes are not guaranteed to have string-like errors.

\begin{figure}[h!]
\centering
\begin{subfigure}[t]{.33\textwidth}
  \centering
  \includegraphics[width=\linewidth]{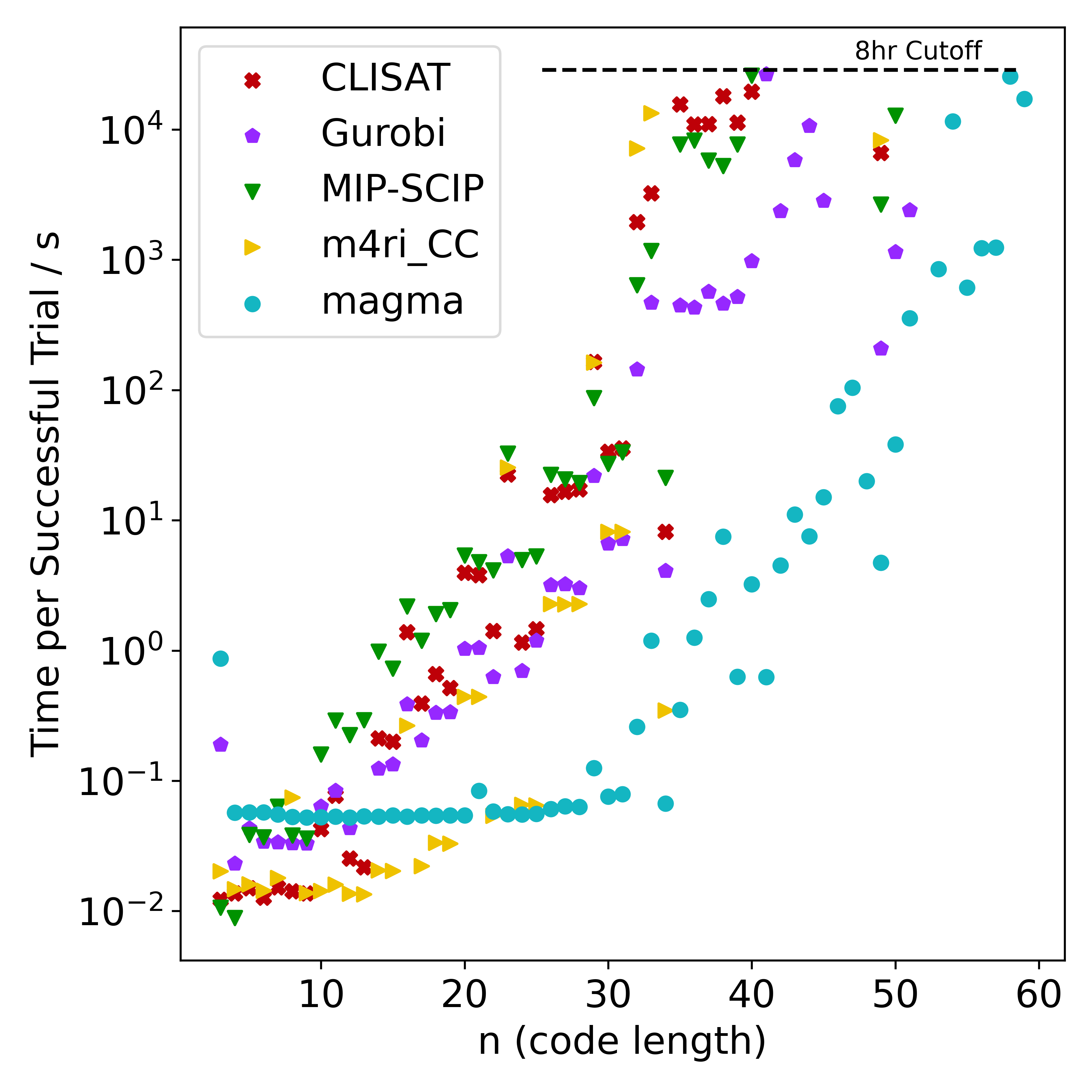}
  \subcaption{Time per Successful Trial \\ (Exact Algorithms)}\label{fig:03_codeTables_QECC_exact_ttd_time_exact}
    \end{subfigure}%
\begin{subfigure}[t]{.33\textwidth}
  \centering
  \includegraphics[width=\linewidth]{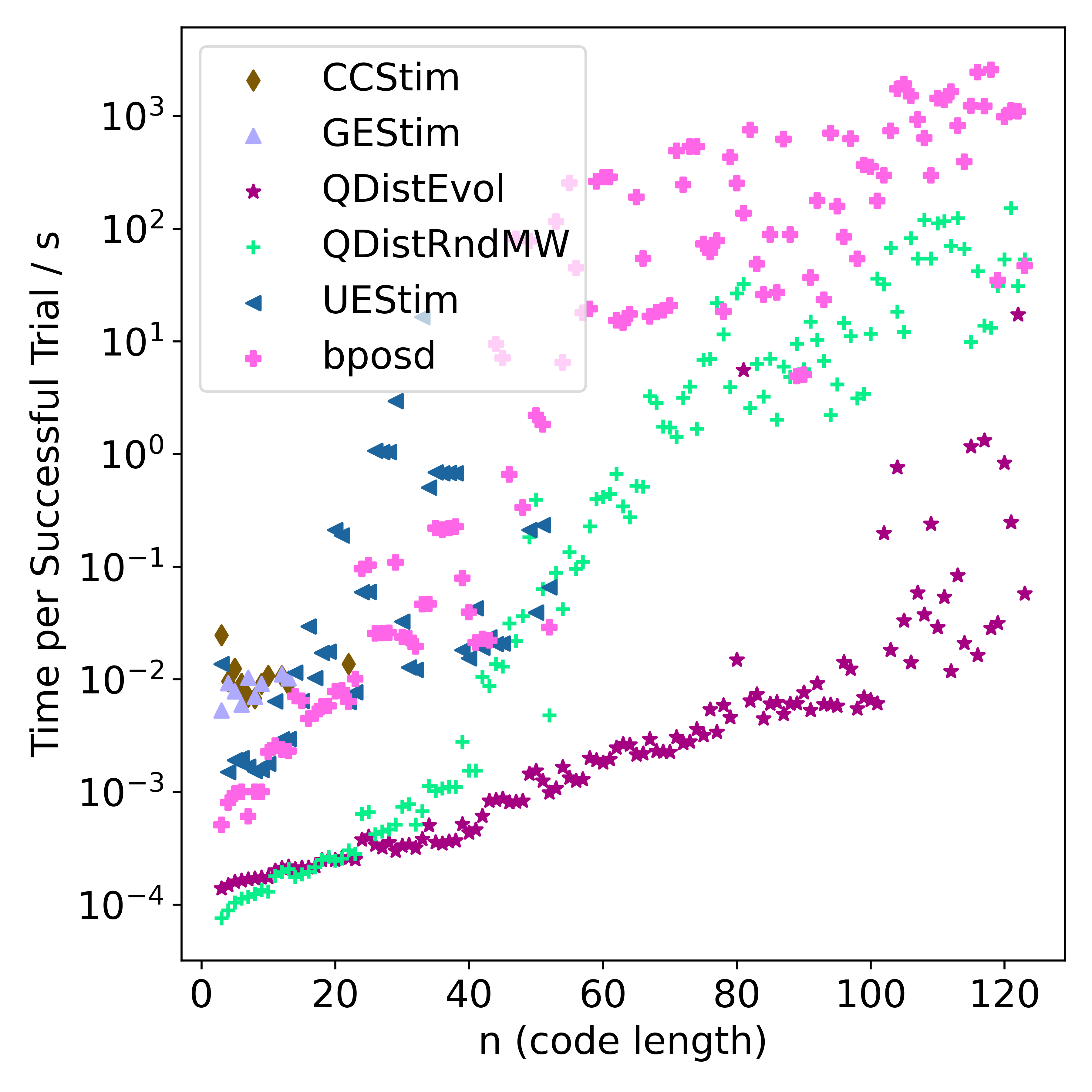}
  \subcaption{Time per Successful Trial \\ (Heuristic Algorithms)}\label{fig:03_codeTables_QECC_heuristic_ttd}
    \end{subfigure}%
\begin{subfigure}[t]{.33\textwidth}
  \centering
  \includegraphics[width=\linewidth]{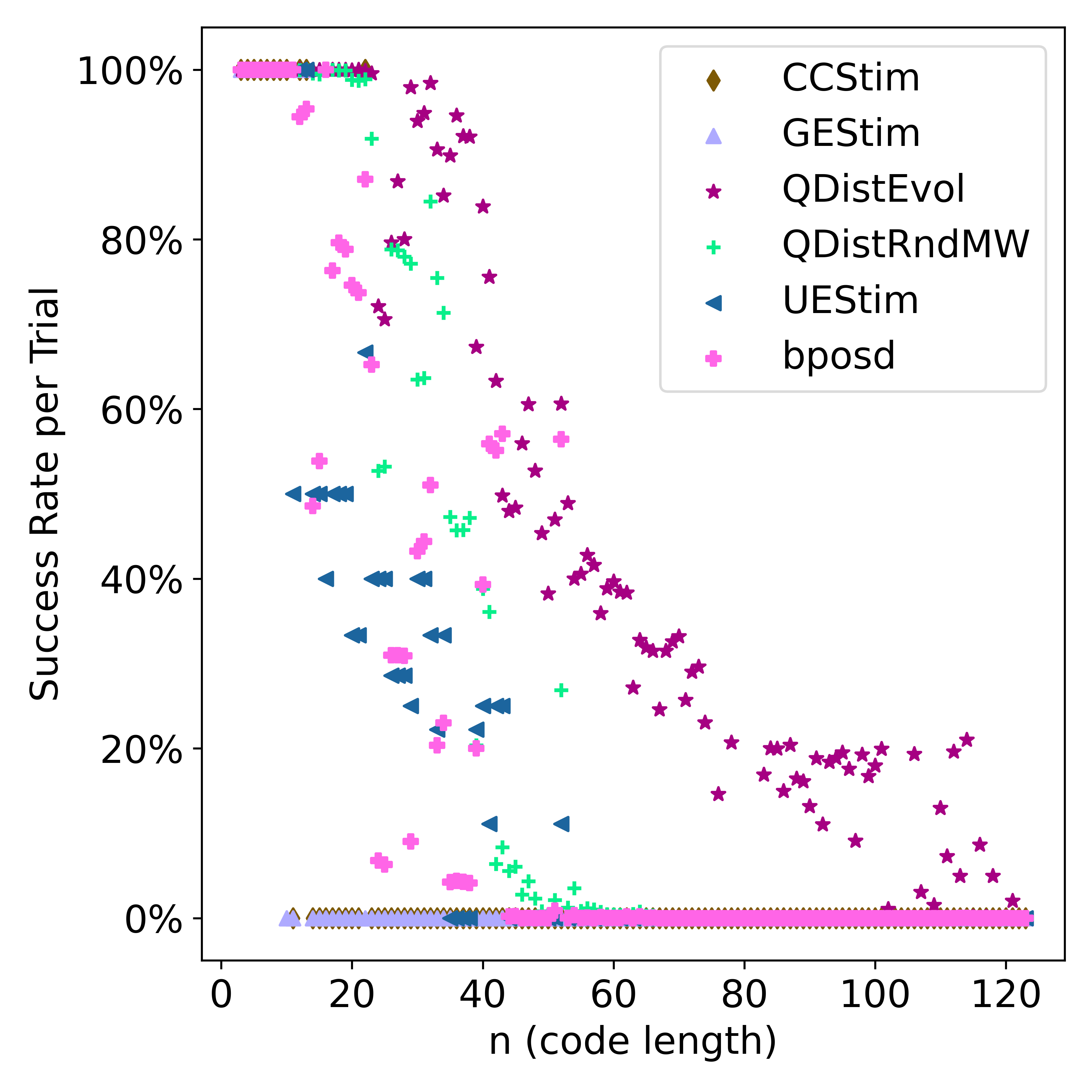}
  \subcaption{Success Rate per Trial \\ (Heuristic Algorithms)}\label{fig:03_codeTables_QECC_heuristic_success_rate}
    \end{subfigure}%
\caption{Benchmark data by code length - codetables.de non-CSS quantum codes}
\label{fig:03_codeTables_QECC}
\end{figure}

Processing time, success rate and time per successful trial for exact methods are plotted in \Cref{fig:03_codeTables_QECC_exact_ttd_time_exact}. We see that Magma has the lowest time per successful trial for medium to large codes with the next best  being the mixed-integer Gurobi and SCIP methods.
For heuristic methods, we see in \Cref{fig:03_codeTables_QECC_heuristic_ttd} that the QDistEvol has better time per successful trial than QDistRnd from code sizes $n > 30$.
Whilst QDistEvol has the highest success rate per trial for heuristic methods, this reduces to close to zero for all methods for the largest codes  (\Cref{fig:03_codeTables_QECC_heuristic_success_rate}).

\begin{table}[h!]
\setlength\tabcolsep{2pt}
\fontfamily{lmss}\fontsize{8}{9}\selectfont{
\begin{center}									
\begin{tabular}{ |l|	r|	r|	r|	r|	r|	r|	r|	}	\hline
 &	\textbf{m4riRW} &	\textbf{QDistRndMW} &	\textbf{QDistEvol} &	\textbf{BP-OSD} &	\textbf{GEStim} &	\textbf{CCStim} &	\textbf{UEStim}	\\	\hline
\textbf{Result Returned} &	121 &	121 &	121 &	121 &	9 &	11 &	47	\\	
\textbf{No Result} &	0 &	0 &	0 &	0 &	112 &	110 &	74	\\	
\textbf{At lowest distance} &	106 &	103 &	109 &	62 &	9 &	11 &	38	\\	
\textbf{Overall Success Rate} &	87.6\% &	85.1\% &	90.1\% &	51.2\% &	7.4\% &	9.1\% &	31.4\%	\\	\hline
\textbf{Total Time (s)} &	2.1E+02 &	3.6E+03 &	1.3E+03 &	6.3E+04 &	3.5E+00 &	3.8E+00 &	6.2E+01	\\	
\textbf{Trials} &	NA &	1,210,000 &	1,210,000 &	1,210,000 &	121 &	121 &	935	\\	
\textbf{Trials at lowest distance} &	 &	322,477 &	543,925 &	242,798 &	9 &	11 &	74	\\	
\textbf{Trial Success Rate} &	 &	26.7\% &	45.0\% &	20.1\% &	7.4\% &	9.1\% &	7.9\%	\\	
\textbf{Time/Trial} &	 &	3.0E-03 &	1.1E-03 &	5.2E-02 &	2.9E-02 &	3.2E-02 &	6.7E-02	\\	
\textbf{Time/Successful Trial} &	 &	1.1E-02 &	2.4E-03 &	2.6E-01 &	3.9E-01 &	3.5E-01 &	8.4E-01	\\	\hline
\end{tabular}									
\end{center}																	         
}
\caption{Distance finding benchmark codetables.de non-CSS quantum codes - Heuristic Methods.}
\label{tab:codetables_QECC_heuristic}
\end{table}

\subsubsection{Two-Dimensional Hyperbolic Surface and Colour CSS Codes}\label{sec:hyperbolic_CSS}
The next family of codes considered are the hyperbolic surface codes of \cite{hastings2013decodinghyperbolicspacesldpc} and the hyperbolic colour codes of \cite{dasilva2018hyperbolicquantumcolorcodes}.
The code families were generated using the reflection group methodology of \cite{Breuckmann_2017}.
We used a GAP script to generate two-dimensional cell complexes where each  plaquette has eight vertices and edges, and each vertex shares 3 plaquettes (an \{8,3\} tessellation).

Surface codes were constructed by placing qubits on the edges of the complex. 
X-type stabilisers are associated with the plaquettes  and Z-type stabilisers are associated with the vertices.
This resulted in a family of 21 surface codes with between 6 and 1221 physical qubits.
In \Cref{tab:hyperbolic_surface_exact} we compare exact methods and see that all algorithms completed within the 8 hour time for all members of the family, apart from Magma. 
The fastest of these was m4riCC - this is likely to be due to the fact that this algorithm works well for codes with Tanner graphs with low degree.
\begin{table}[h!]
\setlength\tabcolsep{2pt}
\fontfamily{lmss}\fontsize{8}{9}\selectfont{
\begin{center}							
\begin{tabular}{ |l|	r|	r|	r|	r|	r|	}	\hline
 &	\textbf{Gurobi} &	\textbf{MIP-SCIP} &	\textbf{CLISAT} &	\textbf{m4riCC} &	\textbf{Magma}	\\	\hline
\textbf{Result Returned} &	21 &	21 &	21 &	21 &	21	\\	
\textbf{Completed < MaxTime} &	21 &	21 &	21 &	21 &	14	\\	
\textbf{Completed > MaxTime} &	0 &	0 &	0 &	0 &	7	\\	
\textbf{No Result} &	0 &	0 &	0 &	0 &	0	\\	
\textbf{At lowest distance} &	21 &	21 &	21 &	21 &	21	\\	
\textbf{Overall success rate} &	100.0\% &	100.0\% &	100.0\% &	100.0\% &	100.0\%	\\	\hline
\textbf{Total Time} &	1.5E+03 &	8.7E+03 &	1.7E+03 &	1.0E+00 &	2.2E+05	\\	
\textbf{Time/Trial} &	7.0E+01 &	4.2E+02 &	7.9E+01 &	4.8E-02 &	1.1E+04	\\	
\textbf{Time/Successful Trial} &	7.0E+01 &	4.2E+02 &	7.9E+01 &	4.8E-02 &	1.1E+04	\\	\hline
\end{tabular}							
\end{center}		
}
\caption{Distance finding benchmark hyperbolic surface codes - Exact Methods.}
\label{tab:hyperbolic_surface_exact}
\end{table}
In \Cref{tab:hyperbolic_surface_heuristic} we compare heuristic methods and see that all methods give good results, with close to perfect accuracy.
The Stim methods, designed to work well in this scenario, had the lowest total processing times of the heuristic methods, though the time per successful trial was lower for the random information set and BP-OSD methods.

\begin{figure}[h!]
\centering
\begin{subfigure}[t]{.33\textwidth}
  \centering
  \includegraphics[width=\linewidth]{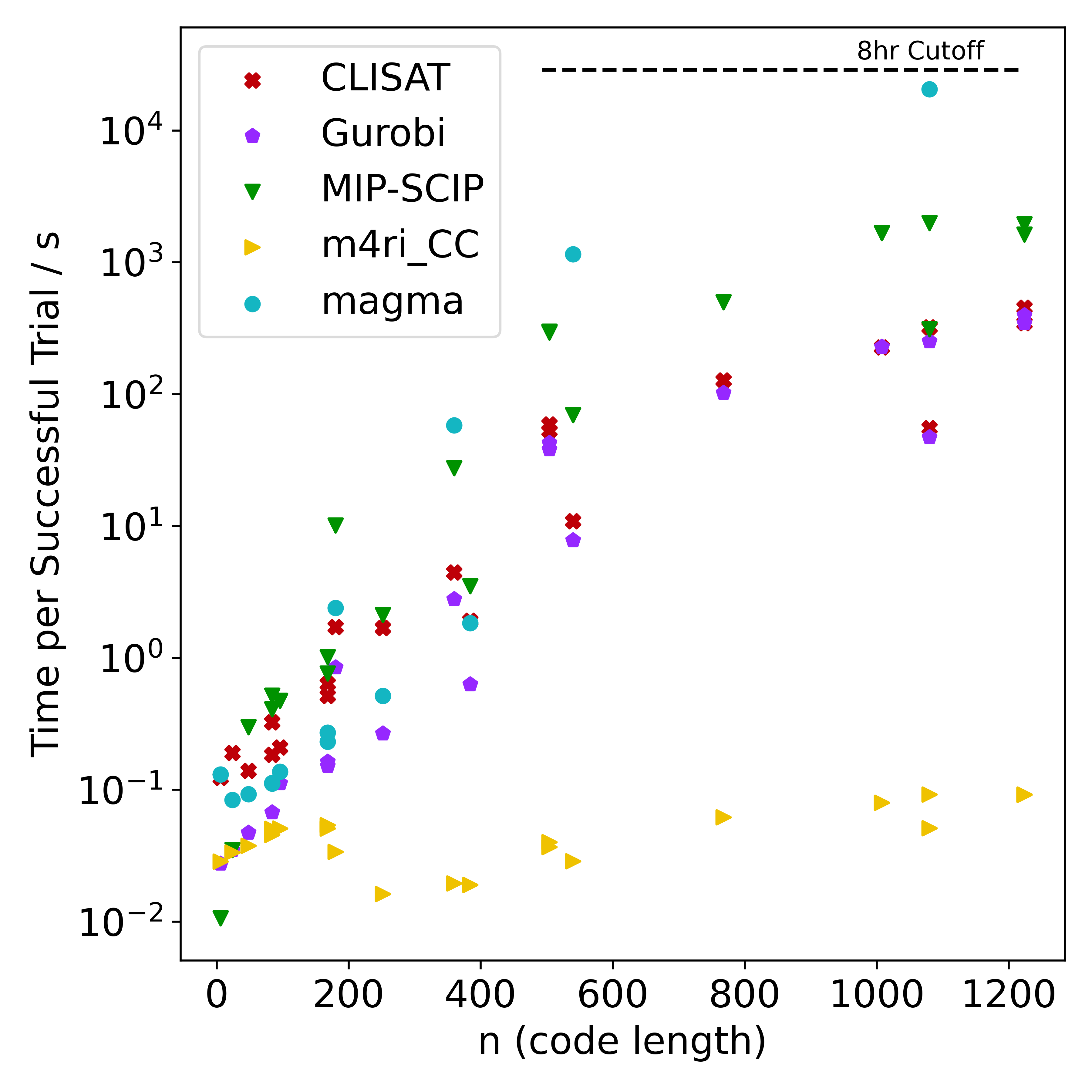}
  \subcaption{Time per Successful Trial \\ (Exact Algorithms)}\label{fig:04_hyperbolic_SC_exact_ttd_time_exact}
    \end{subfigure}%
\begin{subfigure}[t]{.33\textwidth}
  \centering
  \includegraphics[width=\linewidth]{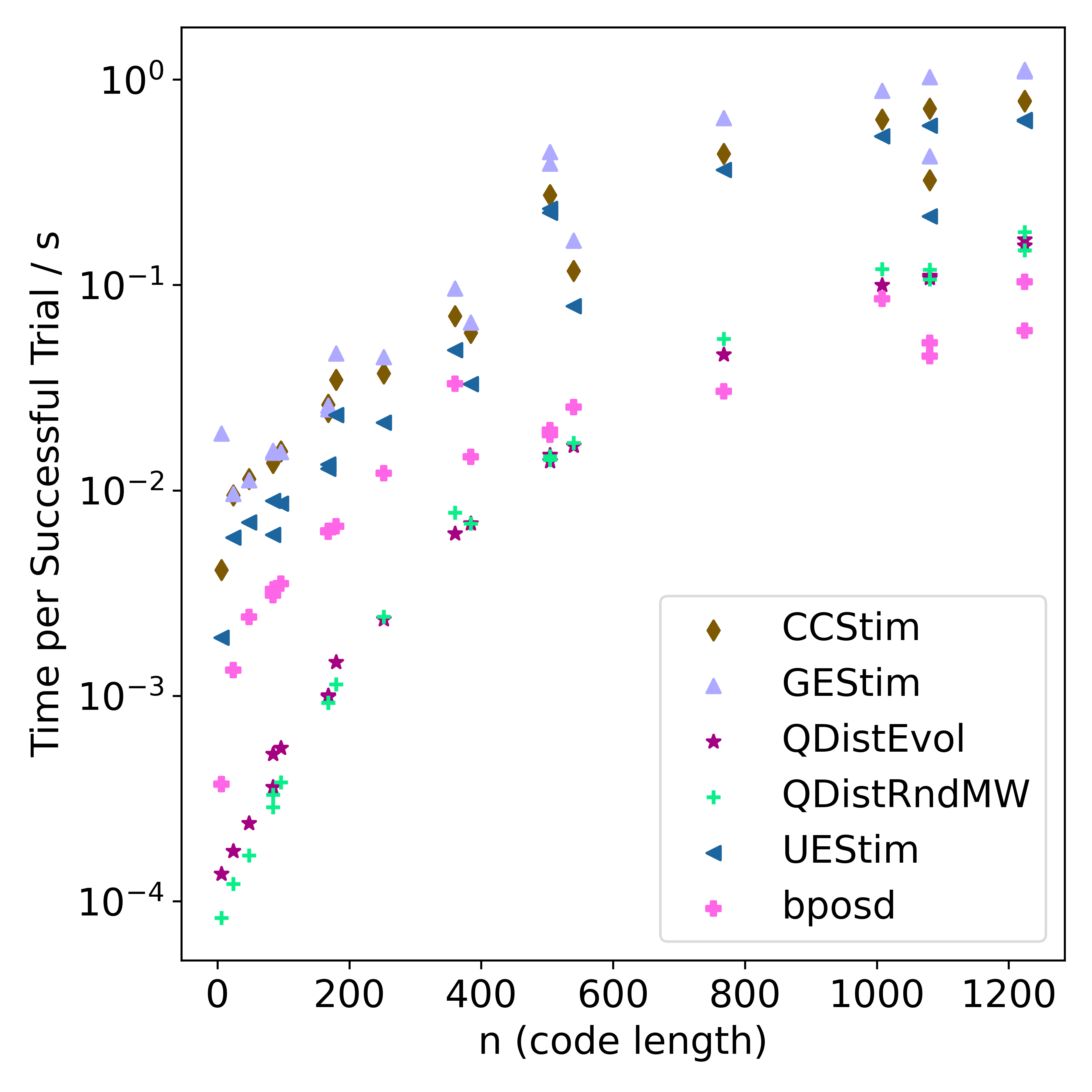}
  \subcaption{Time per Successful Trial \\ (Heuristic Algorithms)}\label{fig:04_hyperbolic_SC_heuristic_ttd}
    \end{subfigure}%
\begin{subfigure}[t]{.33\textwidth}
  \centering
  \includegraphics[width=\linewidth]{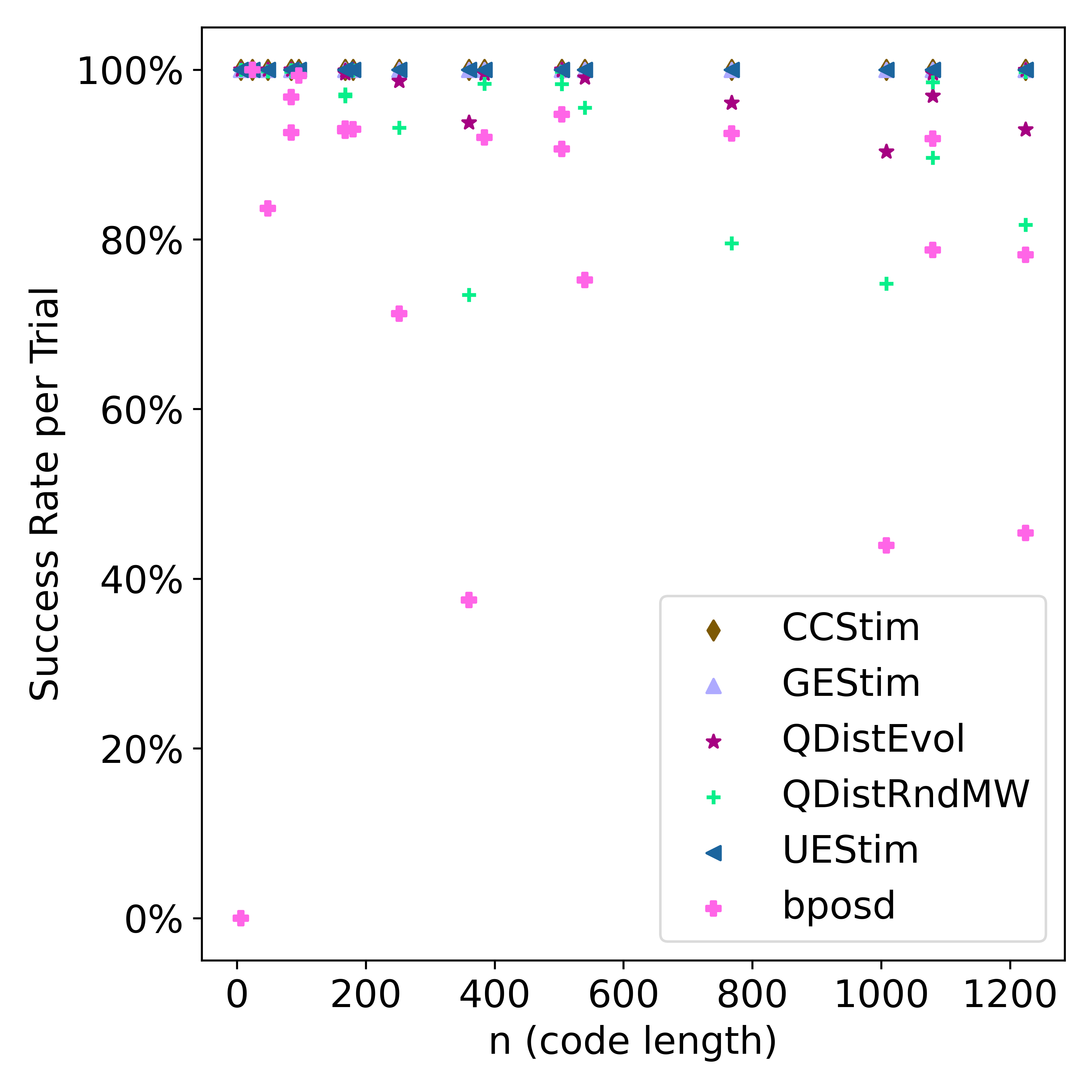}
  \subcaption{Success Rate per Trial \\ (Heuristic Algorithms)}\label{fig:04_hyperbolic_SC_heuristic_success_rate}
    \end{subfigure}%
\caption{Benchmark data by code length - hyperbolic surface codes. We note that GEStim deterministically finds the exact distance for hyperbolic surface codes and can therefore be directly compared to the methods in (a) as well.}
\label{fig:04_hyperbolic_SC}
\end{figure}

In \Cref{fig:04_hyperbolic_SC_exact_ttd_time_exact} we see that the connected cluster algorithm m4riCC has a clear advantage in time per successful trial for exact methods across the whole range. In \Cref{fig:04_hyperbolic_SC_heuristic_ttd} we see that QDistRnd and QDistEvol have the lowest time per successful trial, but for larger codes BP-OSD has better performance. This is most likely due to BP-OSD being well suited to decoding where the check matrix is very sparse. 
In \Cref{fig:04_hyperbolic_SC_heuristic_success_rate} we see that the Stim graphlike error, colour code and undetectable error methods all have 100\% success rate.
The success rate for the BP-OSD method is lower than QDistEvol and QDistRnd meaning that the advantage BP-OSD enjoys for larger codes is due to lower execution time.

\begin{table}[h!]
\setlength\tabcolsep{2pt}
\fontfamily{lmss}\fontsize{8}{9}\selectfont{
\begin{center}									
\begin{tabular}{ |l|	r|	r|	r|	r|	r|	r|	r|	}	\hline
 &	\textbf{m4riRW} &	\textbf{QDistRndMW} &	\textbf{QDistEvol} &	\textbf{BP-OSD} &	\textbf{GEStim} &	\textbf{CCStim} &	\textbf{UEStim}	\\	\hline
\textbf{Result Returned} &	21 &	21 &	21 &	21 &	21 &	21 &	21	\\	
\textbf{No Result} &	0 &	0 &	0 &	0 &	0 &	0 &	0	\\	
\textbf{At lowest distance} &	21 &	21 &	21 &	20 &	21 &	21 &	21	\\	
\textbf{Overall Success Rate} &	100.0\% &	100.0\% &	100.0\% &	95.2\% &	100.0\% &	100.0\% &	100.0\%	\\	\hline
\textbf{Total Time (s)} &	8.9E+01 &	7.0E+03 &	7.2E+03 &	3.6E+03 &	6.6E+00 &	4.7E+00 &	7.4E+00	\\	
\textbf{Trials} &	NA &	210,000 &	210,000 &	210,000 &	21 &	21 &	42	\\	
\textbf{Trials at lowest distance} &	 &	197,591 &	206,628 &	164,316 &	21 &	21 &	42	\\	
\textbf{Trial Success Rate} &	 &	94.1\% &	98.4\% &	78.2\% &	100.0\% &	100.0\% &	100.0\%	\\	
\textbf{Time/Trial} &	 &	3.3E-02 &	3.4E-02 &	1.7E-02 &	3.1E-01 &	2.2E-01 &	1.8E-01	\\	
\textbf{Time/Successful Trial} &	 &	3.6E-02 &	3.5E-02 &	2.2E-02 &	3.1E-01 &	2.2E-01 &	1.8E-01	\\	\hline
\end{tabular}									
\end{center}														
}
\caption{Distance finding benchmark hyperbolic surface codes - Heuristic Methods.}
\label{tab:hyperbolic_surface_heuristic}
\end{table}

We constructed a family of colour codes  by placing qubits on the vertices of the same \{8,3\} complex as the surface code family.
In this construction, both X and Z-type stabilisers are associated with the plaquettes.
This resulted in a family of 21 codes with up to 816 physical qubits. 
In \Cref{tab:hyperbolic_colour_exact} we compare exact methods for this code family.
We again find that exact methods are able to find distances for this code family quite effectively. 
Most methods were noticeably slower than for hyperbolic surface codes due to the higher degree Tanner graph, and only m4riCC was able to complete all codes within the 8 hour maximum execution time.

\begin{table}[h!]
\setlength\tabcolsep{2pt}
\fontfamily{lmss}\fontsize{8}{9}\selectfont{
\begin{center}							
\begin{tabular}{ |l|	r|	r|	r|	r|	r|	}	\hline
 &	\textbf{Gurobi} &	\textbf{MIP-SCIP} &	\textbf{CLISAT} &	\textbf{m4riCC} &	\textbf{Magma}	\\	\hline
\textbf{Result Returned} &	21 &	21 &	21 &	21 &	21	\\	
\textbf{Completed < MaxTime} &	20 &	17 &	20 &	21 &	9	\\	
\textbf{Completed > MaxTime} &	1 &	4 &	1 &	0 &	12	\\	
\textbf{No Result} &	0 &	0 &	0 &	0 &	0	\\	
\textbf{At lowest distance} &	21 &	21 &	21 &	21 &	21	\\	
\textbf{Overall success rate} &	100.0\% &	100.0\% &	100.0\% &	100.0\% &	100.0\%	\\	\hline
\textbf{Total Time} &	3.7E+04 &	1.2E+05 &	3.9E+04 &	1.2E+03 &	3.7E+05	\\	
\textbf{Time/Trial} &	1.7E+03 &	5.7E+03 &	1.8E+03 &	5.7E+01 &	1.8E+04	\\	
\textbf{Time/Successful Trial} &	1.7E+03 &	5.7E+03 &	1.8E+03 &	5.7E+01 &	1.8E+04	\\	\hline
\end{tabular}							
\end{center}							
}
\caption{Distance finding benchmark hyperbolic colour codes - Exact Methods.}
\label{tab:hyperbolic_colour_exact}
\end{table}

Turning now to heuristic methods, in \Cref{tab:hyperbolic_colour_heuristic} we see that the Stim graphlike error algorithm did not complete successfully - this is expected as the logical operators in the hyperbolic surface code are not graphlike in that errors on a single vertex may flip 3 stabilisers. 
The Stim colour code search did perform well because search constraints do allow for it to find errors on colour codes. 
This method had the lowest processing time overall for heuristic methods, though we note that the time per successful trial is significantly lower for the QDistRndMW and QDistEvol methods.

\begin{figure}[h!]
\centering
\begin{subfigure}[t]{.33\textwidth}
  \centering
  \includegraphics[width=\linewidth]{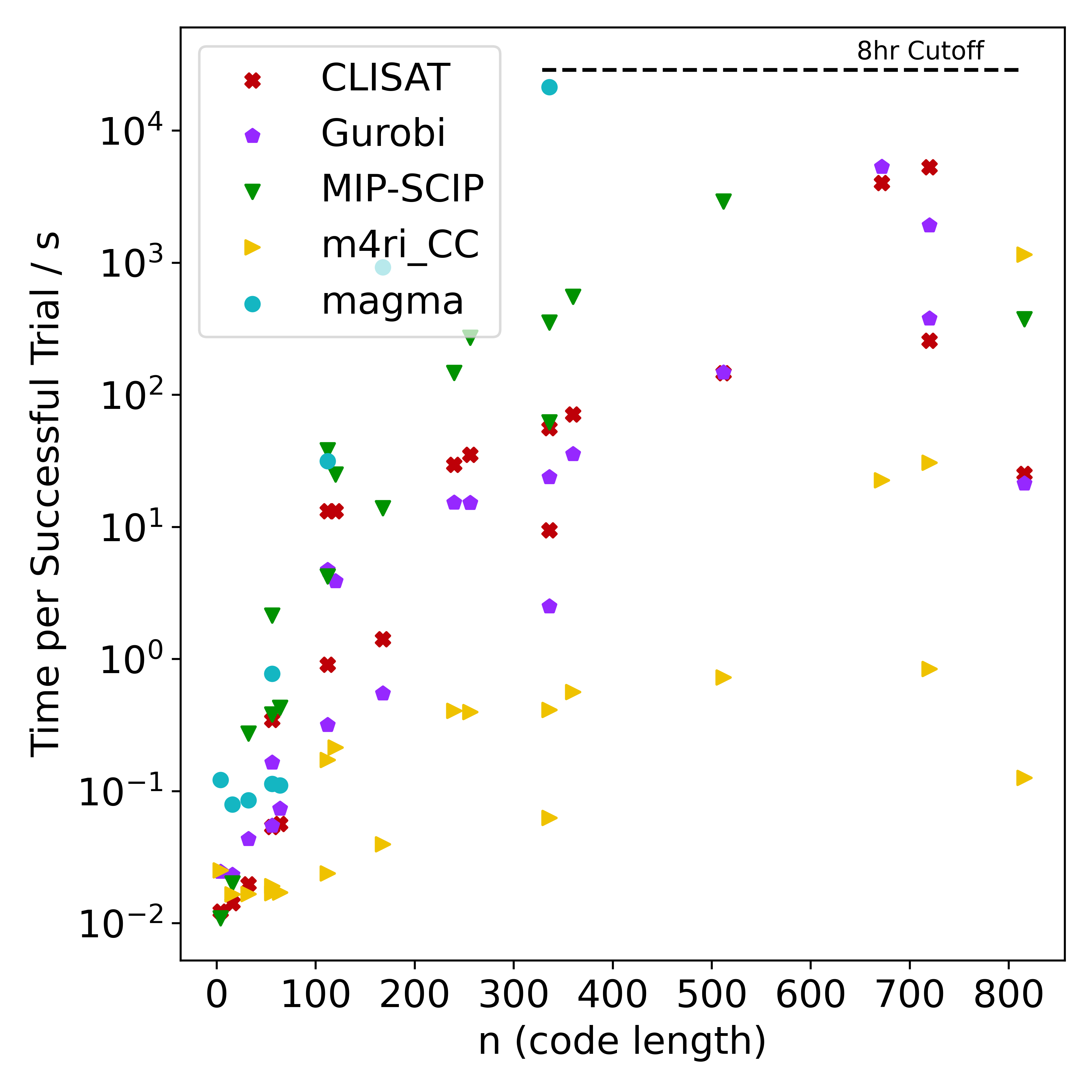}
  \subcaption{Time per Successful Trial \\ (Exact Algorithms)}\label{fig:05_hyperbolic_CC_exact_ttd_time_exact}
    \end{subfigure}%
\begin{subfigure}[t]{.33\textwidth}
  \centering
  \includegraphics[width=\linewidth]{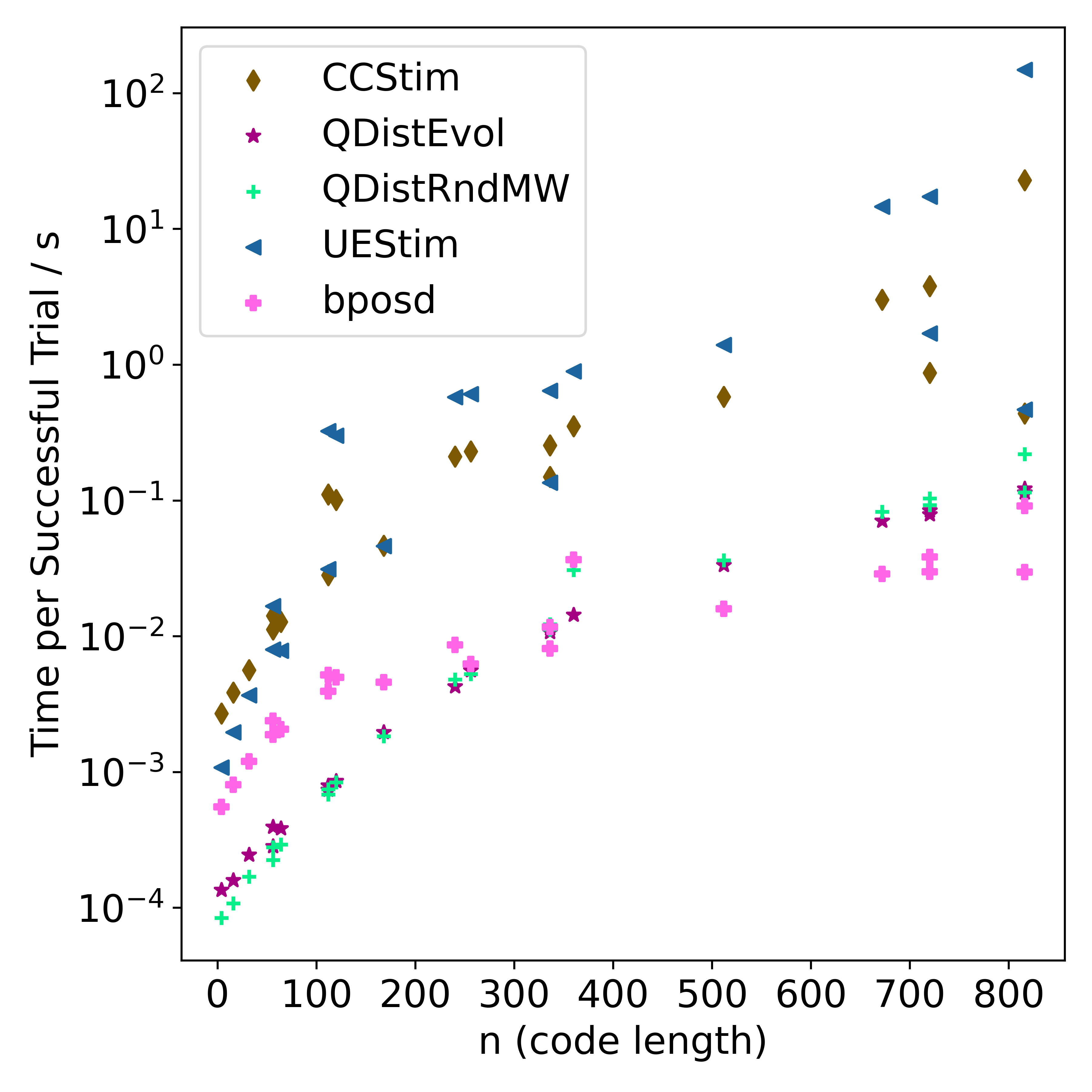}
  \subcaption{Time per Successful Trial \\ (Heuristic Algorithms)}\label{fig:05_hyperbolic_CC_heuristic_ttd}
    \end{subfigure}%
\begin{subfigure}[t]{.33\textwidth}
  \centering
  \includegraphics[width=\linewidth]{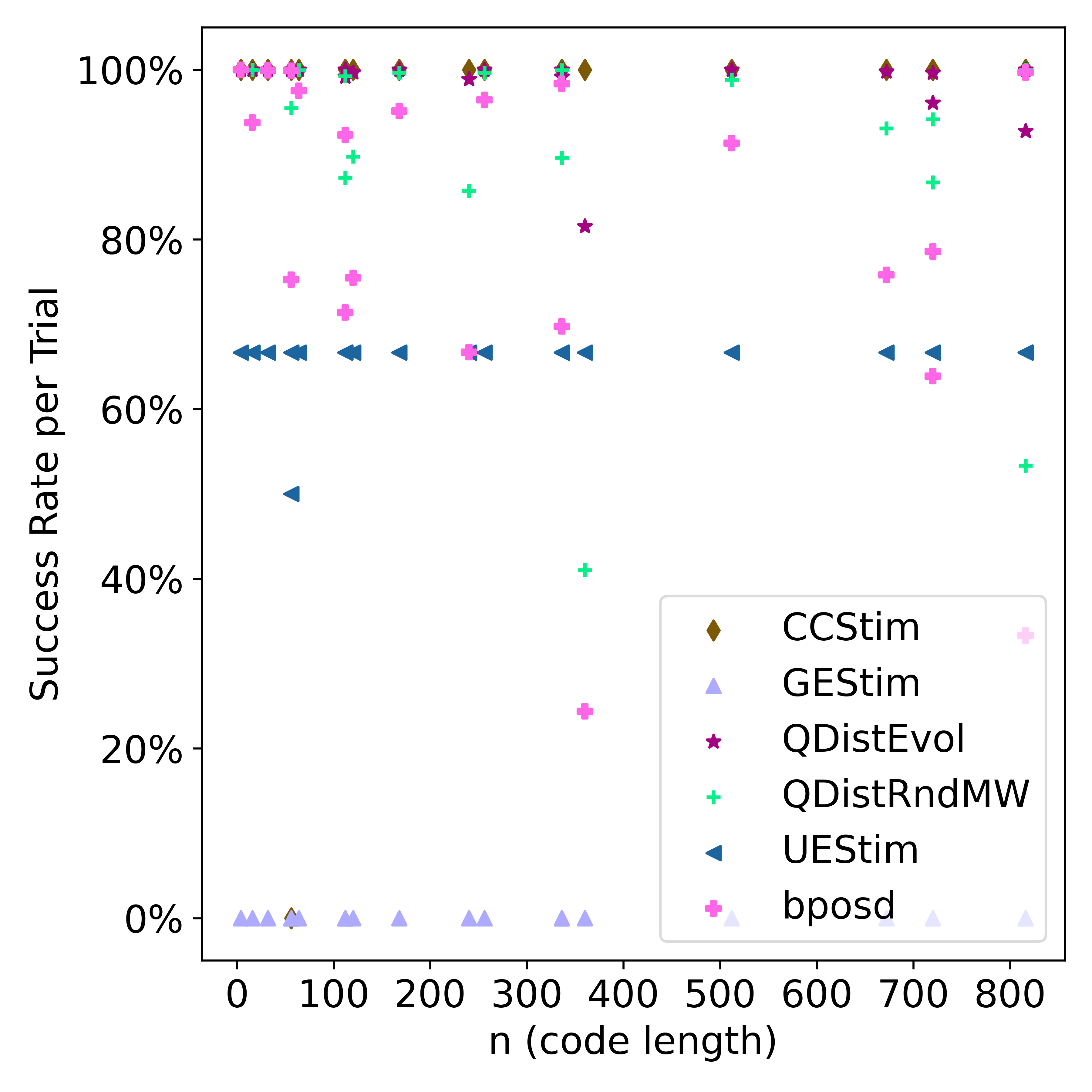}
  \subcaption{Success Rate per Trial \\ (Heuristic Algorithms)}\label{fig:05_hyperbolic_CC_heuristic_success_rate}
    \end{subfigure}%
\caption{Benchmark data by code length - hyperbolic colour codes}
\label{fig:05_hyperbolic_CC}
\end{figure}

In \Cref{fig:05_hyperbolic_CC_exact_ttd_time_exact} we see that the m4riCC method consistently had the lowest time per successful trial of exact methods, but the advantage was not as clear as the surface code case. 
QDistRnd and QDistEvol had the lowest time per successful trial for heuristic methods for small codes, but BP-OSD is competitive for larger members of the code family (\Cref{fig:05_hyperbolic_CC}).
Success rates for heuristic methods are plotted in \Cref{fig:05_hyperbolic_CC_heuristic_success_rate}.

\begin{table}[h!]
\setlength\tabcolsep{2pt}
\fontfamily{lmss}\fontsize{8}{9}\selectfont{
\begin{center}									
\begin{tabular}{ |l|	r|	r|	r|	r|	r|	r|	r|	}	\hline
 &	\textbf{m4riRW} &	\textbf{QDistRndMW} &	\textbf{QDistEvol} &	\textbf{BP-OSD} &	\textbf{GEStim} &	\textbf{CCStim} &	\textbf{UEStim}	\\	\hline
\textbf{Result Returned} &	21 &	21 &	21 &	21 &	0 &	21 &	21	\\	
\textbf{No Result} &	0 &	0 &	0 &	0 &	21 &	0 &	0	\\	
\textbf{At lowest distance} &	21 &	21 &	21 &	21 &	 &	20 &	21	\\	
\textbf{Overall Success Rate} &	100.0\% &	100.0\% &	100.0\% &	100.0\% &	 &	95.2\% &	100.0\%	\\	\hline
\textbf{Total Time (s)} &	3.6E+01 &	5.7E+03 &	5.4E+03 &	2.0E+03 &	3.4E+00 &	3.3E+01 &	3.8E+02	\\	
\textbf{Trials} &	NA &	210,000 &	210,000 &	210,000 &	21 &	21 &	64	\\	
\textbf{Trials at lowest distance} &	 &	191,342 &	206,657 &	169,892 &	 &	20 &	42	\\	
\textbf{Trial Success Rate} &	 &	91.1\% &	98.4\% &	80.9\% &	 &	95.2\% &	65.6\%	\\	
\textbf{Time/Trial} &	 &	2.7E-02 &	2.6E-02 &	9.7E-03 &	1.6E-01 &	1.6E+00 &	5.9E+00	\\	
\textbf{Time/Successful Trial} &	 &	3.0E-02 &	2.6E-02 &	1.2E-02 &	 &	1.7E+00 &	8.9E+00	\\	\hline
\end{tabular}									
\end{center}																								
}
\caption{Distance finding benchmark hyperbolic colour codes - Heuristic Methods.}
\label{tab:hyperbolic_colour_heuristic}
\end{table}

\subsection{Quantum Lifted Product Codes}\label{sec:LP}

The lifted product code family of \cite{LP_Panteleev,LP_Xu} is a high-rate CSS LDPC code which uses the classical LDPC codes of \Cref{sec:LP_classical} as part of its construction. 
We used a set of 19 lifted product codes of between 34 and 3230 physical qubits - these included several codes present in the literature supplemented with randomly generated codes.
The codes had row weights of 8 and column weights of either 3 or 5.
In \Cref{tab:lifted_product_exact} we see that the exact distance-finding methods completed a maximum of 6 of the codes in this family within the 8-hour maximum time frame.
The partial results for the Gurobi algorithm were accurate for all codes in the family, in contrast to the SCIP mixed integer, the SAT solver and Magma methods. 
We note that the minimum distances found were the same as for the classical lifted product code family of \Cref{sec:LP_classical}.
\begin{table}[h!]
\setlength\tabcolsep{2pt}
\fontfamily{lmss}\fontsize{8}{9}\selectfont{          
\begin{center}							
\begin{tabular}{ |l|	r|	r|	r|	r|	r|	}	\hline
 &	\textbf{Gurobi} &	\textbf{MIP-SCIP} &	\textbf{CLISAT} &	\textbf{m4riCC} &	\textbf{Magma}	\\	\hline
\textbf{Result Returned} &	19 &	18 &	14 &	6 &	19	\\	
\textbf{Completed < MaxTime} &	6 &	6 &	6 &	6 &	3	\\	
\textbf{Completed > MaxTime} &	13 &	12 &	8 &	0 &	16	\\	
\textbf{No Result} &	0 &	1 &	5 &	13 &	0	\\	
\textbf{At lowest distance} &	19 &	12 &	6 &	6 &	10	\\	
\textbf{Overall success rate} &	100.0\% &	63.2\% &	31.6\% &	31.6\% &	52.6\%	\\	\hline
\textbf{Total Time} &	3.8E+05 &	3.9E+05 &	4.0E+05 &	3.8E+05 &	4.6E+05	\\	
\textbf{Time/Trial} &	2.0E+04 &	2.1E+04 &	2.1E+04 &	2.0E+04 &	2.4E+04	\\	
\textbf{Time/Successful Trial} &	2.0E+04 &	3.3E+04 &	6.7E+04 &	6.3E+04 &	4.6E+04	\\	\hline
\end{tabular}							
\end{center}																
}
\caption{Distance finding benchmark quantum lifted product codes - Exact Methods.}
\label{tab:lifted_product_exact}
\end{table}

For heuristic methods, we see in \Cref{tab:lifted_product_heuristic} that the best results were given by the random information set methods, with QDistEvol finding the lowest distance in all cases.
In contrast, the BP-OSD method did not complete within the maximum server run time of 48 hours for codes with 2000 physical qubits or more.
The Stim undetectable error algorithm gave a distance estimate for only 9 of the 19 codes in the family, and the graphlike error and colour code versions of the algorithm had an even lower completion rate. This is likely due to the fact that errors in quantum LDPC codes are not guaranteed to be string-like.

\begin{figure}[h!]
\centering
\begin{subfigure}[t]{.33\textwidth}
  \centering
  \includegraphics[width=\linewidth]{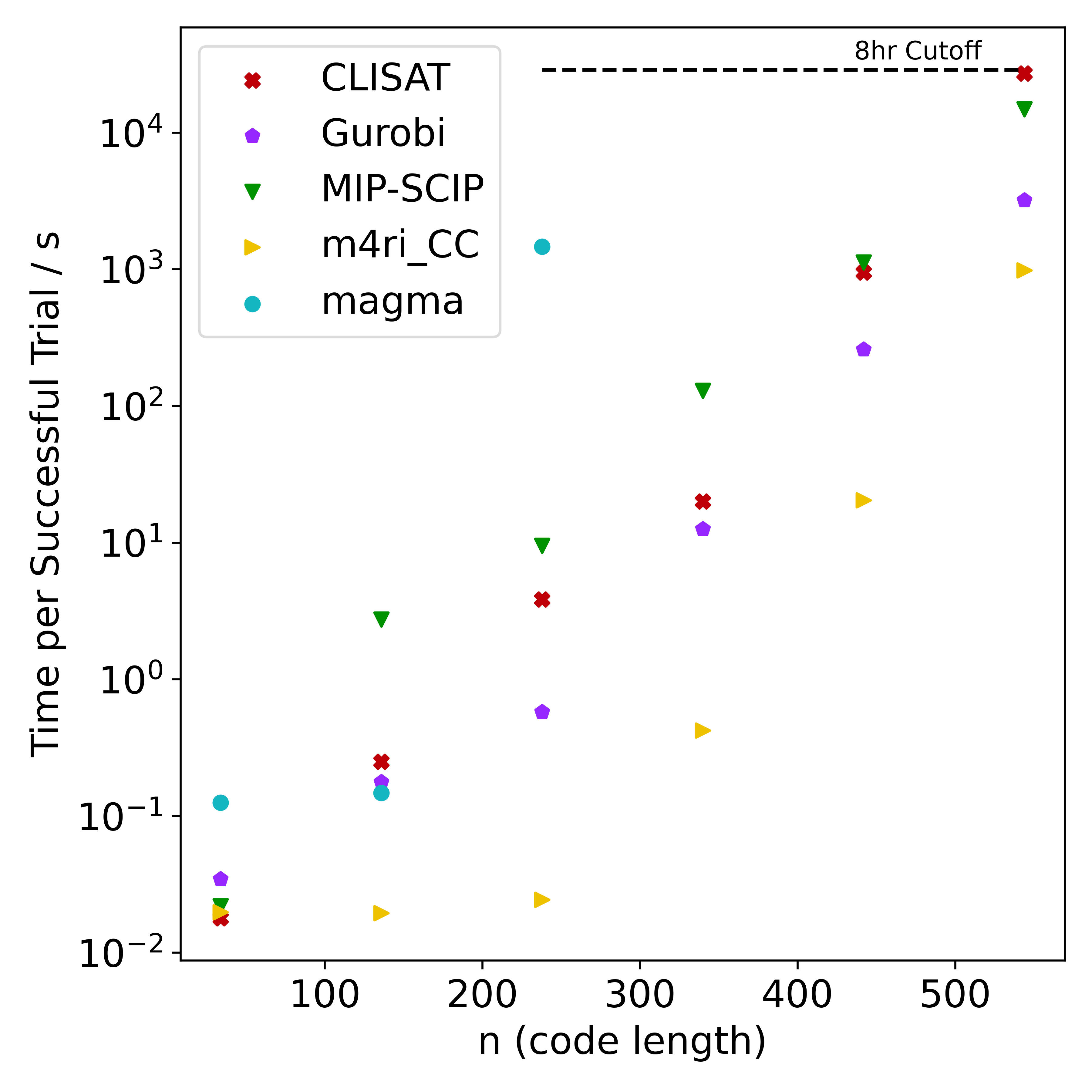}
  \subcaption{Time per Successful Trial \\ (Exact Algorithms)}\label{fig:06_lifted_product_exact_ttd_time_exact}
    \end{subfigure}%
\begin{subfigure}[t]{.33\textwidth}
  \centering
  \includegraphics[width=\linewidth]{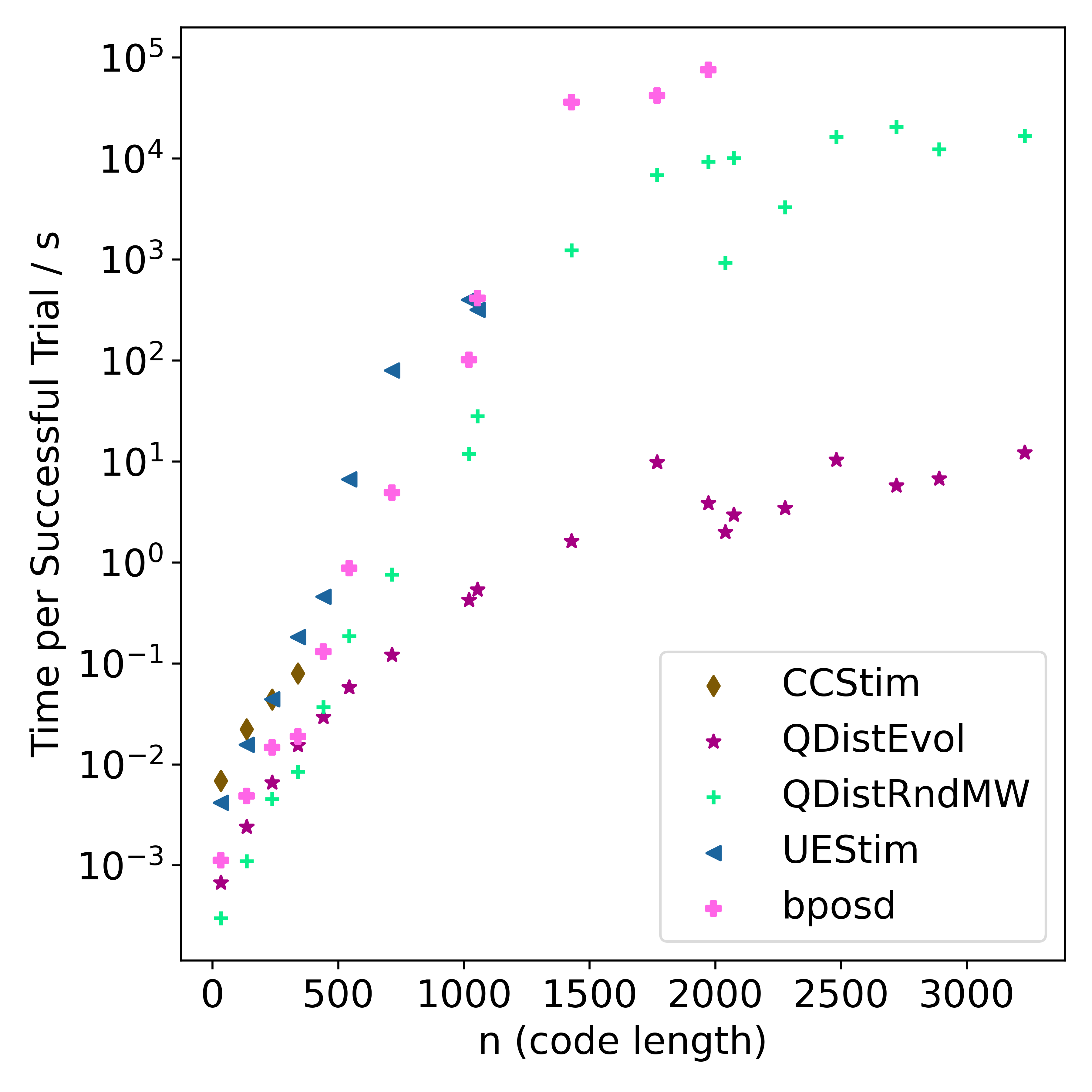}
  \subcaption{Time per Successful Trial \\ (Heuristic Algorithms)}\label{fig:06_lifted_product_heuristic_ttd}
    \end{subfigure}%
\begin{subfigure}[t]{.33\textwidth}
  \centering
  \includegraphics[width=\linewidth]{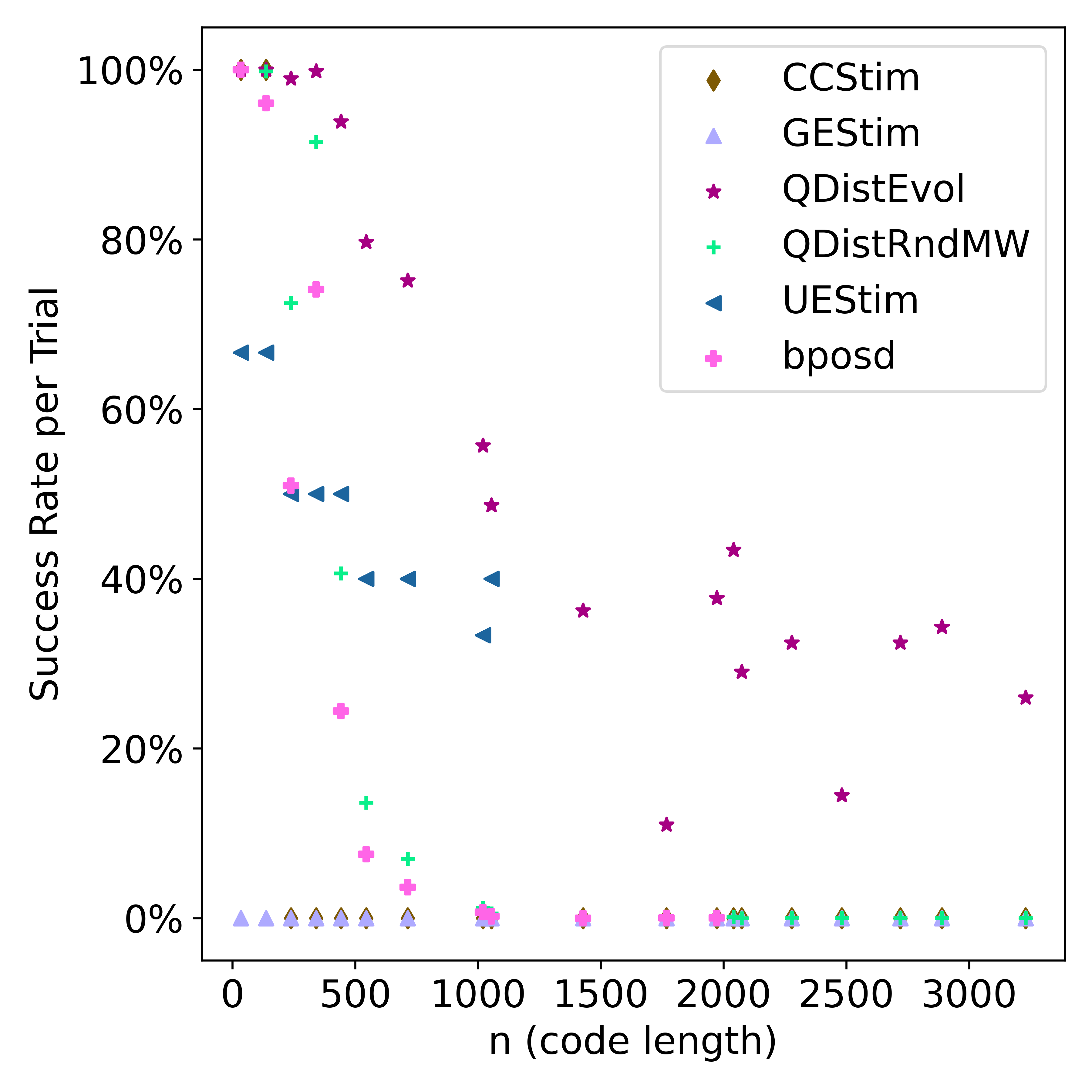}
  \subcaption{Success Rate per Trial \\ (Heuristic Algorithms)}\label{fig:06_lifted_product_heuristic_success_rate}
    \end{subfigure}%
\caption{Benchmark data by code length - lifted product CSS codes}
\label{fig:06_lifted_product}
\end{figure}

In \Cref{fig:06_lifted_product_exact_ttd_time_exact} we see that m4riCC gave the lowest time per successful trial of all the exact algorithms across the whole range, the next best being Gurobi MIP.
In \Cref{fig:06_lifted_product_heuristic_ttd} we see that QDistEvol has a significantly lower time per successful trial compared to other heuristic methods for the data set and this is due to it having a significantly higher success rate per trial (\Cref{fig:06_lifted_product_heuristic_success_rate}).

\begin{table}[h!]
\setlength\tabcolsep{2pt}
\fontfamily{lmss}\fontsize{8}{9}\selectfont{
\begin{center}									
\begin{tabular}{ |l|	r|	r|	r|	r|	r|	r|	r|	}	\hline
 &	\textbf{m4riRW} &	\textbf{QDistRndMW} &	\textbf{QDistEvol} &	\textbf{BP-OSD} &	\textbf{GEStim} &	\textbf{CCStim} &	\textbf{UEStim}	\\	\hline
\textbf{Result Returned} &	19 &	19 &	19 &	12 &	0 &	4 &	9	\\	
\textbf{No Result} &	0 &	0 &	0 &	7 &	19 &	15 &	10	\\	
\textbf{At lowest distance} &	17 &	17 &	19 &	11 &	 &	2 &	9	\\	
\textbf{Overall Success Rate} &	89.5\% &	89.5\% &	100.0\% &	57.9\% &	 &	10.5\% &	47.4\%	\\	\hline
\textbf{Total Time (s)} &	1.3E+03 &	1.5E+05 &	1.5E+05 &	2.9E+05 &	1.7E+01 &	1.7E+01 &	1.6E+03	\\	
\textbf{Trials} &	NA &	190,000 &	190,000 &	120,000 &	19 &	19 &	39	\\	
\textbf{Trials at lowest distance} &	 &	42,702 &	104,876 &	35,765 &	 &	2 &	18	\\	
\textbf{Trial Success Rate} &	 &	22.5\% &	55.2\% &	29.8\% &	 &	10.5\% &	46.2\%	\\	
\textbf{Time/Trial} &	 &	8.0E-01 &	8.1E-01 &	2.4E+00 &	9.0E-01 &	9.1E-01 &	4.1E+01	\\	
\textbf{Time/Successful Trial} &	 &	3.6E+00 &	1.5E+00 &	8.1E+00 &	 &	8.6E+00 &	8.9E+01	\\	\hline
\end{tabular}									
\end{center}																	
}
\caption{Distance finding benchmark quantum lifted product codes - Heuristic Methods.}
\label{tab:lifted_product_heuristic}
\end{table}

\subsubsection{Bivariate Bicycle Codes}\label{sec:BB}
We next benchmarked the distance-finding algorithms on the bivariate bicycle codes of \cite{BB_IBM}.
We constructed a family of 20 bivariate bicycle codes with between 72 and 3024 qubits formed of the codes in \cite{BB_IBM} as well as randomly generated codes.
The check matrix of these codes have row weight 6 and column weight 3.

In \Cref{tab:bivariate_bicycle_exact} we see that exact methods completed at most 5 of the codes within the 8 hour maximum time frame.
Thereafter, the SCIP mixed integer programming method had a higher accuracy for partial results, though even this was relatively low at 60\%.
For this data set lowest distances are not known with certainty.
We used the lowest distances obtained in any trial conducted in testing and benchmarking.

\begin{table}[h!]
\setlength\tabcolsep{2pt}
\fontfamily{lmss}\fontsize{8}{9}\selectfont{
\begin{center}							
\begin{tabular}{ |l|	r|	r|	r|	r|	r|	}	\hline
 &	\textbf{Gurobi} &	\textbf{MIP-SCIP} &	\textbf{CLISAT} &	\textbf{m4riCC} &	\textbf{Magma}	\\	\hline
\textbf{Result Returned} &	20 &	20 &	20 &	5 &	20	\\	
\textbf{Completed < MaxTime} &	5 &	5 &	5 &	5 &	1	\\	
\textbf{Completed > MaxTime} &	15 &	15 &	15 &	0 &	19	\\	
\textbf{No Result} &	0 &	0 &	0 &	15 &	0	\\	
\textbf{At lowest distance} &	9 &	12 &	6 &	5 &	9	\\	
\textbf{Overall success rate} &	45.0\% &	60.0\% &	30.0\% &	25.0\% &	45.0\%	\\	\hline
\textbf{Total Time} &	4.3E+05 &	4.3E+05 &	4.3E+05 &	4.3E+05 &	5.5E+05	\\	
\textbf{Time/Trial} &	2.2E+04 &	2.2E+04 &	2.2E+04 &	2.2E+04 &	2.7E+04	\\	
\textbf{Time/Successful Trial} &	4.8E+04 &	3.6E+04 &	7.2E+04 &	8.6E+04 &	6.1E+04	\\	\hline
\end{tabular}							
\end{center}																	
}
\caption{Distance finding benchmark bivariate bicycle codes - Exact Methods.}
\label{tab:bivariate_bicycle_exact}
\end{table}

Results for heuristic methods in \Cref{tab:bivariate_bicycle_heuristic} demonstrate that the QDistEvol has a clear advantage in accuracy compared to other methods, finding the lowest distance in 13 out of 20 cases.
The QDistRnd and BP-OSD methods had similar accuracy, finding the lowest distance in 8 of the 20 cases.

\begin{figure}[h!]
\centering
\begin{subfigure}[t]{.33\textwidth}
  \centering
  \includegraphics[width=\linewidth]{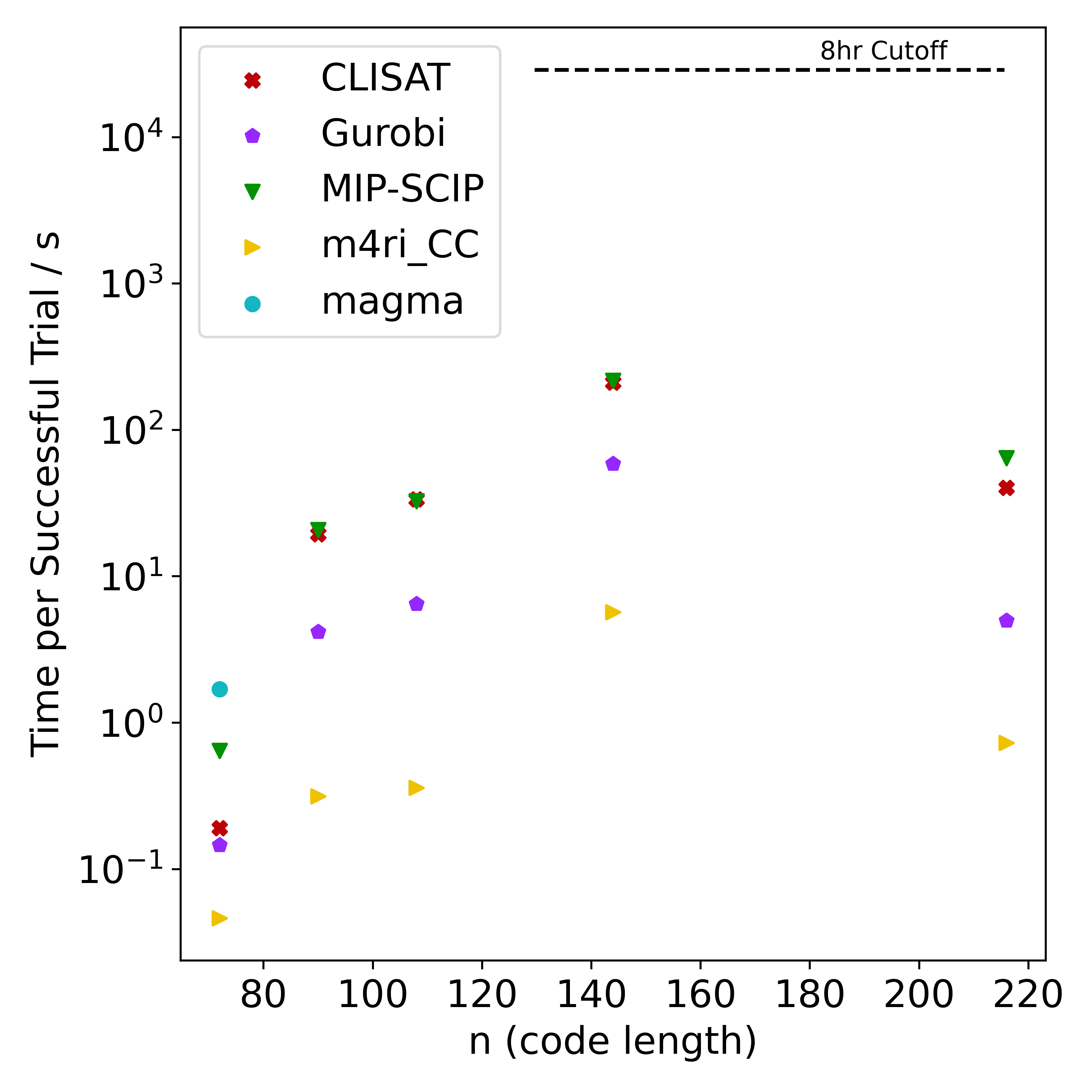}
  \subcaption{Time per Successful Trial \\ (Exact Algorithms)}\label{fig:07_bivariate_bicycle_exact_ttd_time_exact}
    \end{subfigure}%
\begin{subfigure}[t]{.33\textwidth}
  \centering
  \includegraphics[width=\linewidth]{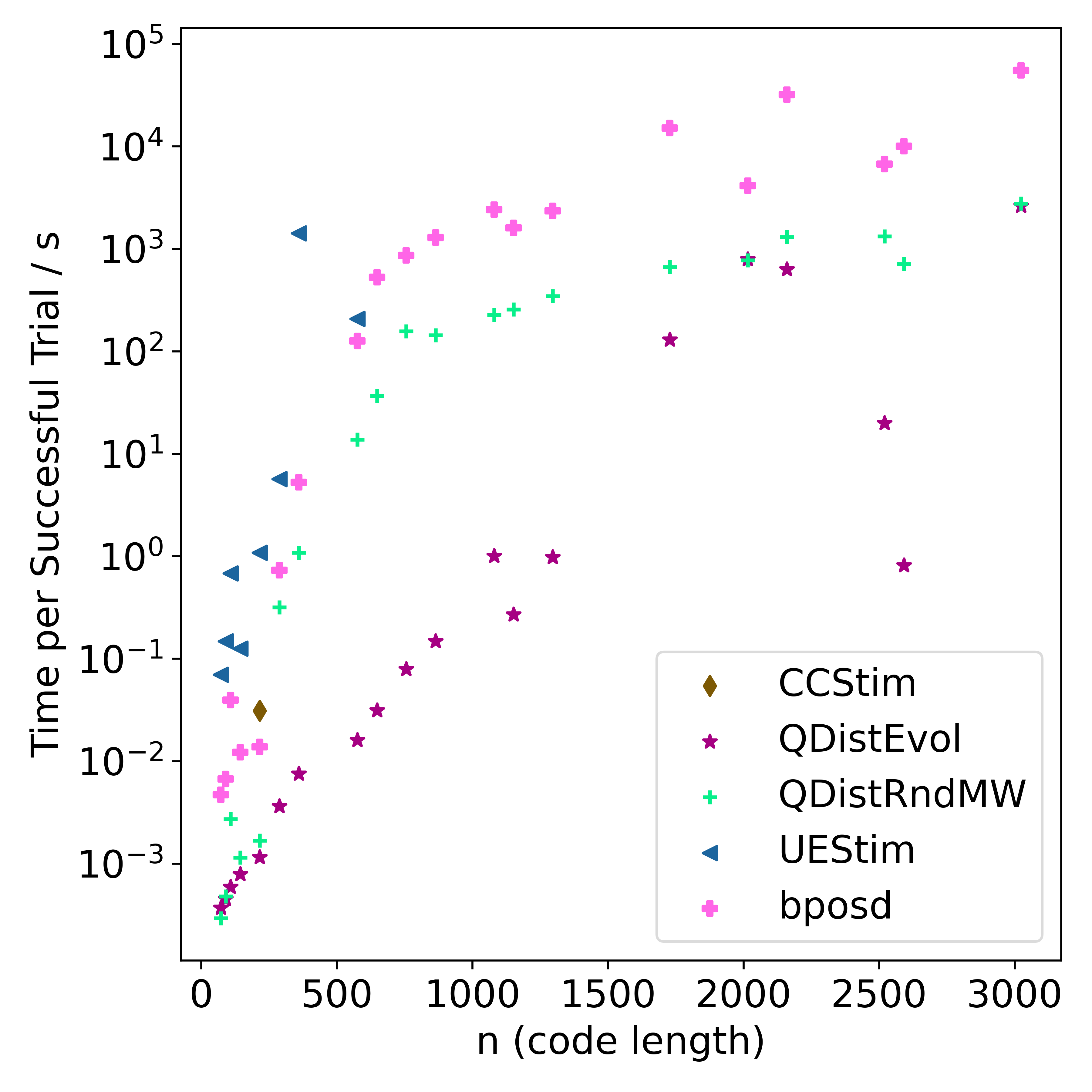}
  \subcaption{Time per Successful Trial \\ (Heuristic Algorithms)}\label{fig:07_bivariate_bicycle_heuristic_ttd}
    \end{subfigure}%
\begin{subfigure}[t]{.33\textwidth}
  \centering
  \includegraphics[width=\linewidth]{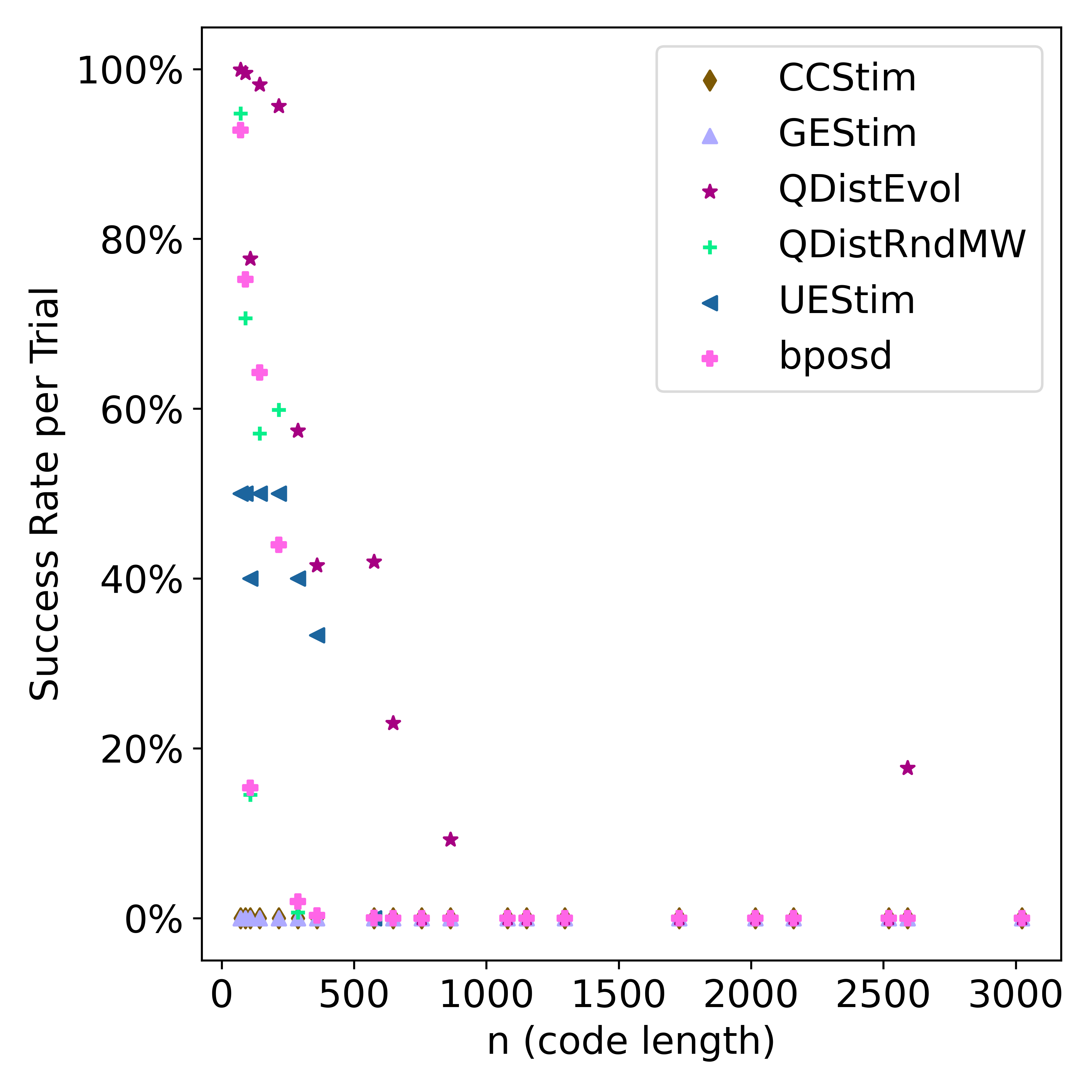}
  \subcaption{Success Rate per Trial \\ (Heuristic Algorithms)}\label{fig:07_bivariate_bicycle_heuristic_success_rate}
    \end{subfigure}%
\caption{Benchmark data by code length - bivariate bicycle CSS codes}
\label{fig:07_bivariate_bicycle}
\end{figure}

In \Cref{fig:07_bivariate_bicycle_exact_ttd_time_exact}, we see that m4riCC has the lowest processing time per successful trial for exact methods across the range. 
In \Cref{fig:07_bivariate_bicycle_heuristic_ttd} we see that QDistEvol has a lower processing time per successful trial compared to other methods, again due to a significantly better success rate per trial (\Cref{fig:07_bivariate_bicycle_heuristic_success_rate}).
                
\begin{table}[h!]
\setlength\tabcolsep{2pt}
\fontfamily{lmss}\fontsize{8}{9}\selectfont{	
\begin{center}									
\begin{tabular}{ |l|	r|	r|	r|	r|	r|	r|	r|	}	\hline
 &	\textbf{m4riRW} &	\textbf{QDistRndMW} &	\textbf{QDistEvol} &	\textbf{BP-OSD} &	\textbf{GEStim} &	\textbf{CCStim} &	\textbf{UEStim}	\\	\hline
\textbf{Result Returned} &	20 &	20 &	20 &	20 &	0 &	1 &	8	\\	
\textbf{No Result} &	0 &	0 &	0 &	0 &	20 &	19 &	12	\\	
\textbf{At lowest distance} &	8 &	8 &	13 &	8 &	 &	 &	7	\\	
\textbf{Overall Success Rate} &	40.0\% &	40.0\% &	65.0\% &	40.0\% &	 &	 &	35.0\%	\\	\hline
\textbf{Total Time (s)} &	7.6E+02 &	9.6E+03 &	9.4E+03 &	1.4E+05 &	3.0E+00 &	3.3E+00 &	3.3E+03	\\	
\textbf{Trials} &	NA &	200,000 &	200,000 &	200,000 &	20 &	20 &	37	\\	
\textbf{Trials at lowest distance} &	 &	29,787 &	66,166 &	29,389 &	 &	 &	14	\\	
\textbf{Trial Success Rate} &	 &	14.9\% &	33.1\% &	14.7\% &	 &	 &	37.8\%	\\	
\textbf{Time/Trial} &	 &	4.8E-02 &	4.7E-02 &	6.9E-01 &	1.5E-01 &	1.6E-01 &	8.8E+01	\\	
\textbf{Time/Successful Trial} &	 &	3.2E-01 &	1.4E-01 &	4.7E+00 &	 &	 &	2.3E+02	\\	\hline
\end{tabular}									
\end{center}									
}
\caption{Distance finding benchmark bivariate bicycle codes - Heuristic Methods.}
\label{tab:bivariate_bicycle_heuristic}
\end{table}

\subsubsection{Quantum Tanner Codes}\label{sec:quantum_tanner}
We next benchmarked the distance-finding algorithms on quantum Tanner codes, introduced in \cite{quantum_tanner}.
We used a selection of 22 codes from the \href{https://aleverrier.github.io/qtanner-search/best_codes/index.html}{Best qTanner codes} website with between 36 and 684 qubits.
These codes generally had row weight of 9, but some had weights of 8 or 16. Column weights were highly variable.

In \Cref{tab:tanner_codes_exact} we see that exact methods completed at most 5 of the codes within the 8 hour maximum time frame, with Gurobi MIP having the fastest processing time.
For partial results, we see that Magma had the highest accuracy, finding the lowest distance for 8 of the codes in the family.

\begin{table}[h!]
\setlength\tabcolsep{2pt}
\fontfamily{lmss}\fontsize{8}{9}\selectfont{
\begin{center}							
\begin{tabular}{ |l|	r|	r|	r|	r|	r|	}	\hline
 &	\textbf{Gurobi} &	\textbf{MIP-SCIP} &	\textbf{CLISAT} &	\textbf{m4riCC} &	\textbf{Magma}	\\	\hline
\textbf{Result Returned} &	22 &	22 &	22 &	4 &	22	\\	
\textbf{Completed < MaxTime} &	5 &	4 &	4 &	4 &	3	\\	
\textbf{Completed > MaxTime} &	17 &	18 &	18 &	0 &	19	\\	
\textbf{No Result} &	0 &	0 &	0 &	18 &	0	\\	
\textbf{At lowest distance} &	6 &	6 &	5 &	4 &	8	\\	
\textbf{Overall success rate} &	27.3\% &	27.3\% &	22.7\% &	18.2\% &	36.4\%	\\	\hline
\textbf{Total Time} &	5.0E+05 &	5.2E+05 &	5.2E+05 &	5.3E+05 &	5.5E+05	\\	
\textbf{Time/Trial} &	2.3E+04 &	2.4E+04 &	2.4E+04 &	2.4E+04 &	2.5E+04	\\	
\textbf{Time/Successful Trial} &	8.4E+04 &	8.7E+04 &	1.0E+05 &	1.3E+05 &	6.9E+04	\\	\hline
\end{tabular}							
\end{center}																
}
\caption{Distance finding benchmark bivariate bicycle codes - Exact Methods.}
\label{tab:tanner_codes_exact}
\end{table}

Results for heuristic methods are set out in \Cref{tab:tanner_codes_heuristic}. 
QDistEvol had the highest success rate of the heuristic methods, also finding the lowest distance for 8 of the 22 codes.

\begin{figure}[h!]
\centering
\begin{subfigure}[t]{.33\textwidth}
  \centering
  \includegraphics[width=\linewidth]{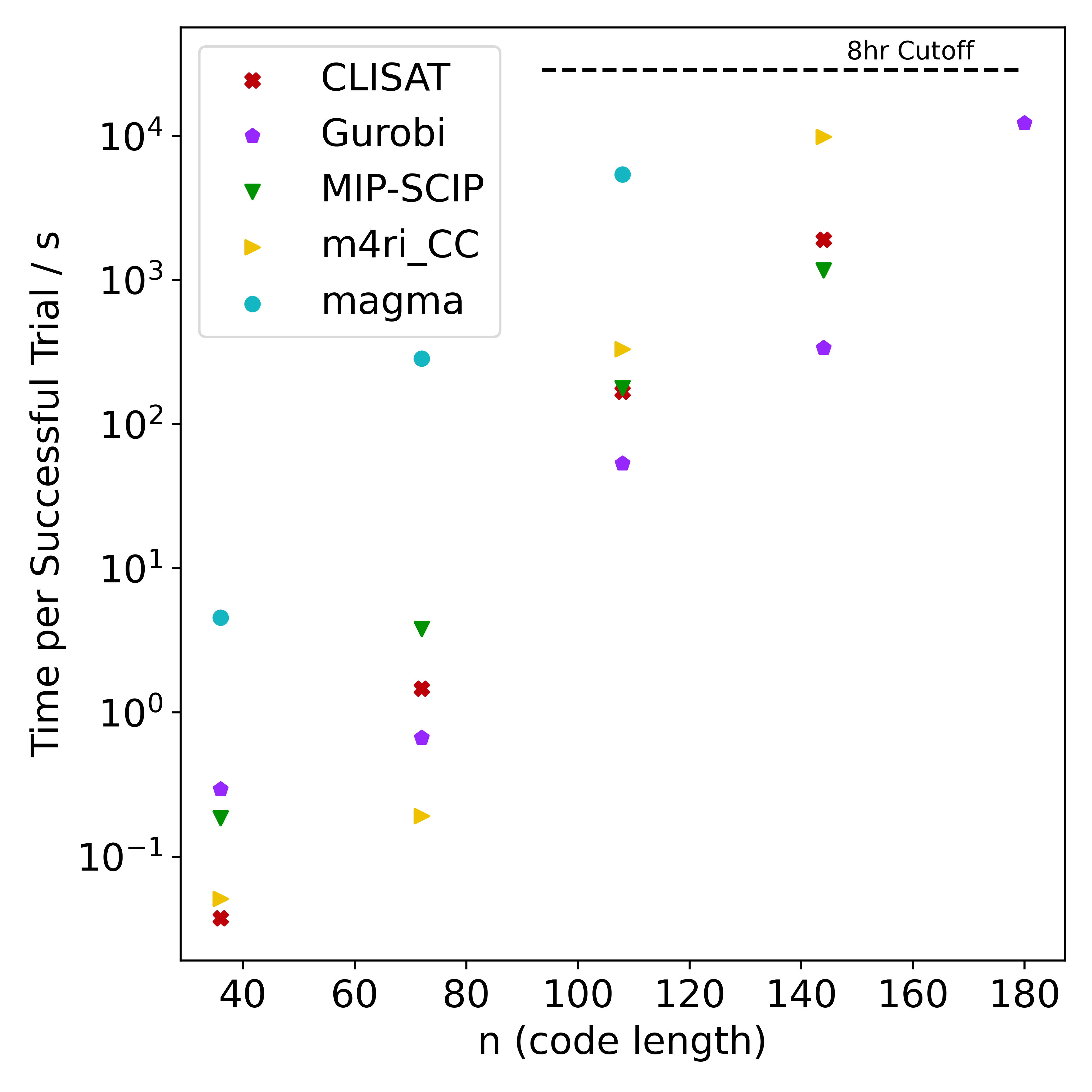}
  \subcaption{Time per Successful Trial \\ (Exact Algorithms)}\label{fig:12_tanner_codes_exact_ttd_time_exact}
    \end{subfigure}%
\begin{subfigure}[t]{.33\textwidth}
  \centering
  \includegraphics[width=\linewidth]{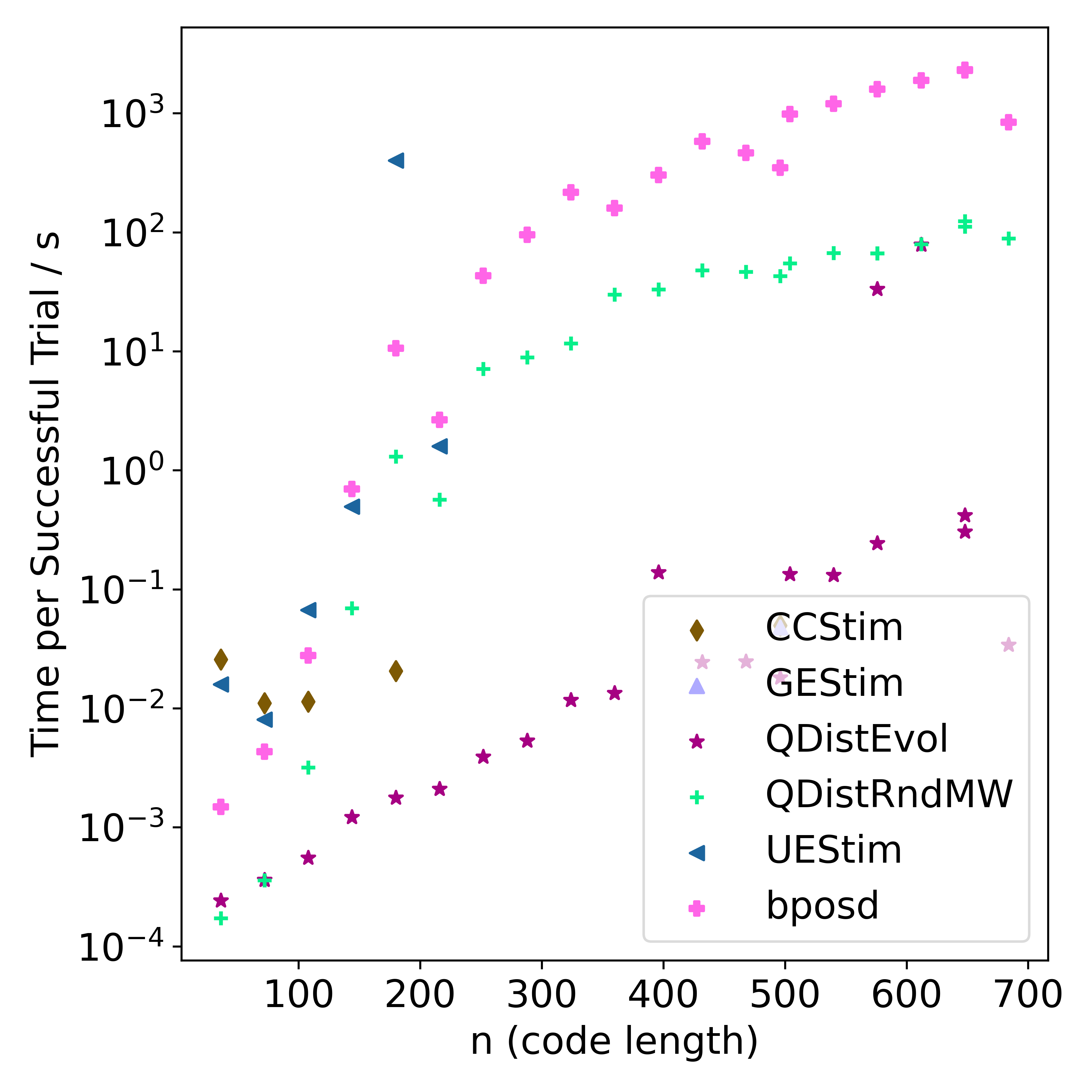}
  \subcaption{Time per Successful Trial \\ (Heuristic Algorithms)}\label{fig:12_tanner_codes_heuristic_ttd}
    \end{subfigure}%
\begin{subfigure}[t]{.33\textwidth}
  \centering
  \includegraphics[width=\linewidth]{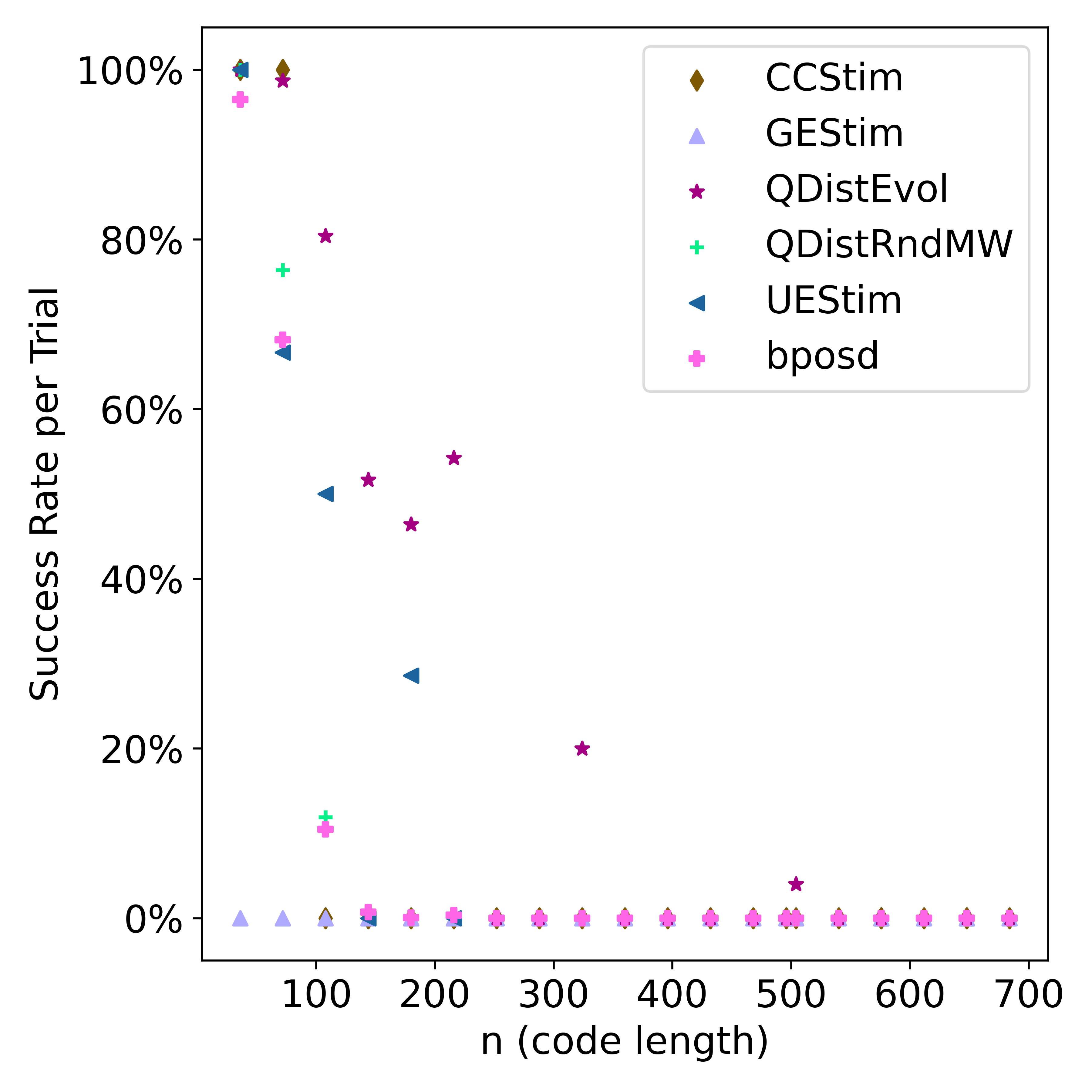}
  \subcaption{Success Rate per Trial \\ (Heuristic Algorithms)}\label{fig:12_tanner_codes_heuristic_success_rate}
    \end{subfigure}%
\caption{Benchmark data by code length - Quantum Tanner CSS codes}
\label{fig:12_tanner_codes}
\end{figure}

When comparing the processing time of exact methods by code size in \Cref{fig:12_tanner_codes_exact_ttd_time_exact}, we see that m4riCC has the lowest processing time per successful trial for smaller codes, but that the SAT and MIP methods are faster for larger codes, with Magma the slowest of the methods.
The same analysis for heuristic methods in \Cref{fig:12_tanner_codes_heuristic_ttd} shows that QDistEvol has a lower processing time per successful trial compared to other methods again due to a significantly better success rate per trial (\Cref{fig:12_tanner_codes_heuristic_success_rate}).
                
\begin{table}[h!]
\setlength\tabcolsep{2pt}
\fontfamily{lmss}\fontsize{8}{9}\selectfont{	
\begin{center}									
\begin{tabular}{ |l|	r|	r|	r|	r|	r|	r|	r|	}	\hline
 &	\textbf{m4riRW} &	\textbf{QDistRndMW} &	\textbf{QDistEvol} &	\textbf{BP-OSD} &	\textbf{GEStim} &	\textbf{CCStim} &	\textbf{UEStim}	\\	\hline
\textbf{Result Returned} &	22 &	22 &	22 &	22 &	1 &	5 &	6	\\	
\textbf{No Result} &	0 &	0 &	0 &	0 &	21 &	17 &	16	\\	
\textbf{At lowest distance} &	6 &	6 &	8 &	6 &	 &	2 &	4	\\	
\textbf{Overall Success Rate} &	27.3\% &	27.3\% &	36.4\% &	27.3\% &	 &	9.1\% &	18.2\%	\\	\hline
\textbf{Total Time (s)} &	8.6E+01 &	9.5E+02 &	9.3E+02 &	1.6E+04 &	8.0E-01 &	8.6E-01 &	8.0E+02	\\	
\textbf{Trials} &	NA &	220,000 &	220,000 &	220,000 &	22 &	22 &	25	\\	
\textbf{Trials at lowest distance} &	 &	18,937 &	45,536 &	17,623 &	 &	2 &	8	\\	
\textbf{Trial Success Rate} &	 &	8.6\% &	20.7\% &	8.0\% &	 &	9.1\% &	32.0\%	\\	
\textbf{Time/Trial} &	 &	4.3E-03 &	4.2E-03 &	7.1E-02 &	3.6E-02 &	3.9E-02 &	3.2E+01	\\	
\textbf{Time/Successful Trial} &	 &	5.0E-02 &	2.1E-02 &	8.8E-01 &	 &	4.3E-01 &	1.0E+02	\\	\hline
\end{tabular}									
\end{center}									}
\caption{Distance finding benchmark Quantum Tanner CSS codes - Heuristic Methods.}
\label{tab:tanner_codes_heuristic}
\end{table}

\subsection{Syndrome Extraction Circuits}\label{sec:benchmark_QC}

We next turn to distance-finding for syndrome extraction circuits.
The circuits used are the test circuits from the GitHub repository for the tesseract decoder of \cite{beni2025tesseractdecoder}.

Certain circuits in the benchmark gave rise to very large detector error models (DEMs) and many of the algorithms struggled to give distance estimates for the circuits.
Accordingly we used a DEM filtering algorithm, set out in \Cref{alg:DEM_filter}, to reduce the number of error mechanisms and detectors.
The circuits used for our benchmarking have \href{https://github.com/quantumlib/chromobius/blob/3ce4454032b1c004a97c4f12b9f96f7f04dc54fa/doc/chromobius_api_reference.md?plain=1#L31-L41}{Chromobius-style coordinates}.
For this circuit type, the fourth coordinate is an integer indicating the basis and colour of the detector and this allows us to identify which of the detectors are of X-type (values 0,1,2) and which are of Z-type (values 3,4,5).
To restrict the DEM to errors of Z-type, we remove any detectors which are not of Z-type and any error mechanism which is not in the support of the filtered detectors. 
To ensure that the observables/logicals are in a single basis, we also filter the DEM by detectors of X-type and check that observables cannot be flipped by both errors of X-type and errors of Z-type.
The method can easily be modified to cater for circuits using different coordinate conventions by modifying the \texttt{coordBasis} function which is used as an input to \Cref{alg:DEM_filter}.

This method leads to a significant reduction in the number of error mechanisms and detectors for the quantum circuits in our benchmark, and this is summarised in \Cref{tab:DEM_filter}.
The results in \Cref{tab:DEM_filter_algorithms} indicate that DEM filtering results in better performance across the board for all distance-finding algorithms in the benchmark for the superdense colour code circuit family.
We restricted the number of iterations for probabilistic algorithms to 1000 for this table because even for this lower number of iterations, the algorithms struggled to complete on the unfiltered DEM.
Note that in the final benchmarking data, we used 10000 iterations on the filtered DEMs as this gave more accurate distance-finding results. 

\begin{figure}[H]
\begin{algorithm}[H]
\caption{DEM Detector/Error Filter}\label{alg:DEM_filter}
\begin{algorithmic}
\State\textbf{Input:} 
\begin{itemize}
\item DEM: A Stim detector error model with coordinates
\item desiredBasis: a Pauli operator type ("X" or "Z")
\item coordBasis: a function taking a Stim coordinate as input and outputting a Pauli basis
\end{itemize}
\State\textbf{Output:}  
\State A new DEM which only includes detectors of type desiredBasis and errors which only flip detectors of this type
\State\textbf{Method:}
\State DEM := DEM.flattened()
\State NewDetectors := set()
\For {D, coord in  DEM.get\_detector\_coordinates().items()}
    \If{coordBasis(coord) == desiredBasis}
        \State NewDetectors.add(D)
    \EndIf
\EndFor
\State NewDEM := stim.DetectorErrorModel()
\For {instruction in DEM}
    \If{instruction.type in \{"detector","error"\}}
        \State myTargets = \{t.val for t in instruction.targets\_copy() if t.is\_relative\_detector\_id()\}
        \If{myTargets.issubset(NewDetectors)}
            \State NewDEM.append(instruction)
        \EndIf
    \Else 
        \State NewDEM.append(instruction)
    \EndIf
\EndFor
\State return NewDEM
\end{algorithmic}
\end{algorithm}
\end{figure}

\begin{table}[h!]
\setlength\tabcolsep{2pt}
\fontfamily{lmss}\fontsize{8}{9}\selectfont{
\begin{center}								
\begin{tabular}{ |r|	r|	r|	r|	r|	r|	r|	r|}	\hline
&	&	\multicolumn{2}{l|}{\textbf{Unfiltered}}	&	\multicolumn{2}{l|}{\textbf{Filtered}}	&	\multicolumn{2}{l|}{\textbf{Decrease}}	\\	\hline
\textbf{n}&	\textbf{k}&	\textbf{Errors}&	\textbf{Detectors}&	\textbf{Errors}&	\textbf{Detectors}&	\textbf{Errors}&	\textbf{Detectors}\\	\hline
\multicolumn{8}{|c|}{\textbf{Surface Code}}\\								\hline
17&	1&	221&	24&	55&	16&	75.1\%&	33.3\%\\	
49&	1&	1,679&	120&	385&	72&	77.1\%&	40.0\%\\	
97&	1&	5,473&	336&	1,141&	192&	79.2\%&	42.9\%\\	
161&	1&	12,707&	720&	2,529&	400&	80.1\%&	44.4\%\\	
241&	1&	24,485&	1,320&	4,741&	720&	80.6\%&	45.5\%\\	\hline
\multicolumn{8}{|c|}{\textbf{Colour Code - Midout}}\\								\hline
9&	1&	143&	12&	33&	8&	76.9\%&	33.3\%\\	
23&	1&	1,028&	55&	171&	33&	83.4\%&	40.0\%\\	
43&	1&	3,186&	147&	475&	84&	85.1\%&	42.9\%\\	
69&	1&	7,197&	306&	1,320&	170&	81.7\%&	44.4\%\\	
101&	1&	13,597&	550&	2,459&	300&	81.9\%&	45.5\%\\	\hline
\multicolumn{8}{|c|}{\textbf{Colour Code - Superdense}}\\								\hline
13&	1&	358&	18&	70&	12&	80.4\%&	33.3\%\\	
37&	1&	2,788&	90&	471&	54&	83.1\%&	40.0\%\\	
73&	1&	9,143&	252&	1,433&	144&	84.3\%&	42.9\%\\	
121&	1&	21,277&	540&	3,215&	300&	84.9\%&	44.4\%\\	
169&	1&	169&	168&	169&	84&	0.0\%&	50.0\%\\	
181&	1&	41,044&	990&	6,069&	540&	85.2\%&	45.5\%\\	\hline
\multicolumn{8}{|c|}{\textbf{Bivariate Bicycle Code}}\\								\hline
144&	12&	16,164&	432&	2,592&	252&	84.0\%&	41.7\%\\	
180&	8&	34,965&	900&	5,400&	495&	84.6\%&	45.0\%\\	
216&	8&	41,958&	1,080&	6,480&	594&	84.6\%&	45.0\%\\	
288&	12&	67,752&	1,728&	10,368&	936&	84.7\%&	45.8\%\\	
576&	12&	206,352&	5,184&	31,104&	2,736&	84.9\%&	47.2\%\\	
720&	12&	346,500&	8,640&	51,840&	4,500&	85.0\%&	47.9\%\\	\hline
\end{tabular}								
\end{center}																						
}
\caption{Quantum Circuit Data Set - DEM Size Reduction due to Circuit Filtering}
\label{tab:DEM_filter}
\end{table}

\begin{table}[h!]
\setlength\tabcolsep{2pt}
\fontfamily{lmss}\fontsize{8}{9}\selectfont{
\begin{center}										
\begin{tabular}{ |l|	l|	r|	r|	r|	r|	r|	r|	r|	r|}	\hline
&	\multicolumn{3}{l|}{\textbf{Completed}}&			\multicolumn{3}{l|}{\textbf{At lowest distance}}&			\multicolumn{3}{l|}{\textbf{Processing time}}\\			\hline
&	\textbf{Unfiltered}&	\textbf{Filtered}&	\textbf{Incr}&	\textbf{Unfiltered}&	\textbf{Filtered}&	\textbf{Incr} &	\textbf{Unfiltered}&	\textbf{Filtered}&	\textbf{Decr}\\	\hline
\multicolumn{10}{|c|}{\textbf{Exact Methods}}\\										\hline
\textbf{Gurobi}&	1&	6&	5&	1&	6&	5&	1.4E+05&	5.9E+04&	59.0\%\\	
\textbf{MIP-SCIP}&	6&	6&	-&	6&	6&	-&	8.8E+04&	6.8E+04&	22.5\%\\	
\textbf{CLISAT}&	6&	6&	-&	6&	6&	-&	8.9E+04&	8.0E+04&	10.8\%\\	
\textbf{m4riCC}&	2&	6&	4&	2&	6&	4&	1.2E+05&	5.8E+04&	50.0\%\\	\hline
\multicolumn{10}{|c|}{\textbf{Heuristic Methods}}\\										\hline
\textbf{m4riRW}&	6&	6&	-&	4&	6&	2&	1.3E+02&	7.0E+00&	94.5\%\\	
\textbf{QDistRndMW}&	6&	6&	-&	5&	6&	1&	9.6E+03&	2.2E+02&	97.7\%\\	
\textbf{QDistEvol}&	6&	6&	-&	6&	6&	-&	9.0E+03&	2.3E+02&	97.4\%\\	
\textbf{BP-OSD}&	4&	6&	2&	4&	6&	2&	1.1E+05&	6.4E+02&	99.4\%\\	
\textbf{GEStim}&	6&	6&	-&	6&	6&	-&	8.4E+00&	3.1E+00&	62.9\%\\	
\textbf{CCStim}&	6&	6&	-&	6&	6&	-&	1.0E+03&	4.6E+01&	95.4\%\\	
\textbf{GEStim}&	6&	6&	-&	6&	6&	-&	8.4E+00&	3.1E+00&	62.9\%\\	\hline
\end{tabular}										
\end{center}										
}
\caption{Superdense Colour Code Circuit Data Set - Benefit of Circuit Filtering - exact and heuristic methods - 1000 trials for probabilistic algorithms.}
\label{tab:DEM_filter_algorithms}
\end{table}

\subsubsection{Surface Code Circuits}\label{sec:QC_surface_codes}
We first calculate distances for syndrome extraction circuits for the rotated surface code of distances 3 to 11 from \cite{surface_code_circuits}. 
In \Cref{tab:qc_SC_exact} we see both mixed integer processing methods gave  correct distances for all members of this family.
The m4riCC algorithm completed 4 of the 5 members of the family within the 8 hour time frame, whereas the SAT method was unable to give any result for 3 of the 5 circuits.

\begin{table}[h!]
\setlength\tabcolsep{2pt}
\fontfamily{lmss}\fontsize{8}{9}\selectfont{
\begin{center}							
\begin{tabular}{ |l|	r|	r|	r|	r|	r|	}	\hline
 &	\textbf{Gurobi} &	\textbf{MIP-SCIP} &	\textbf{CLISAT} &	\textbf{m4riCC} &	\textbf{Magma}	\\	\hline
\textbf{Result Returned} &	5 &	5 &	2 &	4 &	4	\\	
\textbf{Completed < MaxTime} &	3 &	3 &	2 &	4 &	2	\\	
\textbf{Completed > MaxTime} &	2 &	2 &	0 &	0 &	2	\\	
\textbf{No Result} &	0 &	0 &	3 &	1 &	1	\\	
\textbf{At lowest distance} &	5 &	5 &	2 &	4 &	4	\\	
\textbf{Overall success rate} &	100.0\% &	100.0\% &	40.0\% &	80.0\% &	80.0\%	\\	\hline
\textbf{Total Time} &	5.8E+04 &	5.8E+04 &	8.6E+04 &	2.9E+04 &	8.6E+04	\\	
\textbf{Time/Trial} &	1.2E+04 &	1.2E+04 &	1.7E+04 &	5.8E+03 &	1.7E+04	\\	
\textbf{Time/Successful Trial} &	1.2E+04 &	1.2E+04 &	4.3E+04 &	7.3E+03 &	2.2E+04	\\	\hline
\end{tabular}							
\end{center}		
}
\caption{Distance finding benchmark surface code circuits - Exact Methods.}
\label{tab:qc_SC_exact}
\end{table}

In \Cref{tab:qc_SC_heuristic} we see that all heuristic methods gave correct distances for this code family. 
The Stim graphlike error algorithm had the lowest overall processing time, but the random information set algorithms had the lowest processing time per successful trial.

\begin{figure}[h!]
\centering
\begin{subfigure}[t]{.33\textwidth}
  \centering
  \includegraphics[width=\linewidth]{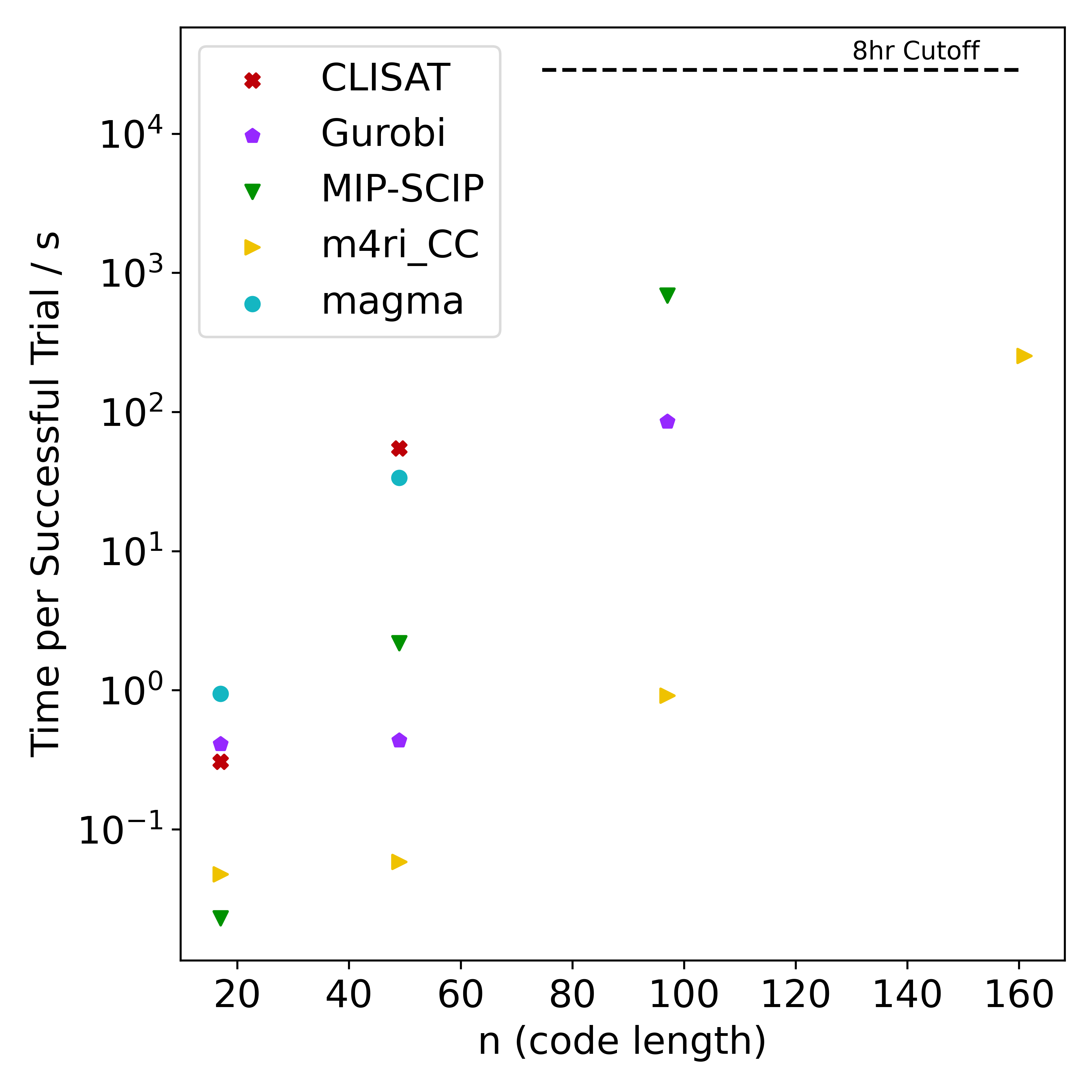}
  \subcaption{Time per Successful Trial \\ (Exact Algorithms)}\label{fig:08_qc_SC_exact_ttd_time_exact}
    \end{subfigure}%
\begin{subfigure}[t]{.33\textwidth}
  \centering
  \includegraphics[width=\linewidth]{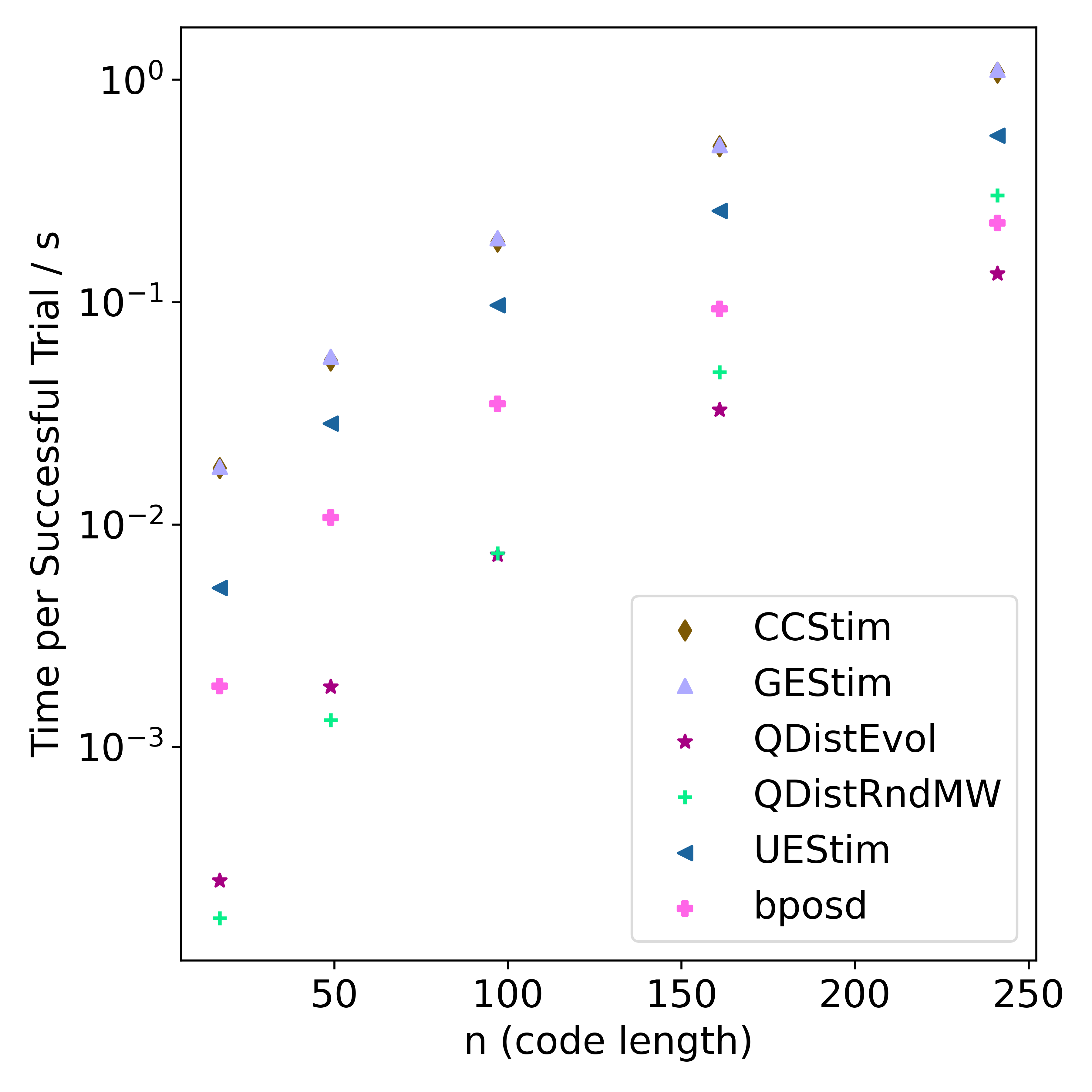}
  \subcaption{Time per Successful Trial \\ (Heuristic Algorithms)}\label{fig:08_qc_SC_heuristic_ttd}
    \end{subfigure}%
\begin{subfigure}[t]{.33\textwidth}
  \centering
  \includegraphics[width=\linewidth]{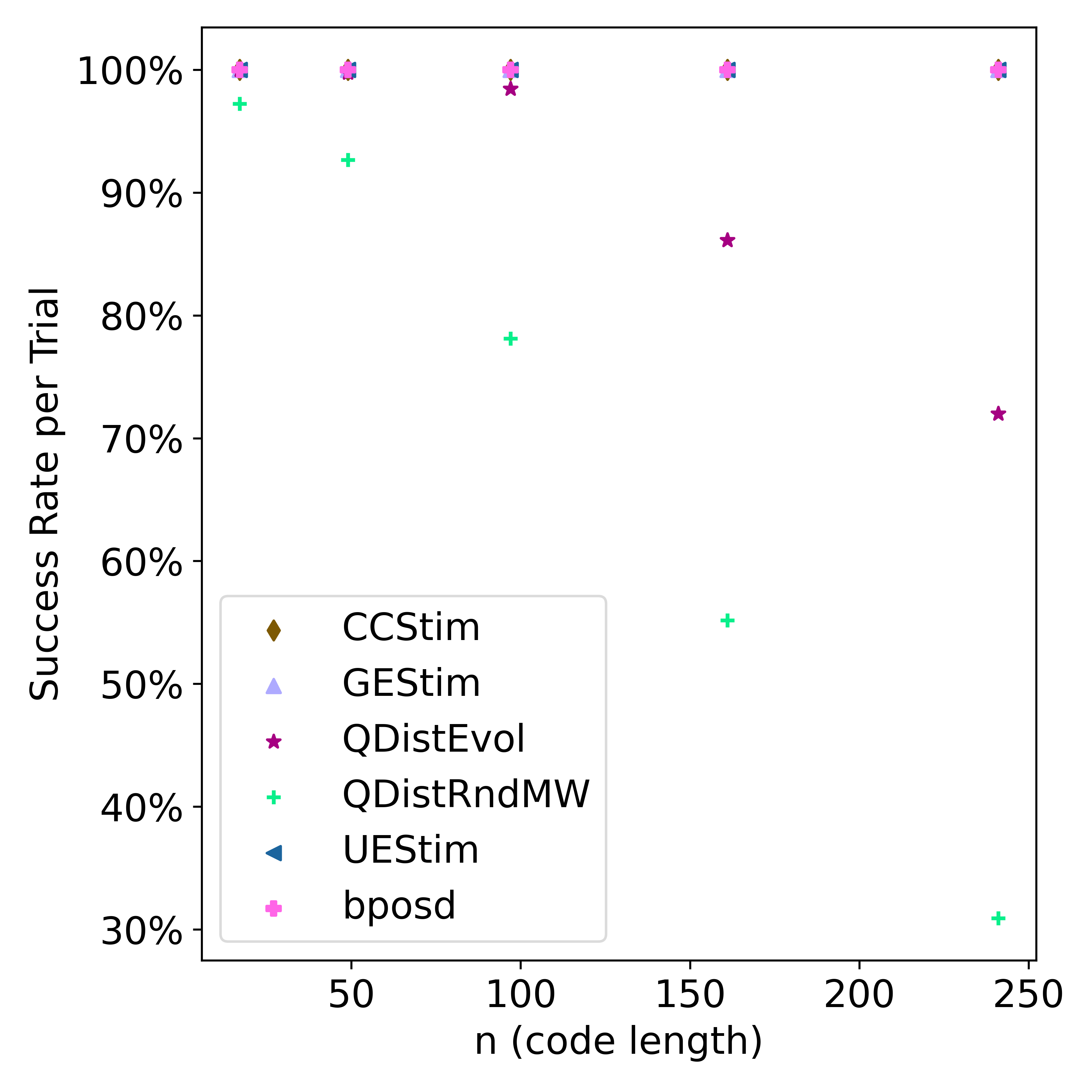}
  \subcaption{Success Rate per Trial \\ (Heuristic Algorithms)}\label{fig:08_qc_SC_heuristic_success_rate}
    \end{subfigure}%
\caption{Benchmark data by code length - surface code syndrome circuits}
\label{fig:08_qc_SC}
\end{figure}

In \Cref{fig:08_qc_SC_exact_ttd_time_exact}, we see that the m4riCC algorithm consistently had the lowest time per successful trial for the exact methods.
In \Cref{fig:08_qc_SC_heuristic_success_rate}, we see that most methods had close to 100\% success rate but that success rate of the QDistRnd and QDistEvol fell steeply for larger codes. QDistRnd and QDistEvol had the lowest time per successful trial for the small circuits in this family, but for larger codes the advantage over BP-OSD reduces and this suggests that BP-OSD may be better for very large members of this family.

\begin{table}[h!]
\setlength\tabcolsep{2pt}
\fontfamily{lmss}\fontsize{8}{9}\selectfont{
\begin{center}									
\begin{tabular}{ |l|	r|	r|	r|	r|	r|	r|	r|	}	\hline
 &	\textbf{m4riRW} &	\textbf{QDistRndMW} &	\textbf{QDistEvol} &	\textbf{BP-OSD} &	\textbf{GEStim} &	\textbf{CCStim} &	\textbf{UEStim}	\\	\hline
\textbf{Result Returned} &	5 &	5 &	5 &	5 &	5 &	5 &	5	\\	
\textbf{No Result} &	0 &	0 &	0 &	0 &	0 &	0 &	0	\\	
\textbf{At lowest distance} &	5 &	5 &	5 &	5 &	5 &	5 &	5	\\	
\textbf{Overall Success Rate} &	100.0\% &	100.0\% &	100.0\% &	100.0\% &	100.0\% &	100.0\% &	100.0\%	\\	\hline
\textbf{Total Time (s)} &	3.2E+01 &	1.3E+03 &	1.3E+03 &	3.7E+03 &	1.9E+00 &	1.8E+00 &	1.9E+00	\\	
\textbf{Trials} &	NA &	50,000 &	50,000 &	50,000 &	5 &	5 &	10	\\	
\textbf{Trials at lowest distance} &	 &	35,411 &	45,627 &	50,000 &	5 &	5 &	10	\\	
\textbf{Trial Success Rate} &	 &	70.8\% &	91.3\% &	100.0\% &	100.0\% &	100.0\% &	100.0\%	\\	
\textbf{Time/Trial} &	 &	2.5E-02 &	2.7E-02 &	7.4E-02 &	3.8E-01 &	3.7E-01 &	1.9E-01	\\	
\textbf{Time/Successful Trial} &	 &	3.6E-02 &	2.9E-02 &	7.4E-02 &	3.8E-01 &	3.7E-01 &	1.9E-01	\\	\hline
\end{tabular}									
\end{center}																	           
}
\caption{Distance finding benchmark surface code circuits - Heuristic Methods.}
\label{tab:qc_SC_heuristic}
\end{table}

\subsubsection{Colour Code Circuits}\label{sec:QC_colour_codes}
We then calculated distances for the superdense and midout colour code circuits of \cite{colour_code_circuits}.
The exact  Gurobi, CLISAT and m4riCC methods completed all members of the family within the 8 hour time frame for this circuit family (see \Cref{tab:qc_CC_midout_exact}) with m4riCC giving the fastest results.
\begin{table}[h!]
\setlength\tabcolsep{2pt}
\fontfamily{lmss}\fontsize{8}{9}\selectfont{
\begin{center}							
\begin{tabular}{ |l|	r|	r|	r|	r|	r|	}	\hline
 &	\textbf{Gurobi} &	\textbf{MIP-SCIP} &	\textbf{CLISAT} &	\textbf{m4riCC} &	\textbf{Magma}	\\	\hline
\textbf{Result Returned} &	5 &	5 &	5 &	5 &	5	\\	
\textbf{Completed < MaxTime} &	5 &	4 &	5 &	5 &	4	\\	
\textbf{Completed > MaxTime} &	0 &	1 &	0 &	0 &	1	\\	
\textbf{No Result} &	0 &	0 &	0 &	0 &	0	\\	
\textbf{At lowest distance} &	5 &	5 &	5 &	5 &	5	\\	
\textbf{Overall success rate} &	100.0\% &	100.0\% &	100.0\% &	100.0\% &	100.0\%	\\	\hline
\textbf{Total Time} &	1.9E+03 &	2.9E+04 &	2.1E+04 &	2.7E+00 &	4.5E+04	\\	
\textbf{Time/Trial} &	3.9E+02 &	5.8E+03 &	4.3E+03 &	5.3E-01 &	9.0E+03	\\	
\textbf{Time/Successful Trial} &	3.9E+02 &	5.8E+03 &	4.3E+03 &	5.3E-01 &	9.0E+03	\\	\hline
\end{tabular}							
\end{center}																						
}
\caption{Distance finding benchmark colour code midout circuits - Exact Methods.}
\label{tab:qc_CC_midout_exact}
\end{table}

For heuristic methods, we see in \Cref{tab:qc_CC_midout_heuristic}  that all methods found the same lowest distances.
QDistRnd and QDistEvol had the lowest time per successful trial for this data set.
Surprisingly, the Stim Graphlike Error algorithm also performed extremely well - despite the fact that it did not perform well in calculating the code capacity distance of colour codes (see \Cref{sec:hyperbolic_CSS}).

\begin{figure}[h!]
\centering
\begin{subfigure}[t]{.33\textwidth}
  \centering
  \includegraphics[width=\linewidth]{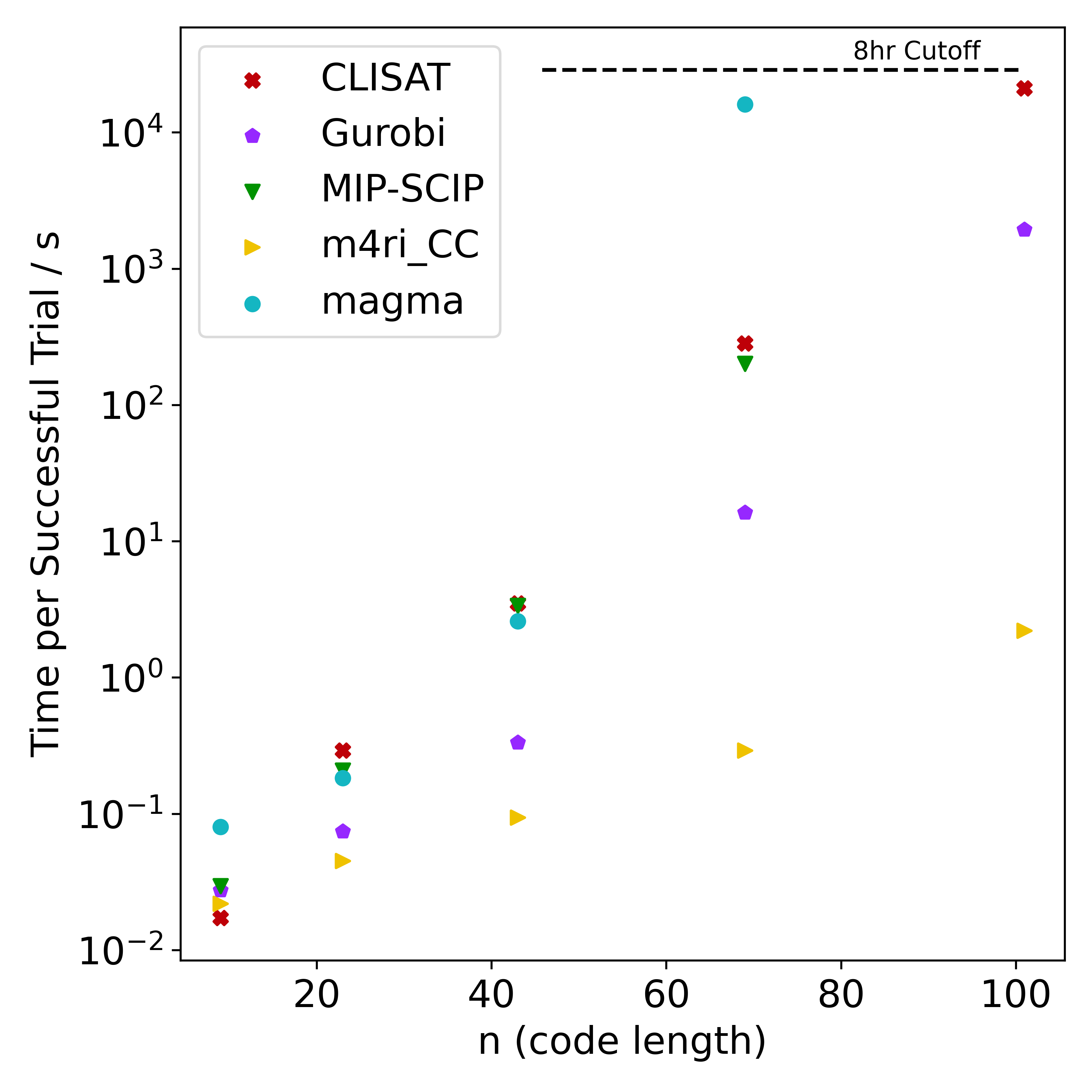}
  \subcaption{Time per Successful Trial \\ (Exact Algorithms)}\label{fig:09_qc_CC_midout_exact_ttd_time_exact}
    \end{subfigure}%
\begin{subfigure}[t]{.33\textwidth}
  \centering
  \includegraphics[width=\linewidth]{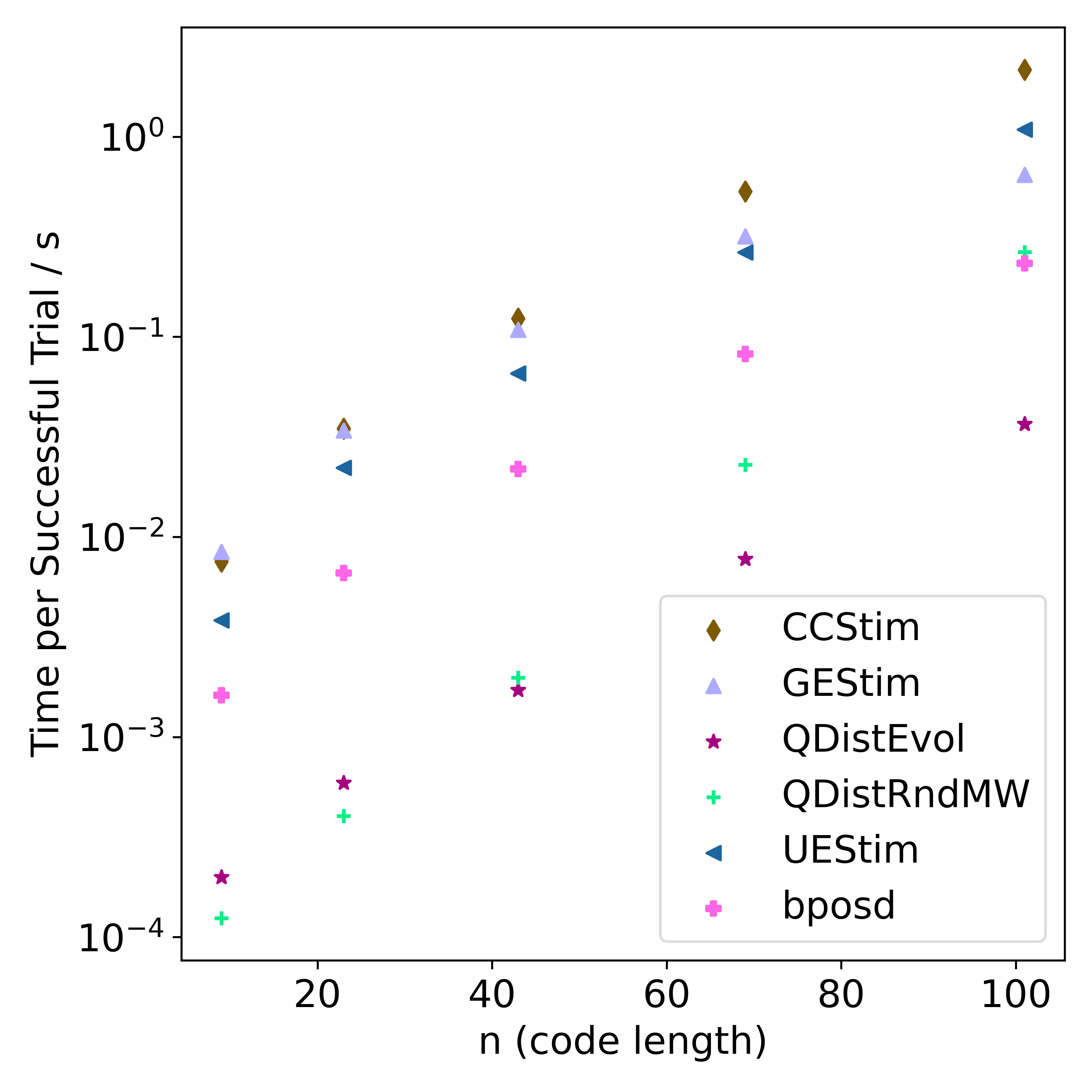}
  \subcaption{Time per Successful Trial \\ (Heuristic Algorithms)}\label{fig:09_qc_CC_midout_heuristic_ttd}
    \end{subfigure}%
\begin{subfigure}[t]{.33\textwidth}
  \centering
  \includegraphics[width=\linewidth]{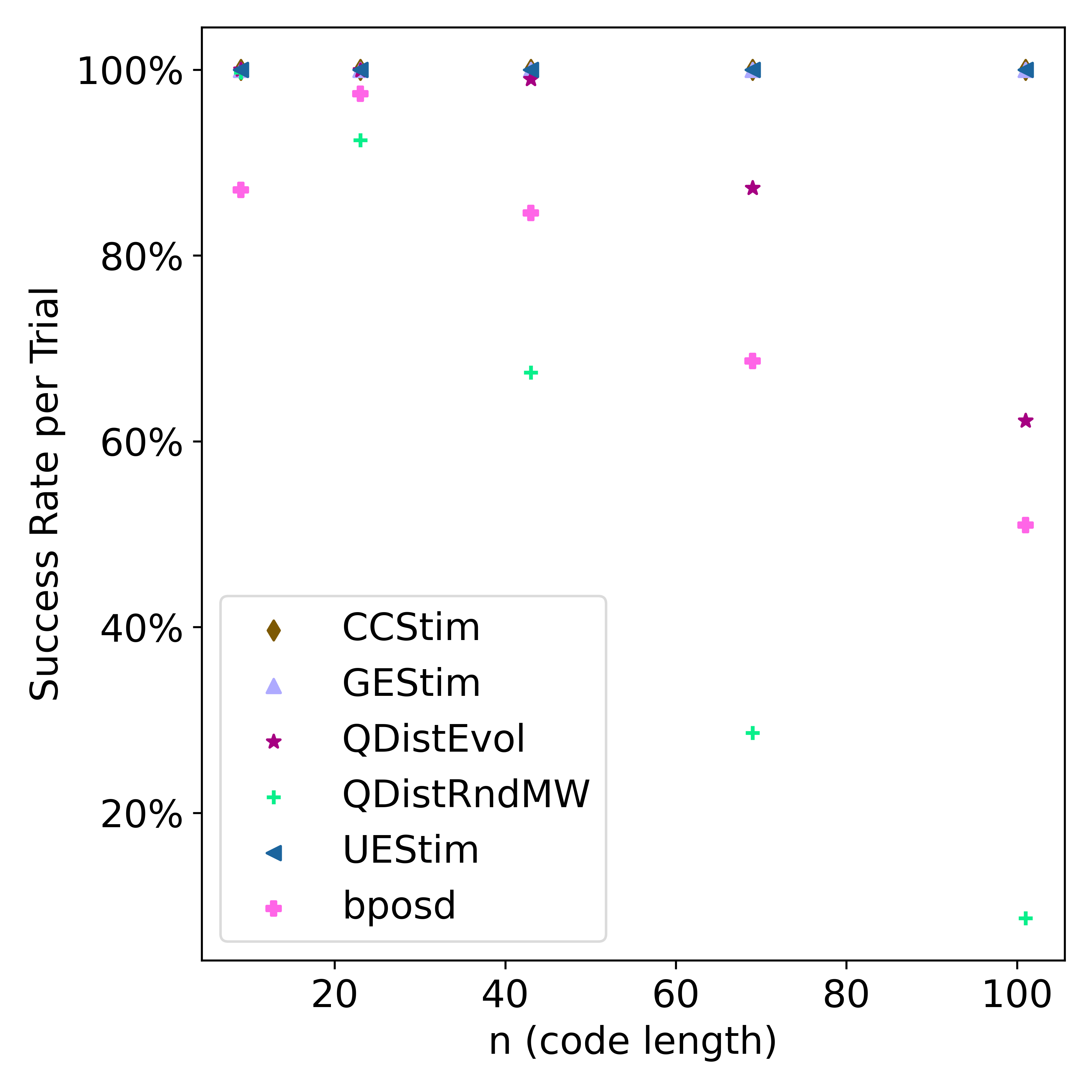}
  \subcaption{Success Rate per Trial \\ (Heuristic Algorithms)}\label{fig:09_qc_CC_midout_heuristic_success_rate}
    \end{subfigure}%
\caption{Benchmark data by code length - colour code midout syndrome circuits}
\label{fig:09_qc_CC_midout}
\end{figure}

In \Cref{fig:09_qc_CC_midout_exact_ttd_time_exact}, we see that m4riCC had the lowest time per successful trial for the exact algorithms across all circuits in the family. 
In \Cref{fig:09_qc_CC_midout_heuristic_success_rate} we see that that the success rate for  the QDistRnd, QDistEvol and BP-OSD algorithms fell sharply as the size of the circuit increased. In \Cref{fig:08_qc_SC_heuristic_ttd} we see that QDistRnd has the best time per successful trial for small members of this family but that QDistEvol had better time per successful trial for codes on 60 qubits or more.

\begin{table}[h!]
\setlength\tabcolsep{2pt}
\fontfamily{lmss}\fontsize{8}{9}\selectfont{				      
\begin{center}									
\begin{tabular}{ |l|	r|	r|	r|	r|	r|	r|	r|	}	\hline
 &	\textbf{m4riRW} &	\textbf{QDistRndMW} &	\textbf{QDistEvol} &	\textbf{BP-OSD} &	\textbf{GEStim} &	\textbf{CCStim} &	\textbf{UEStim}	\\	\hline
\textbf{Result Returned} &	5 &	5 &	5 &	5 &	5 &	5 &	5	\\	
\textbf{No Result} &	0 &	0 &	0 &	0 &	0 &	0 &	0	\\	
\textbf{At lowest distance} &	5 &	5 &	5 &	5 &	5 &	5 &	5	\\	
\textbf{Overall Success Rate} &	100.0\% &	100.0\% &	100.0\% &	100.0\% &	100.0\% &	100.0\% &	100.0\%	\\	\hline
\textbf{Total Time (s)} &	1.4E+01 &	3.1E+02 &	3.2E+02 &	2.0E+03 &	1.1E+00 &	2.9E+00 &	2.9E+00	\\	
\textbf{Trials} &	NA &	50,000 &	50,000 &	50,000 &	5 &	5 &	10	\\	
\textbf{Trials at lowest distance} &	 &	29,680 &	44,837 &	38,871 &	5 &	5 &	10	\\	
\textbf{Trial Success Rate} &	 &	59.4\% &	89.7\% &	77.7\% &	100.0\% &	100.0\% &	100.0\%	\\	
\textbf{Time/Trial} &	 &	6.3E-03 &	6.4E-03 &	4.0E-02 &	2.2E-01 &	5.7E-01 &	2.9E-01	\\	
\textbf{Time/Successful Trial} &	 &	1.1E-02 &	7.2E-03 &	5.2E-02 &	2.2E-01 &	5.7E-01 &	2.9E-01	\\	\hline
\end{tabular}									
\end{center}									
}
\caption{Distance finding benchmark colour code midout circuits - Heuristic Methods.}
\label{tab:qc_CC_midout_heuristic}
\end{table}

\begin{figure}[h!]
\centering
\begin{subfigure}[t]{.33\textwidth}
  \centering
  \includegraphics[width=\linewidth]{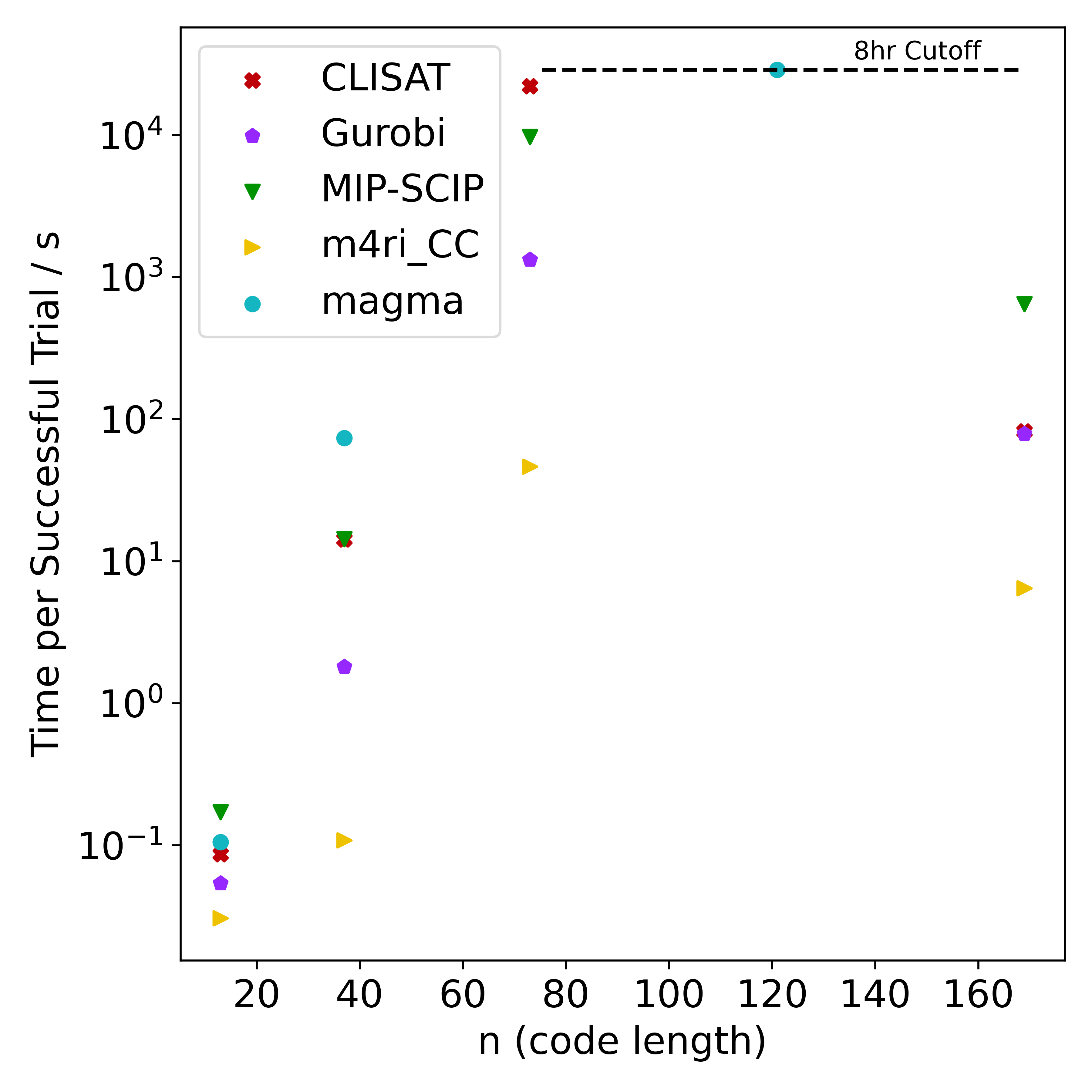}
  \subcaption{Time per Successful Trial \\ (Exact Algorithms)}\label{fig:10_qc_CC_superdense_exact_ttd_time_exact}
    \end{subfigure}%
\begin{subfigure}[t]{.33\textwidth}
  \centering
  \includegraphics[width=\linewidth]{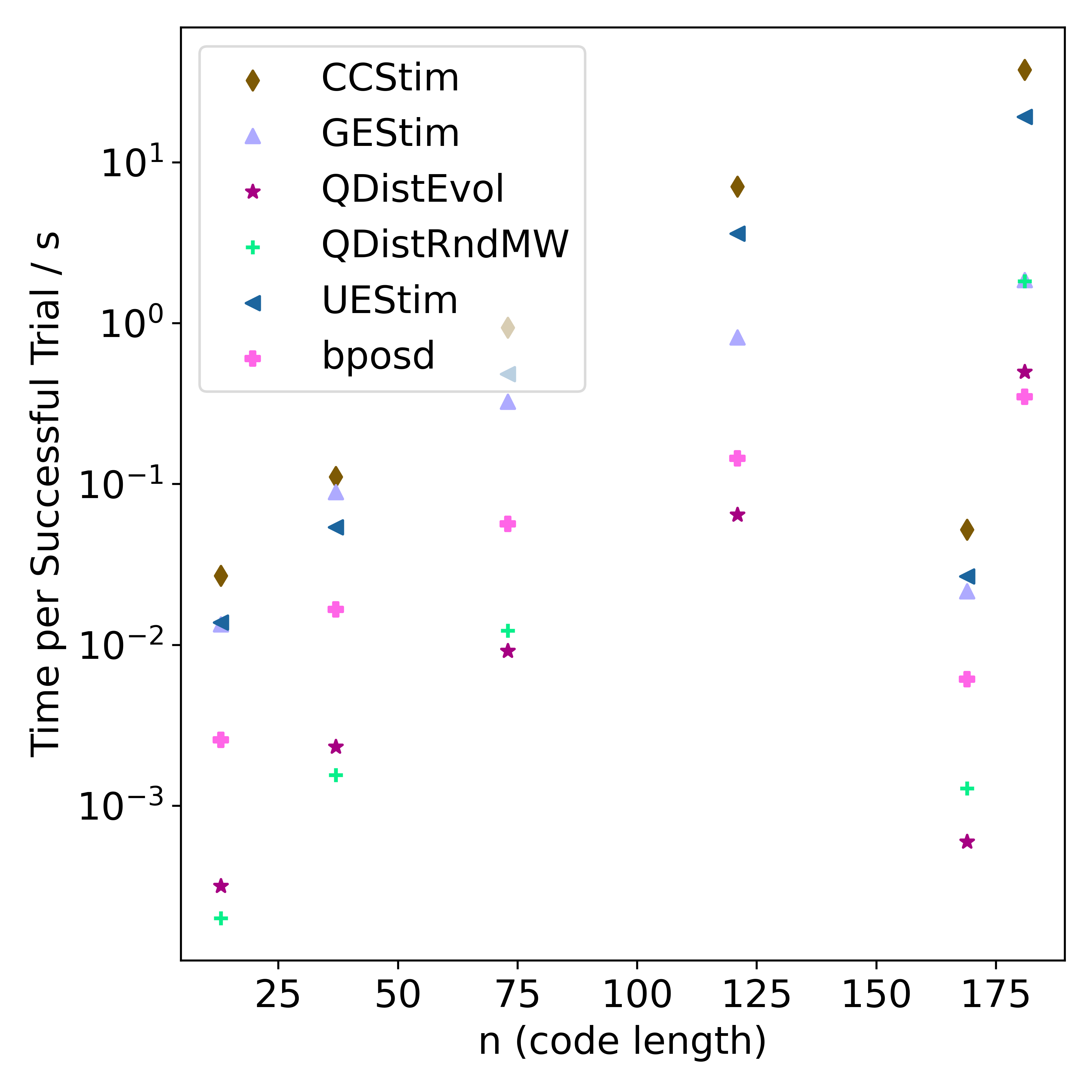}
  \subcaption{Time per Successful Trial \\ (Heuristic Algorithms)}\label{fig:10_qc_CC_superdense_heuristic_ttd}
    \end{subfigure}%
\begin{subfigure}[t]{.33\textwidth}
  \centering
  \includegraphics[width=\linewidth]{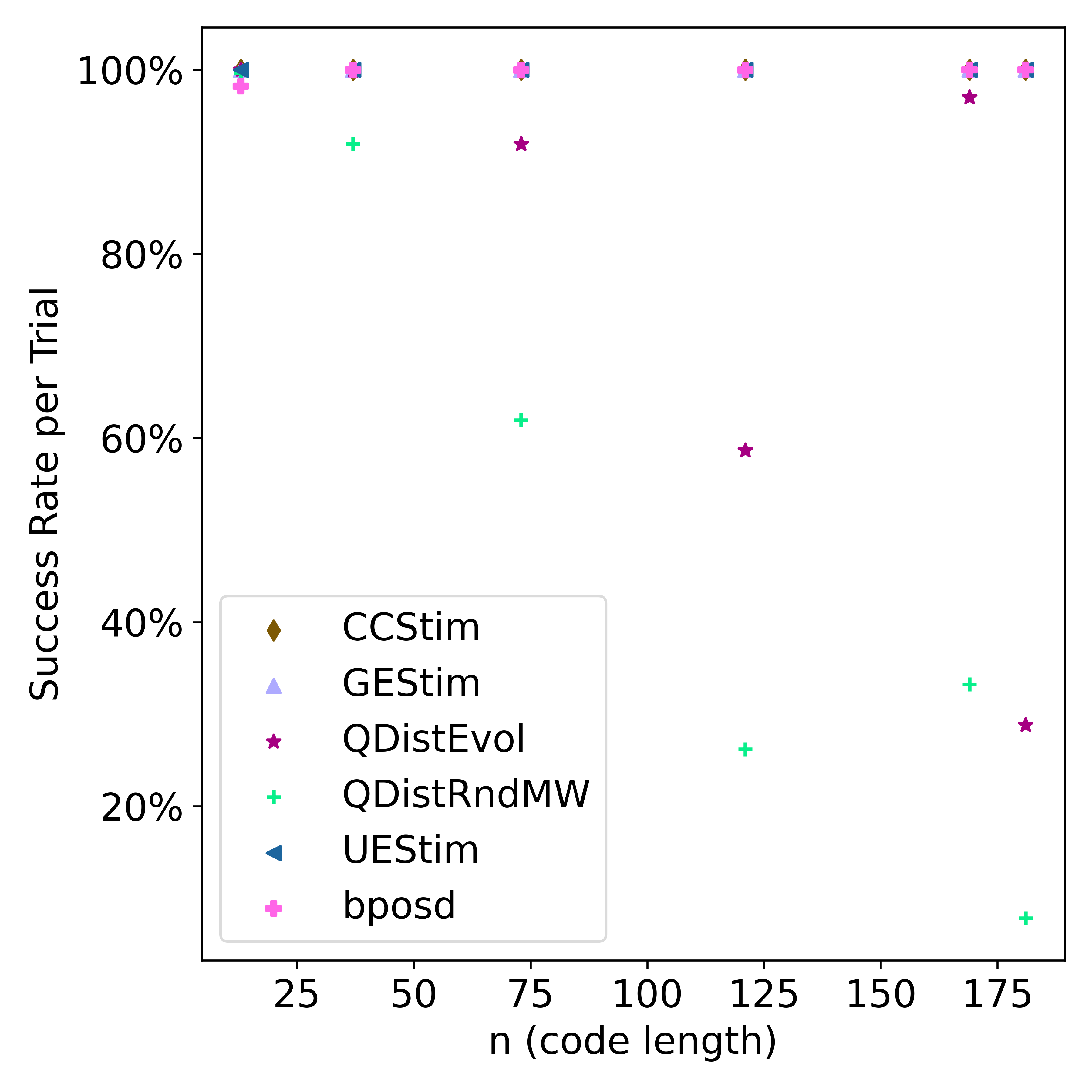}
  \subcaption{Success Rate per Trial \\ (Heuristic Algorithms)}\label{fig:10_qc_CC_superdense_heuristic_success_rate}
    \end{subfigure}%
\caption{Benchmark data by code length - colour code superdense syndrome circuits}
\label{fig:10_qc_CC_superdense}
\end{figure}

We see a similar pattern for superdense colour code circuits (\Cref{tab:qc_CC_superdense_exact} \Cref{tab:qc_CC_superdense_heuristic}, \Cref{fig:10_qc_CC_superdense}).
Notably, the success rate for the BP-OSD algorithm remained close to 100\% even for large circuits.
        
\begin{table}[h!]
\setlength\tabcolsep{2pt}
\fontfamily{lmss}\fontsize{8}{9}\selectfont{
\begin{center}							
\begin{tabular}{ |l|	r|	r|	r|	r|	r|	}	\hline
 &	\textbf{Gurobi} &	\textbf{MIP-SCIP} &	\textbf{CLISAT} &	\textbf{m4riCC} &	\textbf{Magma}	\\	\hline
\textbf{Result Returned} &	6 &	6 &	6 &	4 &	5	\\	
\textbf{Completed < MaxTime} &	4 &	4 &	4 &	4 &	3	\\	
\textbf{Completed > MaxTime} &	2 &	2 &	2 &	0 &	2	\\	
\textbf{No Result} &	0 &	0 &	0 &	2 &	1	\\	
\textbf{At lowest distance} &	6 &	6 &	4 &	4 &	5	\\	
\textbf{Overall success rate} &	100.0\% &	100.0\% &	66.7\% &	66.7\% &	83.3\%	\\	\hline
\textbf{Total Time} &	5.9E+04 &	6.8E+04 &	8.0E+04 &	5.8E+04 &	8.7E+04	\\	
\textbf{Time/Trial} &	9.9E+03 &	1.1E+04 &	1.3E+04 &	9.6E+03 &	1.4E+04	\\	
\textbf{Time/Successful Trial} &	9.9E+03 &	1.1E+04 &	2.0E+04 &	1.4E+04 &	1.7E+04	\\	\hline
\end{tabular}							
\end{center}														
}
\caption{Distance finding benchmark superdense colour code circuits - Exact Methods.}
\label{tab:qc_CC_superdense_exact}
\end{table}

\              
\begin{table}[h!]
\setlength\tabcolsep{2pt}
\fontfamily{lmss}\fontsize{8}{9}\selectfont{
\begin{center}									
\begin{tabular}{ |l|	r|	r|	r|	r|	r|	r|	r|	}	\hline
 &	\textbf{m4riRW} &	\textbf{QDistRndMW} &	\textbf{QDistEvol} &	\textbf{BP-OSD} &	\textbf{GEStim} &	\textbf{CCStim} &	\textbf{UEStim}	\\	\hline
\textbf{Result Returned} &	6 &	6 &	6 &	6 &	6 &	6 &	6	\\	
\textbf{No Result} &	0 &	0 &	0 &	0 &	0 &	0 &	0	\\	
\textbf{At lowest distance} &	6 &	6 &	6 &	6 &	6 &	6 &	6	\\	
\textbf{Overall Success Rate} &	100.0\% &	100.0\% &	100.0\% &	100.0\% &	100.0\% &	100.0\% &	100.0\%	\\	\hline
\textbf{Total Time (s)} &	3.9E+01 &	1.9E+03 &	1.9E+03 &	5.7E+03 &	3.1E+00 &	4.6E+01 &	4.7E+01	\\	
\textbf{Trials} &	NA &	60,000 &	60,000 &	60,000 &	6 &	6 &	12	\\	
\textbf{Trials at lowest distance} &	 &	32,077 &	47,630 &	59,809 &	6 &	6 &	12	\\	
\textbf{Trial Success Rate} &	 &	53.5\% &	79.4\% &	99.7\% &	100.0\% &	100.0\% &	100.0\%	\\	
\textbf{Time/Trial} &	 &	3.2E-02 &	3.2E-02 &	9.6E-02 &	5.2E-01 &	7.6E+00 &	3.9E+00	\\	
\textbf{Time/Successful Trial} &	 &	5.9E-02 &	4.0E-02 &	9.6E-02 &	5.2E-01 &	7.6E+00 &	3.9E+00	\\	\hline
\end{tabular}									
\end{center}															
}
\caption{Distance finding benchmark superdense colour code circuits - Heuristic Methods.}
\label{tab:qc_CC_superdense_heuristic}
\end{table}

\subsubsection{Bivariate Bicycle Code Circuits}\label{sec:QC_bivariate_bicycle}
Finally we applied the algorithms to syndrome extraction circuits for the bivariate bicycle codes of \cite{BB_IBM} and these proved to be the most challenging circuits for distance-finding.
We see in \Cref{tab:qc_bivariate_bicycle_exact} that only a few of the exact algorithms yielded a distance estimate within the 8 hour time frame. 
Gurobi performed very well giving the lowest distance of all algorithms in 4 of the 7 members of the circuit family, though it completed within 8 hours only for the first member of the family.

In  \Cref{tab:qc_bivariate_bicycle_heuristic}, we see that m4riRW was the only heuristic algorithms able to complete 10,000 trials for all seven circuits in the family.
This algorithm gave accurate distances for the first four members of the family, but gave distances larger than the code capacity distance for the remaining three. 
The BP-OSD algorithm only completed three of the circuits in the family within the  48 hour maximum server runtime whilst QDistRndMW and QDistEvol only completed four each.
Unlike the other circuit families, the Stim Undetectable Error family of algorithms was not effective on this dataset, with the UEStim algorithm only giving a result for the smallest of the circuits.

\begin{figure}[h!]
\centering
\begin{subfigure}[t]{.33\textwidth}
  \centering
  \includegraphics[width=\linewidth]{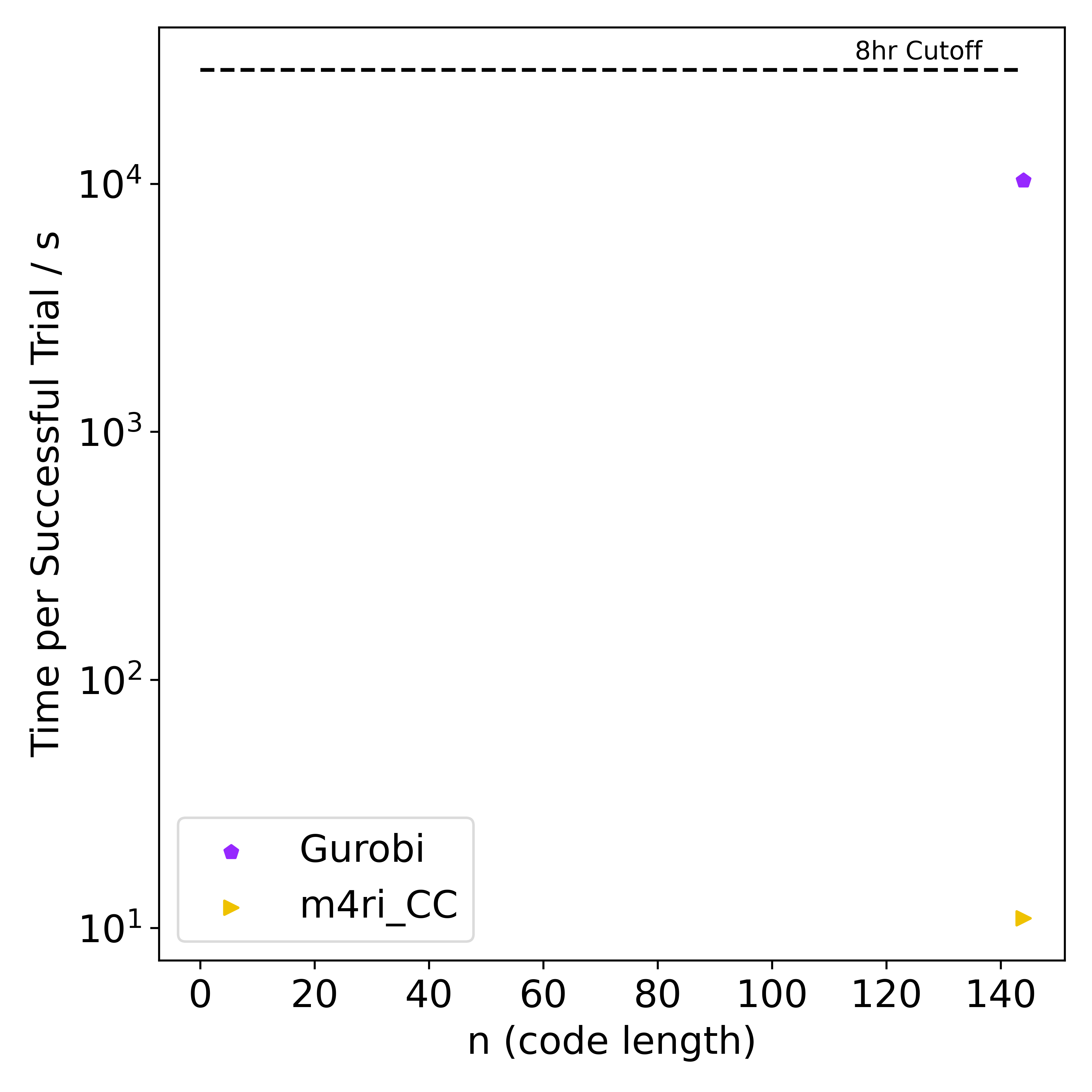}
  \subcaption{Time per Successful Trial \\ (Exact Algorithms)}\label{fig:11_qc_BB_exact_ttd_time_exact}
    \end{subfigure}%
\begin{subfigure}[t]{.33\textwidth}
  \centering
  \includegraphics[width=\linewidth]{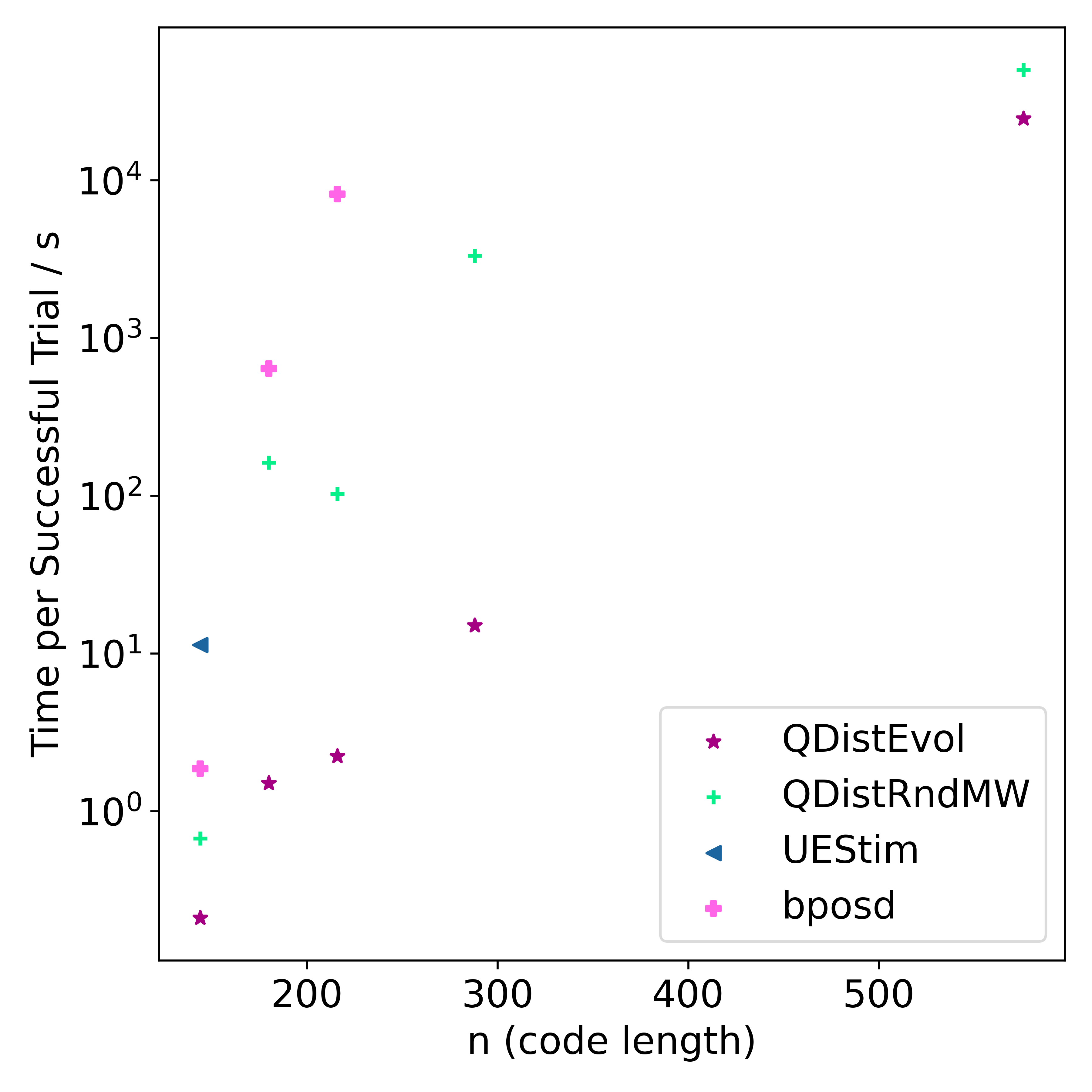}
  \subcaption{Time per Successful Trial \\ (Heuristic Algorithms)}\label{fig:11_qc_BB_heuristic_ttd}
    \end{subfigure}%
\begin{subfigure}[t]{.33\textwidth}
  \centering
  \includegraphics[width=\linewidth]{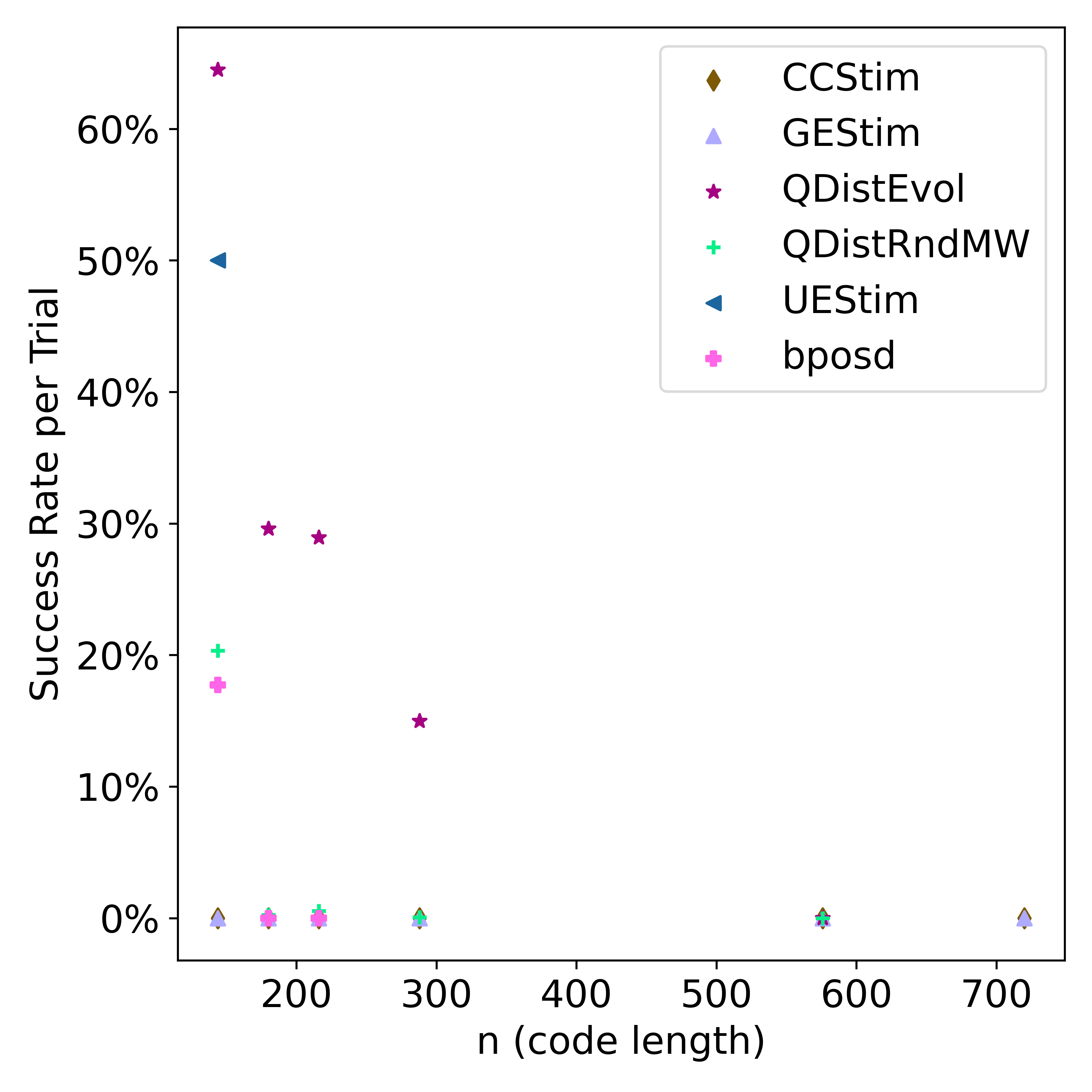}
  \subcaption{Success Rate per Trial \\ (Heuristic Algorithms)}\label{fig:11_qc_BB_heuristic_success_rate}
    \end{subfigure}%
\caption{Benchmark data by code length - bivariate bicycle code syndrome circuits}
\label{fig:11_qc_BB}
\end{figure}

Exact algorithms were only able to give distance estimates for a few members of this data set.
It is also difficult to draw any conclusions for  heuristic algorithms but it does seem that QDistEvol has the lowest time per successful trial for this family (see \Cref{fig:11_qc_BB}).

\begin{table}[h!]
\setlength\tabcolsep{2pt}
\fontfamily{lmss}\fontsize{8}{9}\selectfont{
\begin{center}							
\begin{tabular}{ |l|	r|	r|	r|	r|	r|	}	\hline
 &	\textbf{Gurobi} &	\textbf{MIP-SCIP} &	\textbf{CLISAT} &	\textbf{m4riCC} &	\textbf{Magma}	\\	\hline
\textbf{Result Returned} &	5 &	3 &	1 &	1 &	1	\\	
\textbf{Completed < MaxTime} &	1 &	0 &	0 &	1 &	0	\\	
\textbf{Completed > MaxTime} &	4 &	3 &	1 &	0 &	1	\\	
\textbf{No Result} &	2 &	4 &	6 &	6 &	6	\\	
\textbf{At lowest distance} &	4 &	2 &	1 &	1 &	1	\\	
\textbf{Overall success rate} &	57.1\% &	28.6\% &	14.3\% &	14.3\% &	14.3\%	\\	\hline
\textbf{Total Time} &	1.8E+05 &	2.1E+05 &	2.0E+05 &	1.7E+05 &	2.0E+05	\\	
\textbf{Time/Trial} &	2.6E+04 &	3.0E+04 &	2.9E+04 &	2.5E+04 &	2.9E+04	\\	
\textbf{Time/Successful Trial} &	4.6E+04 &	1.0E+05 &	2.0E+05 &	1.7E+05 &	2.0E+05	\\	\hline
\end{tabular}							
\end{center}																		 
}
\caption{Distance finding benchmark bivariate bicycle code circuits - Exact Methods.}
\label{tab:qc_bivariate_bicycle_exact}
\end{table}

\begin{table}[h!]
\setlength\tabcolsep{2pt}
\fontfamily{lmss}\fontsize{8}{9}\selectfont{
\begin{center}									
\begin{tabular}{ |l|	r|	r|	r|	r|	r|	r|	r|	}	\hline
 &	\textbf{m4riRW} &	\textbf{QDistRndMW} &	\textbf{QDistEvol} &	\textbf{BP-OSD} &	\textbf{GEStim} &	\textbf{CCStim} &	\textbf{UEStim}	\\	\hline
\textbf{Result Returned} &	7 &	5 &	5 &	3 &	0 &	0 &	1	\\	
\textbf{No Result} &	0 &	2 &	2 &	4 &	7 &	7 &	6	\\	
\textbf{At lowest distance} &	4 &	4 &	4 &	1 &	 &	 &	1	\\	
\textbf{Overall Success Rate} &	57.1\% &	57.1\% &	57.1\% &	14.3\% &	 &	 &	14.3\%	\\	\hline
\textbf{Total Time (s)} &	6.4E+04 &	8.1E+04 &	5.9E+04 &	6.4E+04 &	1.2E+02 &	2.0E+02 &	2.3E+01	\\	
\textbf{Trials} &	NA &	41,000 &	41,000 &	30,000 &	6 &	6 &	4	\\	
\textbf{Trials at lowest distance} &	 &	2,119 &	13,798 &	1,773 &	 &	 &	2	\\	
\textbf{Trial Success Rate} &	 &	5.2\% &	33.7\% &	5.9\% &	 &	 &	50.0\%	\\	
\textbf{Time/Trial} &	 &	2.0E+00 &	1.4E+00 &	2.1E+00 &	2.0E+01 &	3.4E+01 &	5.7E+00	\\	
\textbf{Time/Successful Trial} &	 &	3.8E+01 &	4.3E+00 &	3.6E+01 &	 &	 &	1.1E+01	\\	\hline
\end{tabular}									
\end{center}									
}
\caption{Distance finding benchmark benchmark bivariate bicycle code circuits - Heuristic Methods.}
\label{tab:qc_bivariate_bicycle_heuristic}
\end{table}

\section{Conclusions and Future Work}\label{sec:conclusions}
In this section, we draw conclusions from our benchmarking results and make recommendations about which algorithms give the best performance for each data set.
We first assess the exact distance-finding methods.
We found that Magma performed very well for small classical and quantum codes. 
When the algorithm was terminated early due to the 8 hour maximum run time, the distance bounds for these code families were very accurate.
We were also able to use Magma for large code families which in general did not complete within the 8 hour limit. 
The accuracy of distance bounds was good for the hyperbolic surface and colour code families, but significantly less accurate for the larger lifted product and bivariate bicycle codes. 
Magma unexpectedly terminated when attempting to process the larger members of the circuit data sets.

The mixed-integer programming (MIP) methods (Gurobi and MIP-SCIP) gave good all-round performance on a wide range of codes. 
Gurobi generally had lower processing time than MIP-SCIP but accuracy after early termination was sometimes higher for MIP-SCIP than Gurobi. 
We found that the MIP methods generally outperformed the SAT solver method both in terms of processing time and accuracy.

For LDPC codes with low degree Tanner graphs, we found that the connected cluster algorithm m4riCC performed very well and this also extended to some of the quantum circuit families. 
We summarise the recommended exact algorithms by data set in \Cref{tab:recommendations_exact2}.

Now turning to heuristic distance-finding algorithms, we found that for general code families the random information set algorithms m4riRW and QDistRndMW were quite effective both in terms of accuracy and run-time.
The QDistEvol algorithm gave significantly more accurate results on the high-rate LDPC codes in our benchmark. The QDistEvol algorithm may also benefit from further refinement and a C implementation using asymptotically optimal Gaussian elimination as in the dist-m4ri package. 

In contrast, the BP-OSD syndrome decoder method did not clearly outperform the random information set algorithms for any data sets in our benchmarking study. 
On the other hand, for the larger members of the hyperbolic surface and colour code data sets the BP-OSD method did have better time per successful trial metrics than the random information set algorithms - a faster implementation of BP-OSD may make it more competitive versus random information set algorithms.

The Stim Undetectable Error family of algorithms performed well for codes and circuits with low degree Tanner graphs and in this respect are quite complementary to the m4riCC algorithm.
We summarise the  heuristic algorithms recommended for each data set in \Cref{tab:recommendations_heuristic2}.

In conclusion, there is currently significant interest in exploring families of quantum LDPC codes for use in fault-tolerant quantum computing architectures. 
The distance of these new code families is critical to their performance, but can be difficult to calculate due to their size and structure. 
Determining the distance of the syndrome extraction circuits associated with these codes is even more challenging but is also crucial.
We hope that the insights of this work make choosing appropriate distance-finding methods more straightforward and so assist with the exploration of code families.

\begin{table}[h!]
\setlength\tabcolsep{2pt}
\fontfamily{lmss}\fontsize{8}{9}\selectfont{
\begin{center}			
\begin{tabular}{ |l|	l|	l|}	\hline
&	\textbf{Recommendation}&	\textbf{Also Effective}\\	\hline
\multicolumn{3}{|l|}{\textbf{Classical Codes}}\\			\hline
\textbf{CodeTables GF(2)}&	Magma &	Gurobi, MIP-SCIP\\	
\textbf{Lifted Product GF(2)}&	Magma&	MIP-SCIP, Gurobi\\	\hline
\multicolumn{3}{|l|}{\textbf{Quantum Codes}}\\			\hline
\textbf{CodeTables Non-CSS}&	Magma&	Gurobi, MIP-SCIP\\	
\textbf{Hyperbolic Surface CSS}&	m4riCC&	Gurobi, MIP-SCIP, SAT\\	
\textbf{Hyperbolic Colour CSS}&	m4riCC&	Gurobi, MIP-SCIP, SAT\\	
\textbf{Lifted Product CSS}&	m4riCC&	Gurobi\\	
\textbf{Bivariate Bicycle CSS}&	m4riCC,Gurobi,MIP-SCIP&	-\\	\hline
\textbf{Quantum Tanner CSS}&	Gurobi&	m4riCC, Magma, MIP-SCIP, SAT\\	\hline
\multicolumn{3}{|l|}{\textbf{Quantum Syndrome Extraction Circuits}}\\			\hline
\textbf{Surface Code}&	m4riCC&	Gurobi, MIP-SCIP\\	
\textbf{Colour Code - Midout}&	m4riCC&	Gurobi, MIP-SCIP, SAT\\	
\textbf{Colour Code - Superdense}&	m4riCC&	Gurobi, MIP-SCIP, SAT\\	
\textbf{Bivariate Bicycle}&	Gurobi&	-\\	\hline
\end{tabular}			
\end{center}			
}
\caption{Recommended Distance-Finding Algorithms by Data Set - Exact Methods}
\label{tab:recommendations_exact2}
\end{table}

\begin{table}[h!]
\setlength\tabcolsep{2pt}
\fontfamily{lmss}\fontsize{8}{9}\selectfont{
\begin{center}			
\begin{tabular}{ |l|	l|	l|}	\hline
&	\textbf{Recommendation}&	\textbf{Also Effective}\\	\hline
\multicolumn{3}{|l|}{\textbf{Classical Codes}}\\			\hline
\textbf{CodeTables GF(2)}&	QDistEvol&	QDistRnd\\	
\textbf{Lifted Product GF(2)}&	QDistEvol&	QDistRnd, BP-OSD\\	\hline
\multicolumn{3}{|l|}{\textbf{Quantum Codes}}\\			\hline
\textbf{CodeTables Non-CSS}&	QDistRnd&	QDistEvol\\	
\textbf{Hyperbolic Surface CSS}&	QDistEvol,BP-OSD&	Stim GE, Stim UE, Stim CC\\	
\textbf{Hyperbolic Colour CSS}&	QDistEvol,BP-OSD&	Stim UE, Stim CC, BP-OSD\\	
\textbf{Lifted Product CSS}&	QDistEvol&	QDistRnd\\	
\textbf{Bivariate Bicycle CSS}&	QDistEvol&	-\\	\hline
\textbf{Quantum Tanner CSS}&	QDistEvol&	-\\	\hline
\multicolumn{3}{|l|}{\textbf{Quantum Syndrome Extraction Circuits}}\\			\hline
\textbf{Surface Code}&	QDistEvol, QDistRnd, BP-OSD&	Stim GE\\	
\textbf{Colour Code - Midout}&	QDistEvol, QDistRnd, BP-OSD&	Stim GE\\	
\textbf{Colour Code - Superdense}&	QDistEvol, QDistRnd, BP-OSD&	Stim GE\\	
\textbf{Bivariate Bicycle}&	QDistEvol&	-\\	\hline
\end{tabular}			
\end{center}			
}
\caption{Recommended Distance-Finding Algorithms by Data Set - Heuristic Methods}
\label{tab:recommendations_heuristic2}
\end{table}

\section{Acknowledgements}
MW is supported by the Engineering and Physical Sciences Research Council [grant number EP/W032635/1 and EP/S005021/1] and  Innovate UK [grant number 10179725]. AJ is supported by the Engineering and Physical Sciences Research Council [grant number EP/S021582/1].
MW thanks Stergios Koutsioumpas, Michael Vasmer, Anthony Leverrier and George Umbrarescu for helpful discussions.
OH thanks Nikolas Breuckmann for explaining his variant of the MILP formulation for distance finding, used in \Cref{sec:MIP}.
The authors thank Dan Browne, Noah Shutty and Adam Zalcman for helpful feedback on drafts of this paper.
This work used computing equipment funded by the Research Capital Investment Fund (RCIF) provided by UKRI, and partially funded by the UCL Cosmoparticle Initiative.
\medskip
\bibliography{references_new}

\appendix

\section{SAT Solver Model Construction}\label{sec:SAT_solver_model}
In this section, we present an algorithm for constructing a WNCF file which which can be used for distance finding SAT solvers.
SAT solvers solve for constraints on a series of binary variables. 
For distance finding, we supply a file in weighted canonical normal form (WCNF) to the SAT solver. 
Here, we briefly explain the format of a WCNF file.

Sample text from a WCNF file is below:
\begin{align*}
&\texttt{wcnf 10 54 55}\\
    &\texttt{1 5 0}\\
&\texttt{55 3 -4 5 0}
\end{align*}
The first line gives details of the number of variables (10) and the number of clauses (54). The number 55 is the weight we give to hard constraints. 
Subsequent lines are constraint clauses and the first entry is the weight of the clause.
In the second line, the weight 1 indicates that this is a soft constraint and that we should minimise variable $v_5$. The last entry of a constraint is always zero.
In the third line, the weight 55 indicates this is a hard constraint. The text \texttt{3 -4 5} corresponds to the condition $v_3 \lor \lnot v_4 \lor v_5$.
To specify the entire set of constraints, the conditions in each line are combined using $\land$. 
The SAT solver searches for solutions which satisfy all hard constraints, and prioritises those which satisfy as many soft constraints as possible.

The process for constructing a WCNF file for distance finding for a given detector error model is given in \Cref{alg:SAT}.

\begin{algorithm}[H]
\caption{Distance via SAT Solver}\label{alg:SAT}
\begin{algorithmic}
\State \textbf{Inputs:}
\State $H_X$: $r\times n$ X-check matrix for a CSS code
\State $L_X$: $k \times n$ X-logical basis
\State \textbf{Outputs:}
\State A WCNF file for the SAT Z-distance finding model for the code
\State \textbf{Method:}
\State HLT := transpose(vstack([$H_X,L_X$]))
\Comment{HLParities - parity variables for each row of HL}
\State HLParities := [0] * (r+k+1)
\Comment{v tracks the number of variables - initially number of errors}
\State v := n
\Comment{SAT clauses - stored as list of lists}
\State clauses := list()
\For{e in [1..n]}
\Comment{Add a soft clause for each error mechanism to minimise weight}
    \State clauses.append([1,e])
    \For{d in support(HLT[e-1])}
        \State u := HLParities[d]
        \If{u = 0}
            \Comment{e is the leftmost nonzero entry - set parity variable to e}
            \State HLParities[d] := e
        \Else
            \Comment{Add a new variable and update row parity constr}
            \State v := v+1
            \State \Call{RowConstr}{e,u,v,clauses}
            \State HLParities[d] := v
        \EndIf
    \EndFor
\EndFor
\Comment{Hard constraint - $H_X$ parities}
\For{d in [1..r]}
    \State clauses.append([0,-HLParities[d]]
\EndFor
\Comment{Hard constraint - anticommute with at least one row of $L_X$}
\State logicalClause := [0]
\For{d in [r+1..r+k]}
    \State logicalClause.append(HLParities[d])
\EndFor
\State clauses.append(logicalClause)
\Comment{convert clauses to WCNF format}
\State return \Call{clauses2WCNF}{clauses,v}
\end{algorithmic}
\end{algorithm}

\begin{algorithm}
\begin{algorithmic}
\Function {RowConstr}{e,u,v,clauses}
\Comment{Construct Parity Constraint for SAT Solver Model}
    \State clauses.append([0, -u, v, e])
    \State clauses.append([0, u, -v, e])
    \State clauses.append([0, u, v, -e])
    \State clauses.append([0, -u, -v, -e])
\EndFunction
\end{algorithmic}
\end{algorithm}
\begin{algorithm}
\begin{algorithmic}
\Function {clauses2WCNF}{clauses,v}
\Comment{Convert SAT clauses to WCNF text format}
\State wMax := len(clauses)
\State WCNF := list()
\State WCNF.append(" ".join(["p wcnf", v, wMax, wMax+1])
\For{clause in clauses}
    \If{clause[0] == 0}
        \Comment Hard constraint - set weight to maximum
        \State clause[0] := wMax + 1
    \EndIf
    \State clause.append(0)
    \State WCNF.append(" ".join(clause))
    \State return "\textbackslash n".join(WCNF)
\EndFor
\EndFunction
\end{algorithmic}
\end{algorithm}

\end{document}